# Static and dynamic characteristics of protein contact networks

Susan Khor[1]

**Abstract** The principles underlying protein folding remains one of Nature's puzzles with important practical consequences for Life. An approach that has gathered momentum since the late 1990's, looks at protein hetero-polymers and their folding process through the lens of complex network analysis. Consequently, there is now a body of empirical studies describing topological characteristics of protein macro-molecules through their contact networks and linking these topological characteristics to protein folding. The present paper is primarily a review of this rich area. But it delves deeper into certain aspects by emphasizing short-range and long-range links, and suggests unconventional places where "power-laws" may be lurking within protein contact networks. Further, it considers the dynamical view of protein contact networks. This closer scrutiny of protein contact networks raises new questions for further research, and identifies new regularities which may be useful to parameterize a network approach to protein folding. Preliminary experiments with such a model confirm that the regularities we identified cannot be easily reproduced through random effects. Indeed, the grand challenge of protein folding is to elucidate the process(es) which not only generates the specific and diverse linkage patterns of protein contact networks, but also reproduces the dynamic behavior of proteins as they fold.

**Keywords:** network analysis, protein contact networks, protein folding

## 1. Introduction

Breaking the code underlying protein folding has remained an intellectually tantalizing puzzle as well as a problem of great practical significance. Everything a protein requires for correct folding under normal circumstances appears to be embedded in its amino-acid sequence (Afinsen 1973), although a minority rely on the aid of chaperone molecules to fulfill their destiny. Due to the large sizes that amino-acid sequences can take, a random search approach to protein folding is deemed infeasible for practical biological purposes (Levinthal 1969). However, an argument based on separability of the protein folding problem, i.e. that the problem can be separated into parts which can be solved independently and assembled into an optimal solution[2], has been conceived as a way out of Levinthal's paradox (Zwanzig *et al* 1992; Karplus 1997). This argument is supported by the observation that some sections of protein sequences have a propensity to fold to their secondary structures. In medicine, protein misfolding has been identified as a causative factor in diseases such as cystic fibrosis, ALS and Alzheimer's (Chen *et al* 2008). Proteins play many roles in biological cells, e.g. as structural material, catalysts, adaptors, hormones, transporters and regulators (Tramontano 2006). Proteins attain their functionality through their unique (though not necessarily static) native three dimensional states and are the ultimate expression of genes. Thus, the ability to predict the three dimensional structure of a protein sequence is useful to comprehend genomic data. Current approaches to the protein structure prediction problem such as comparative modeling, fold-recognition methods (profile-based and sequence threading) and fragment-based methods rely heavily on existing knowledge of protein sequences and their folds (Tramontano 2006). Notwithstanding the success of these methods, understanding the protein folding process from first principles is a more complete and satisfying solution, and may prove invaluable for protein design and therapies targeting protein misfolding.

[1] Contact info.: slc.khor@gmail.com

[2] For a more whimsical description, see Herbert A. Simon's parable of the two watchmakers in The Architecture of Complexity in *The Sciences of the Artificial*, 1969 MIT Press.

In general, protein folding is a process that occurs in stages. What essentially begins as a linear hetero-polymer (organized as a backbone with protruding side chain groups) obtains local structure in the form of secondary alpha helices and beta sheets and finally global structure as the secondary structures arrange themselves compactly in three dimensions. For a long time, this spontaneous biological self-organization has been attributed to various inter-atomic physical forces and chemical constraints impacting a protein molecule. However, in the last decade or so, another theory based on the network topology of a protein's native state has blossomed. Alm and Baker (1999) even suggest that "protein folding mechanisms and landscapes are largely determined by the topology of the native state and are relatively insensitive to details of the interatomic interactions." In this other theory, a network view of protein molecules (mostly in their native states) is adopted. The general recipe to transform a protein molecule into a network is to represent amino acid residues ($C_\alpha$ or $C_\beta$) as nodes, and contact (spatial, non-covalent) distances between pairs of amino acid residues below a certain threshold as links. Such *protein contact networks* or *maps* (Vendruscolo *et al* 2002) are constructed from the Cartesian coordinates of amino acid residues of protein molecules stored in the Protein Data Bank (PDB) (Berman *et al* 2000).

By examining protein contact networks (PCN), researchers have compiled a list of topological characteristics shared by a diverse (in terms of structural class, homology and taxon) set of proteins and speculated on the reasons for the observed topological characteristics in relation to the protein folding mechanism. For example, a common feature of protein contact networks is their small-world nature, i.e. they have lattice like clustering coefficients but random graph like diameters and characteristic path lengths (Watts and Strogatz 1998). The need for rapid communication between amino acid residues to facilitate *interaction cooperativity* crucial for protein folding is frequently cited as the reason for the small-world feature of PCNs (Vendruscolo *et al* 2002; Dokholyan *et al* 2002; Atilgan *et al* 2004; Del Sol *et al* 2006). Protein contact networks are also reported to exhibit high assortativity values which can be related to protein folding speeds (Bagler and Sinha 2007).

In this paper, we collected the PCN structural characteristics that have been reported in literature on various sets of proteins and using different methods of PCN construction, and applied them on a single set of proteins – the GH64 dataset – with a single PCN construction method. We find that in general, the reported network characteristics hold for GH64. We also carry a line of inquiry found in the literature (Green and Higman 2003) of probing the effects of short-range links and of long-range links on network characteristics of PCNs throughout the paper. Our investigations led us to new research questions, and to new quantitative and qualitative observations about PCNs which could be helpful to finally realize a network approach to protein folding. These new observations include: about 30% of all links are long-range, a power-law like distribution of link sequence distance, and for proteins which are not too large, the maximum link sequence distance is about 70% of a protein's size. However our model of a network approach to protein folding is far from complete as the problem of choosing endpoints for long-range links remains elusive (about 60% of a protein's amino acids are involved in at least one long-range interaction).

## 2. Protein Contact Map (PCN)

A protein contact network or *PCN* (Vendruscolo *et al* 2002) is an adjacency matrix representing a network whose nodes are the Cα atom of amino acid residues of a protein, and where a link is placed between a pair of nodes when the node pair is situated less than some distance threshold (TH) apart from each other. Distance between node pairs is the Euclidean distance between their 3D Cartesian coordinates obtained from the Protein Data Bank or *PDB* (Berman *et al* 2000). For this paper, TH is 7Å. Structural characteristics of PCNs do not seem overly sensitive to the choice of the threshold value (Bartoli *et al* 2007). We build a PCN for each protein in the GH64 dataset. Examples of four such PCNs can be found in Figure 1. The proteins were randomly chosen for their differences in size: 1aep, 1agd, 1psd and 1CVJ have 153, 385, 813 and 1261 nodes in their PCNs respectively.

So far, we have referred to PCNs as uniform entities. In actuality, there are variations in the way PCNs are constructed. For example, a single PCN may represent several non-homologous proteins rather than a single protein. PCNs may also represent different aspects (e.g. surface or core), states (e.g. native or transitional), structural classes (e.g. *α*, *β*, *α*+*β* or *α*/*β*), or types (e.g. globular or fibrous) of proteins. Further, the nodes and links of PCNs may carry different meanings, e.g. atoms of the side chain group of an amino acid may be included so that a node may represent more than one atom and multiple links between nodes or weighted links are allowed. Examples of these alternative PCN constructions can be found in the references for this paper. Incidentally, an earlier use of a matrix to depict protein folding can be found in (Tanaka and Scheraga 1977).

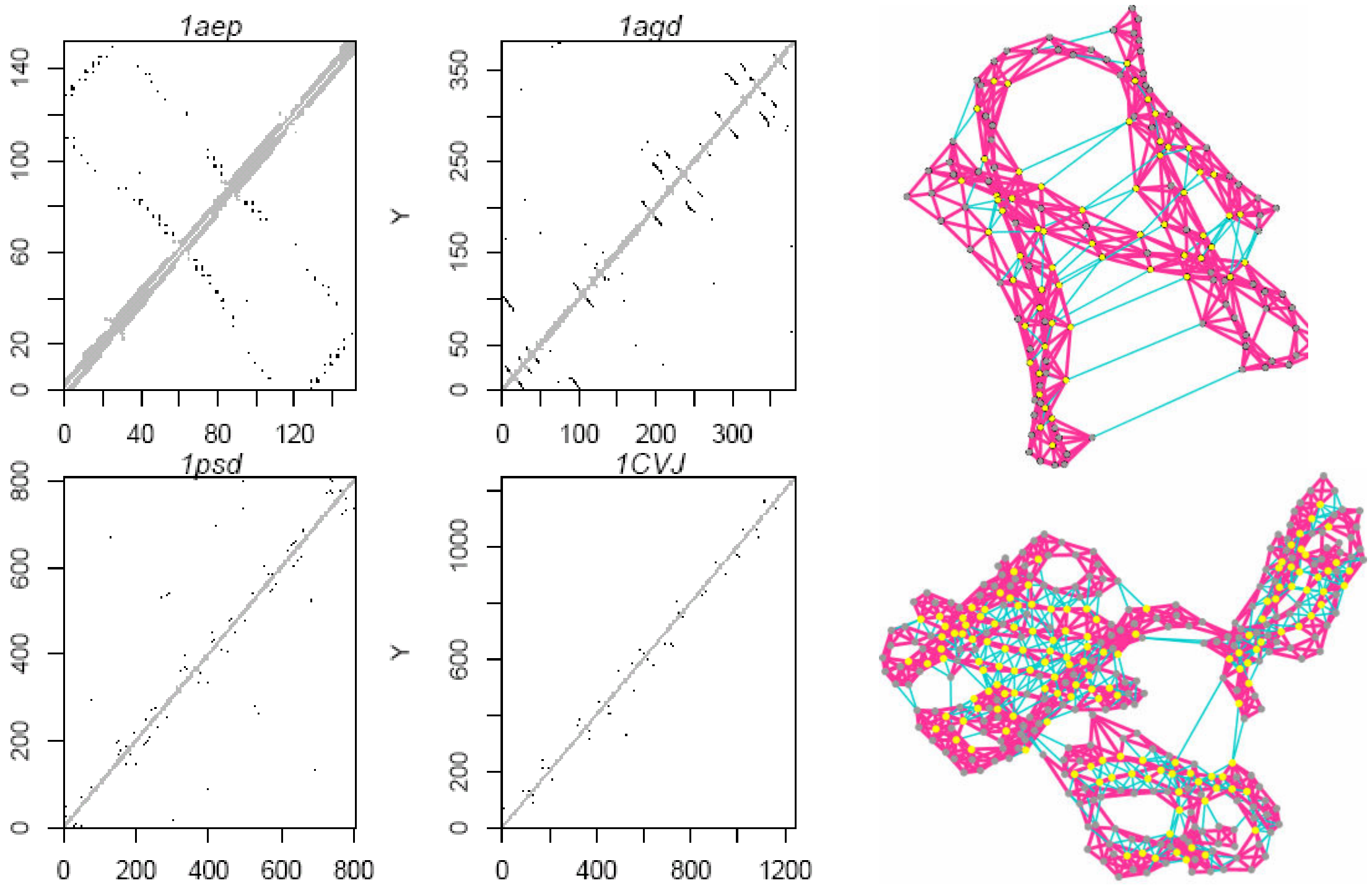

**Figure 1 Left: Protein Contact Networks (7Å) for four proteins in the GH64 dataset. A shaded cell (*x*, *y*) denotes a link between nodes *x* and *y*. Right: Networks produced from two of the PCNs: 1aep (top) and 1psd (bottom). SE links are pink, LE links are turquoise, yellow highlights nodes with more than 8 direct neighbours. Sections 4 and 5 of the text provide further explanation.**

## 3. The GH64 Protein Dataset

The 64 proteins in the GH64 dataset (Table 1) were selected from literature surveyed, specifically (Green and Higman 2003). The dataset encompass proteins from different protein classes, fold types and branches of life. Their sizes range from 69 to 1734 residues (Figure 2). Proteins which did not form a single component (1cuk and 1ho4), or had unusually high link density (1feo) in the PCNs we constructed were excluded from the dataset. So too were proteins with more nodes in their PCNs than their DSSP output (Kabsch and Sander 1983) (2hmz and 1epf). For the reverse situation, we truncated the DSSP output. Another protein dataset, EVA132, is described in Appendix A and it is used as a check to increase confidence in the main observations made in this paper.

**Table 1 Proteins in the GH64 Dataset**

| |
|---|
| 1mjc 1gvp 1ten 1ris 2acy 1tlk 1ayc 1sha 1CD8 1d4t 1e86 2fgf 1eif 1pdo 1h7i 1amx 1bj7 1aep 1gm6 |
| 3rab 1wba 1rbp 1eyl 153L 1fap 1nsj 1hro 1jr8 256b 1ICE 1arb 1vlt 1urn 1amp 1j8m 1cjl 1beb 1OBP |
| 1b7f 1hng 1agd 1aye 1g4t 1eov 1bmt 7tim 1ce7 1hwn 2AAI 1fbv 1bf5 1jly 1dar 1eun 1rpx 1bbp 1bih |
| 1psd 1b8a 1ava 1CVJ 3eca 3kbp 1dio |

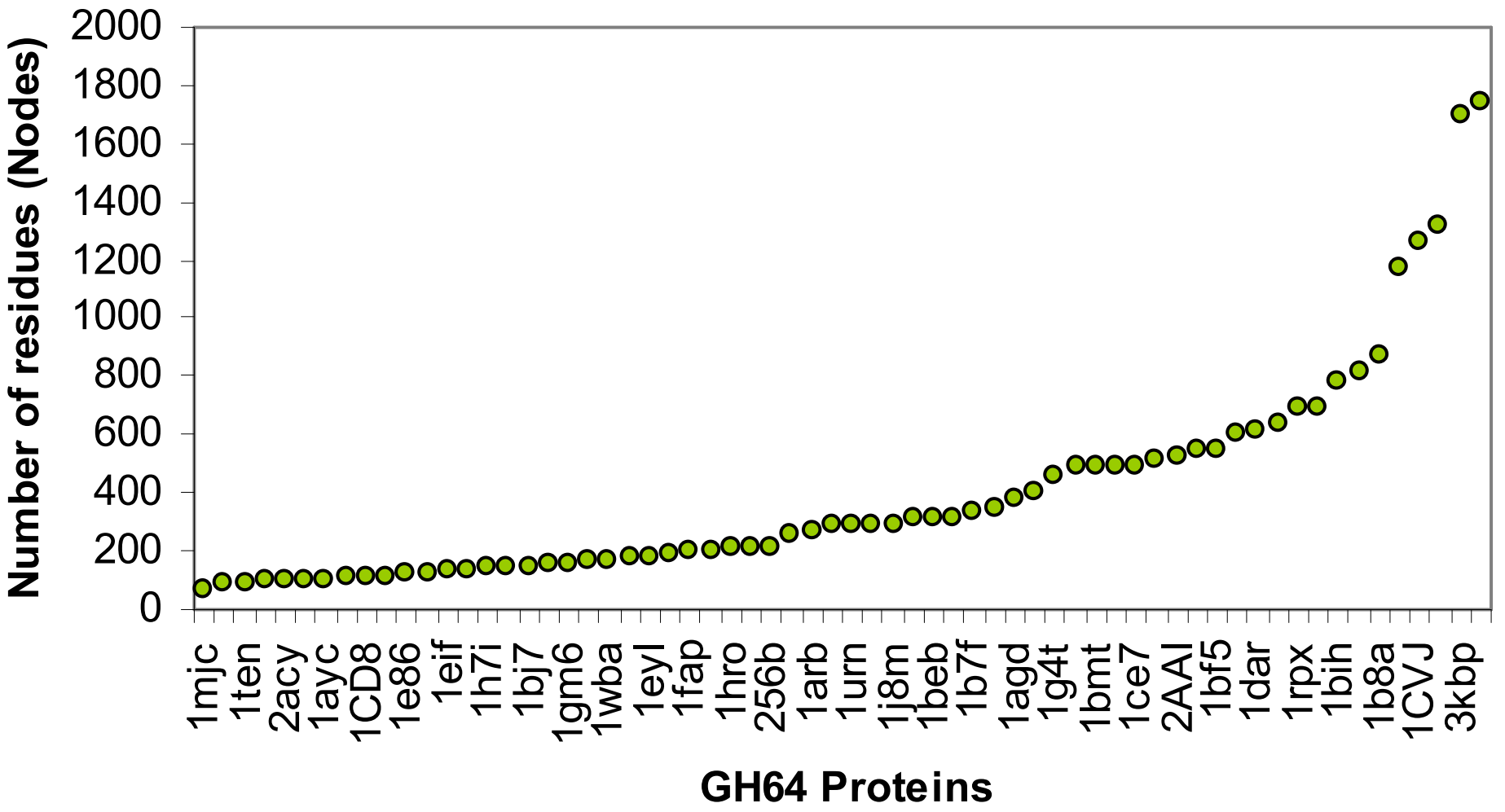

**Figure 2 Size of GH64 PCNs in terms of the number of Cα atoms documented in their PDB files.**

## 4. Links: short-range (SE) and long-range (LE)

A link is placed between a pair of nodes representing the Cα atom of amino acids when the node pair is situated less than 7Å apart from each other (section 2). We partitioned the set of links in a PCN into two sets: long-range links (*LE*) and short-range links (*SE*). A link between nodes *x* and *y* is classified as *long-range* if their absolute distance on the amino acid sequence chain is more than some distance threshold (LE_TH). Long-range links connect amino acids which are far apart from each other in the primary structure but are in close spatial proximity in the tertiary structure. In this study, LE_TH is 9 (Green and Higman 2003), although we also consider LE_TH = 4. Unless otherwise stated, LE links refer to long-range links with LE_TH = 9. Figures 3(a)-(c) examine various aspects of PCN links in the GH64 dataset. Similar observations are made on the EVA132 dataset (Appendix A).

The number of links M in a PCN increases linearly with protein size (number of nodes in a PCN) (Figure 3a). The number of long-range links |LE|, regardless of LE_TH, also increases linearly with protein size, albeit at a slower rate (Figure 3a). For comparison, we cut in half the sequence distance threshold for long-range links (LE_TH) from 9 to 4 (Figures 3a &b). On average, 30% of all links in a PCN are LE (when LE_TH = 9, mean |LE| / M = 0.2929 with standard deviation of 0.0834; when LE_TH = 4, mean |LE| / M = 0.3549 with std. dev. = 0.1005).

Instead of the entire PCN, Gaci and Balev (2009) considered the subgraph induced by the set of amino acids participating in secondary structures (links belonging to the less evolutionary conserved loop regions between SSEs are excluded). They call such a subgraph Secondary Structure Elements Interaction Network or *SSE-IN*. Gaci and Balev (2009) observed that the ratio of the number of interactions between secondary structures to the total number of interactions in SSE-INs does not exceed 20%.

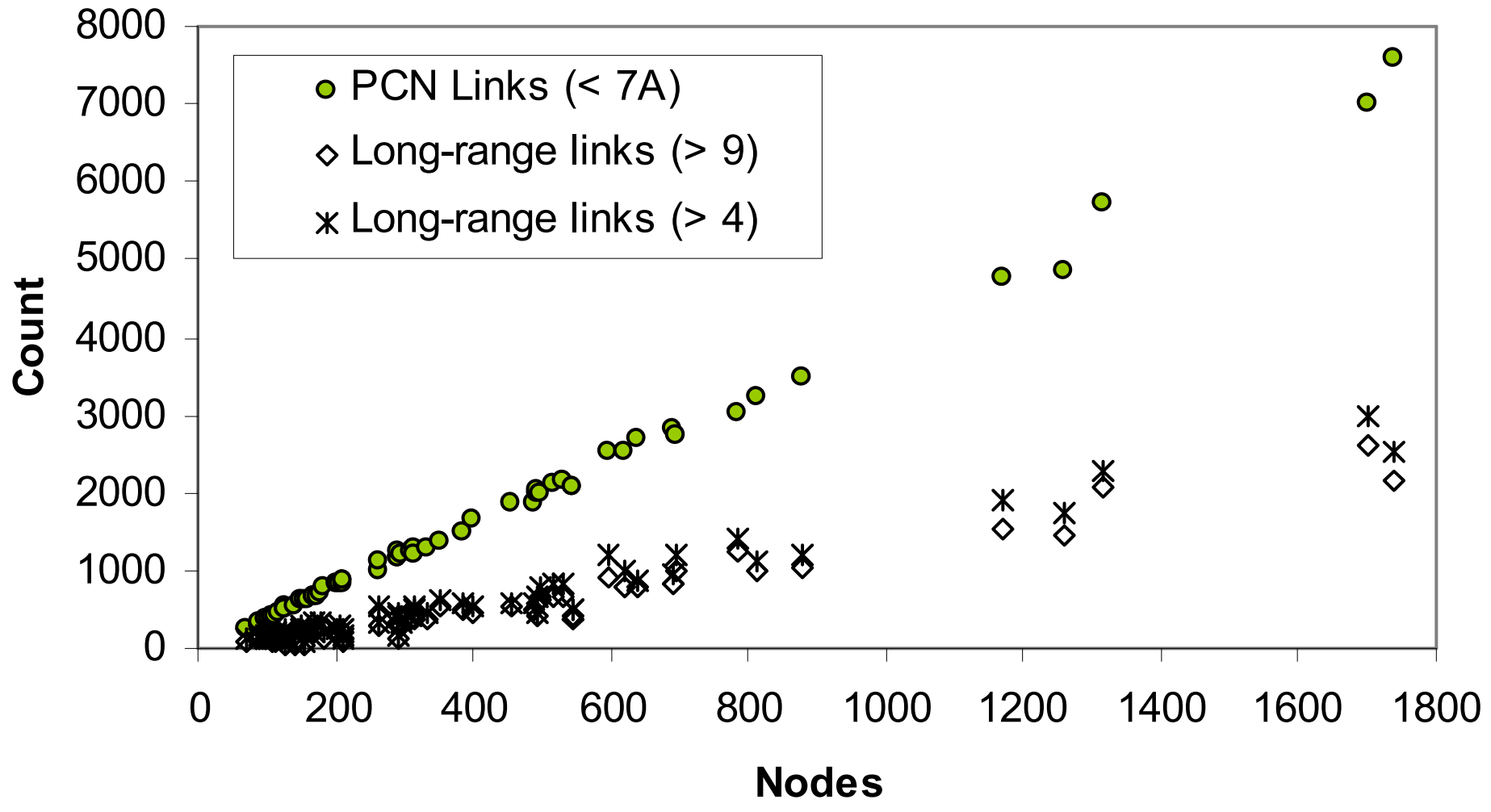

**Figure 3a Link count M, by protein size.**

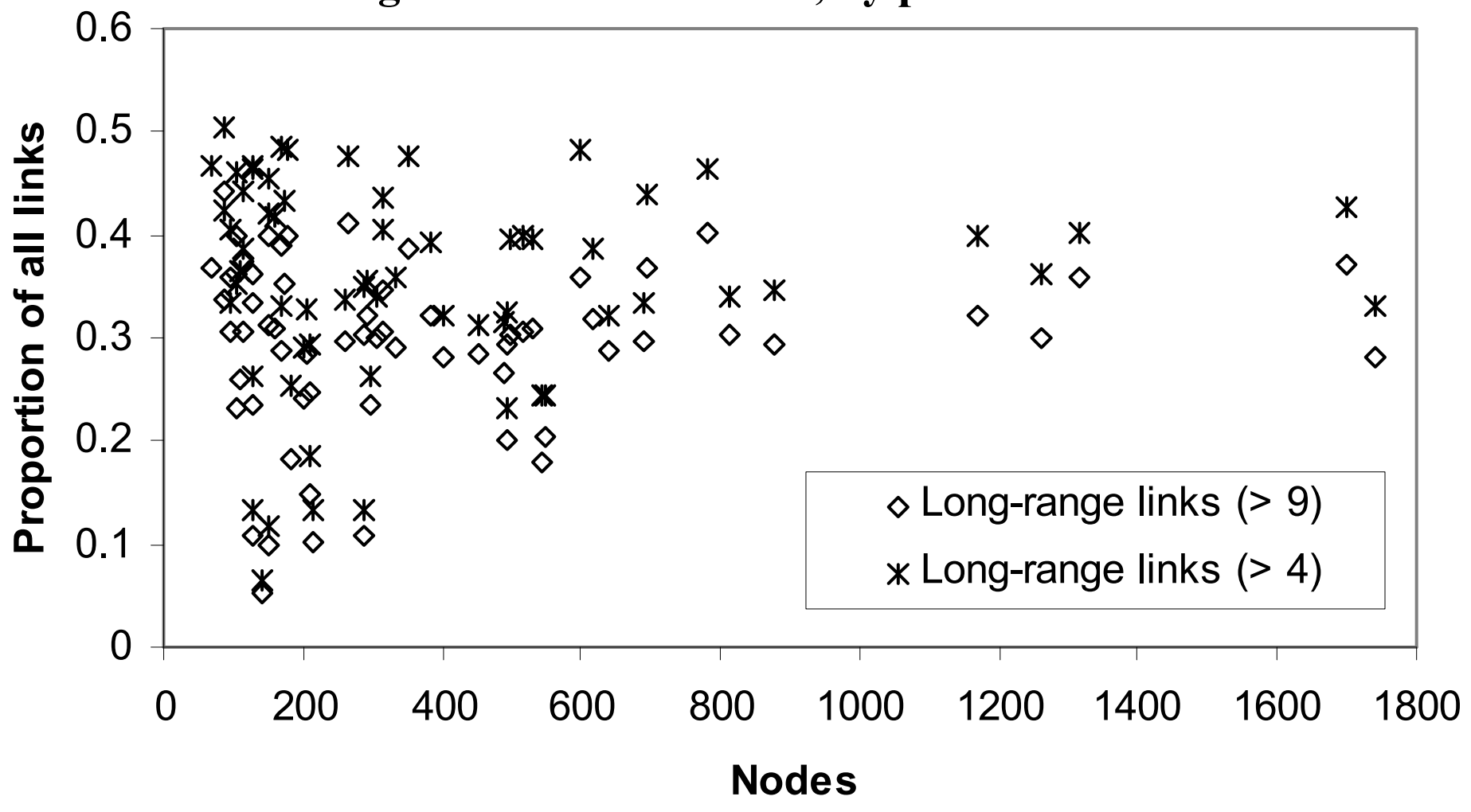

**Figure 3b Long-range links as a proportion of all links (|LE|/M), by protein size. When LE_TH is 9, the fraction of links in LE averages at 0.2929 with a standard deviation of 0.0834. When LE_TH is 4, the average is 0.3549 with a standard deviation of 0.1005**

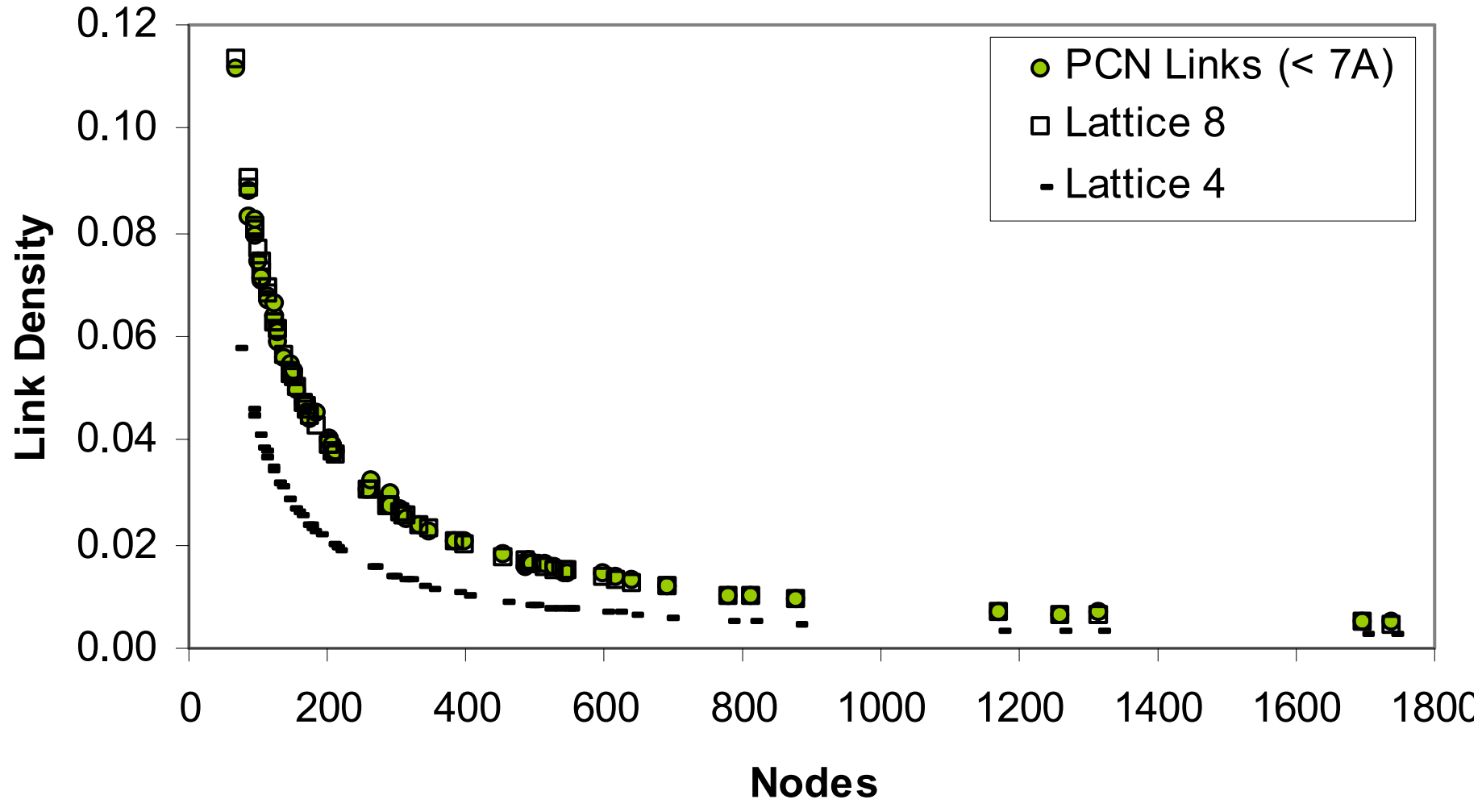

**Figure 3c Link density by protein size.**

The GH64 PCNs have low link densities, and link density decreases rapidly with protein size (Figure 3c). Link density is the fraction of actual links out of all possible links: 2M / [N (N-1)]. On a log-log plot, link density for PCN in Figure 3c straightens out into a line (section 9). For reasons which will become clearer in section 5, Figure 3c compares the link densities of GH64 PCNs with those of two linear lattices: Lattice8 and Lattice4. Lattice*V* is a linear lattice with *V*/2 nearest neighbours to the left and to the right where possible (*V*/2 nodes at each of the two ends of the lattice chain will have fewer links than the rest of the nodes in the middle which will have *V* links each). For a given network size (number of nodes), PCN link densities are indistinguishable from Lattice8 link densities (Figure 3c).

## 5. Node Degree and Degree Distribution

Typically, a node in a PCN represents the Cα atom of amino acids (section 2). The degree of a node is the number of links attached to the node. In the present context, node degree measures the number of contacts a node has in a PCN. Gaci and Balev (2009) remarked on the homogeneity of node degree in their SSE-INs. The mean degree of SSE-INs increases very slightly with protein size and falls within the range of 5 and 8. The absence of nodes with much higher degrees is attributed to the excluded volume effect which imposes a physical limit on the number of residues that can reside within a given radius around another amino acid.

Node degree distribution is the probability that a node has a particular degree. Node degree distributions for a variety of PCNs (i.e. PCNs constructed in different ways and from different protein sets) are bell-shaped, characteristic of a Gaussian distribution (Green and Higman 2003; Atilgan *et al* 2004; Bagler and Sinha 2005). Although, Green and Higman (2003) found that node degree distributions of PCNs which consider only long-range interactions exhibit if not scale-free tendencies, then exponential or highly right skewed distributions, irrespective of fold type. However, their PCN is unlike our PCN and other PCNs referenced in this paper. Each amino acid residue is still represented by a single node in the Green and Higman PCN. However, all atoms of an amino acid are considered when deciding on contact or link establishment, and multiple links between nodes are allowed. Therefore the Green and Higman PCN can register much larger node degrees, with reported modes of 100 (Green and Higman 2003, p. 783). Gaci and Balev (2009) also observed scale-free tendencies in the node degree distributions of their SSE-INs, with the caveat of a sharp cut-off above the mean node degree. Finding scale-free connectivity patterns in PCNs is attractive since such connectivity pattern has direct consequences on efficiency of information transfer and robustness to random attacks in proteins (Green and Higman 2003; Atilgan *et al* 2004).

Node degree statistics for PCNs in GH64 is presented in Figure4a. The mean node degrees are independent of protein size and hover around 8 (mean node degree averaged over all PCNs in GH64 is 7.9696 with a standard deviation of 0.3126). Little difference is observed between the corresponding mean node degrees and the median node degrees. The standard deviation values are about four times smaller than the corresponding mean node degrees. All these statistics point towards bell shaped degree distributions as reported in the literature, and as shown in Figure 4b for four proteins in GH64.

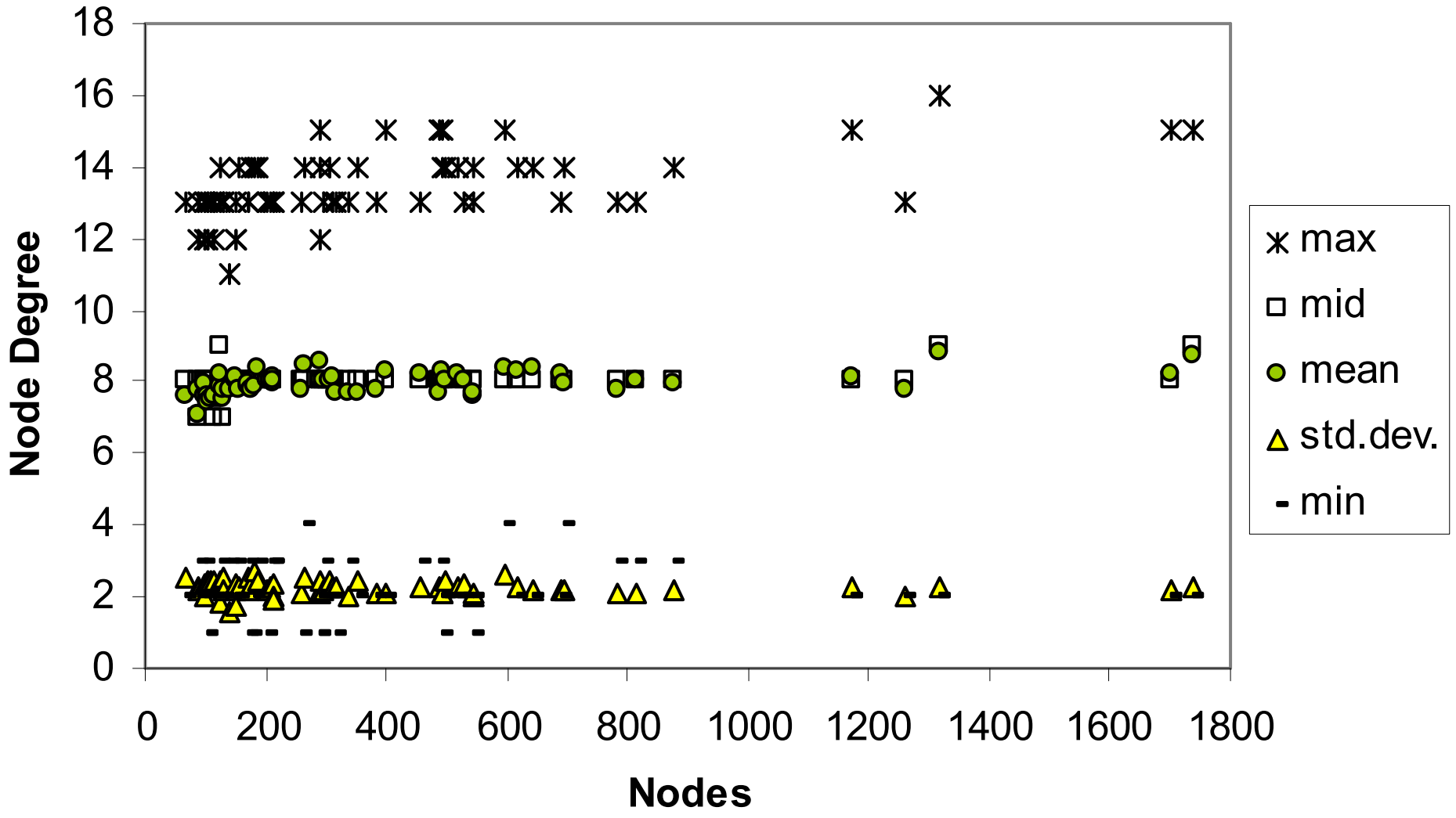

**Figure 4a Node degree summary statistics for GH64 PCNs. The average mean node degree (K) is 7.9696 with a standard deviation of 0.3126.**

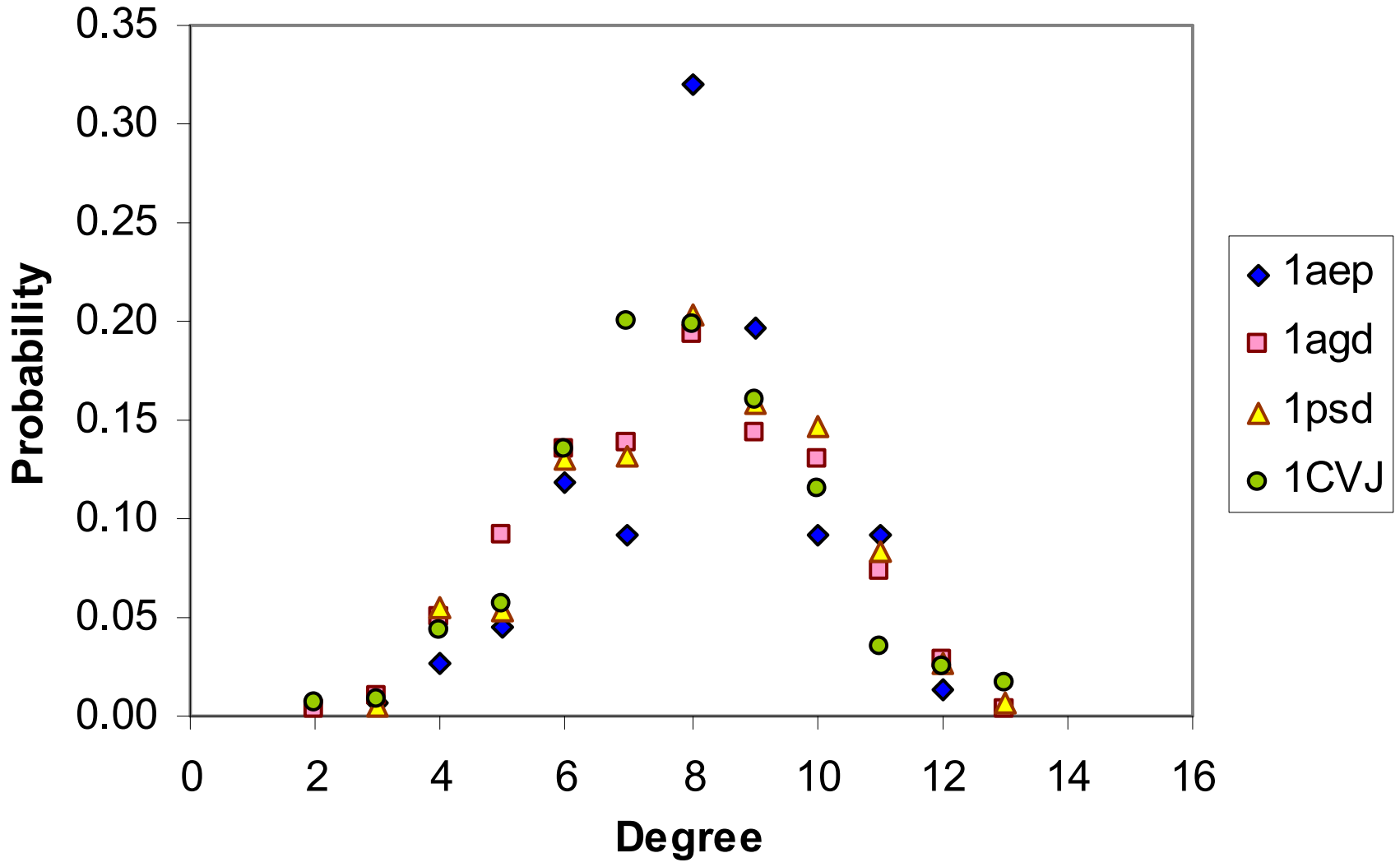

**Figure 4b Node degree distributions of four PCNs in the GH64 dataset.**

Since we have partition the set of links into SE and LE (section 4), we can similarly classify the set of nodes into two subsets (although these two subsets are not necessarily disjoint): *SE_nodes* for the set of nodes which are endpoints of links in SE, and *LE_nodes* for the set of nodes which are endpoints of links in LE. Figure 5a compares the sizes of these subsets and we find that LE_nodes is generally slightly smaller than SE_nodes for proteins in GH64. This is not entirely surprising since we expect almost all nodes to be in SE_nodes. For GH64, we found that the mean |SE_nodes| / N is 0.9970 with a std. dev. of 0.0039, and the mean |LE_nodes| / N is 0.6659 with a std. dev. of 0.1093 where N = number of nodes. Thus there is significant overlap between SE_nodes and LE_nodes, almost a subset

relationship. For GH64, we found that mean |SE_nodes ∩ LE_nodes| / N is 0.6628 with a std. dev. of 0.1103. Green and Higman (2003) found that between 5% to 25% of nodes within each of their PCNs do not have long-range contacts, i.e. are not in LE_nodes. For the GH64 PCNs, the mean (N - |LE_Nodes|) / N is 0.3341 with a std. dev. of 0.1093 (Figure 5b). Similar observations are made on the EVA132 dataset (Appendix A).

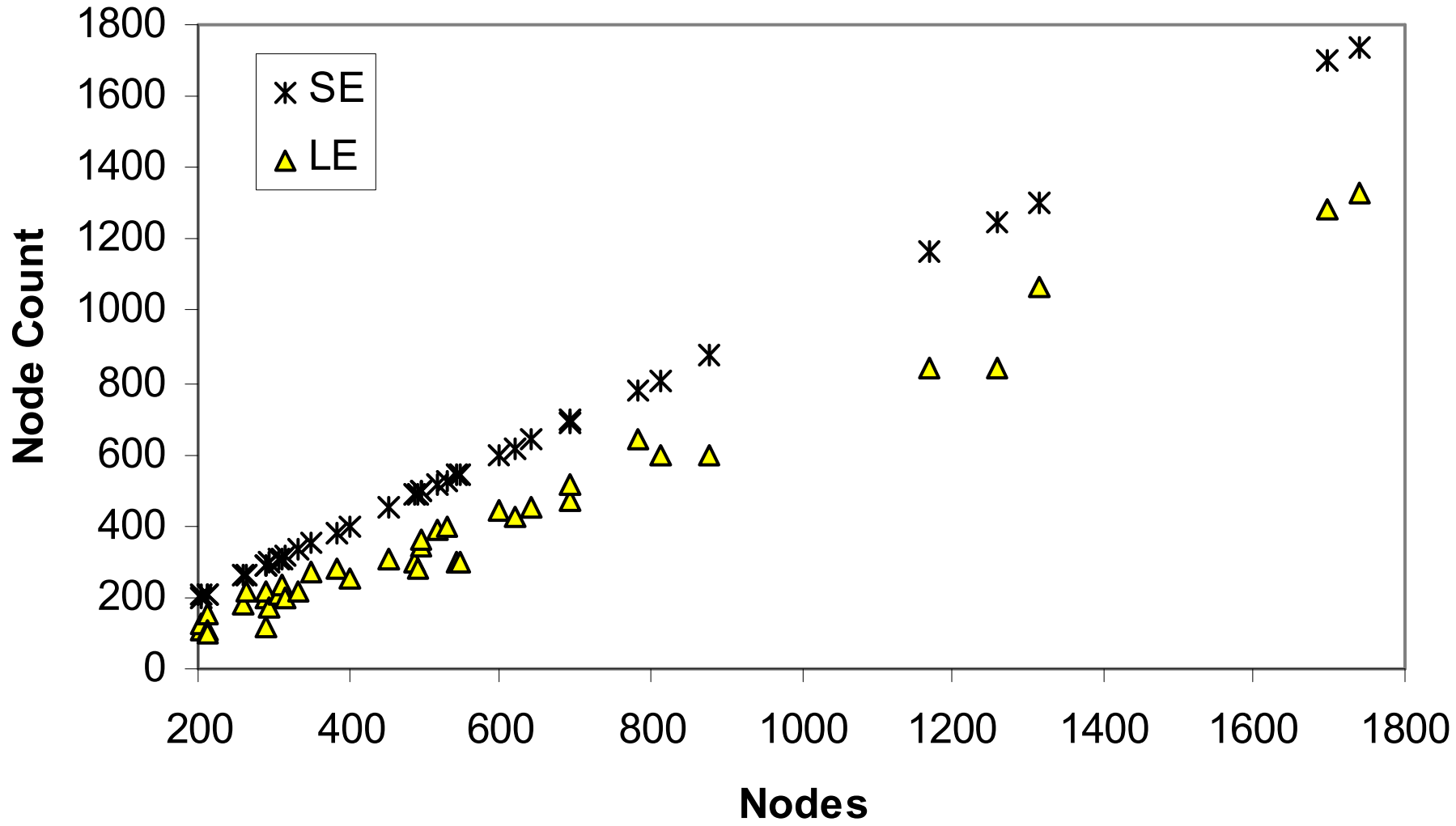

**Figure 5a Number of nodes which are endpoints for links in SE (SE_nodes), and number of nodes which are endpoint for links in LE (LE_nodes), by protein size for GH64. SE_nodes as a fraction of all nodes averages at 0.9970 with a standard deviation of 0.0039. LE_nodes as a fraction of all nodes averages at 0.6659 with a standard deviation of 0.1093.**

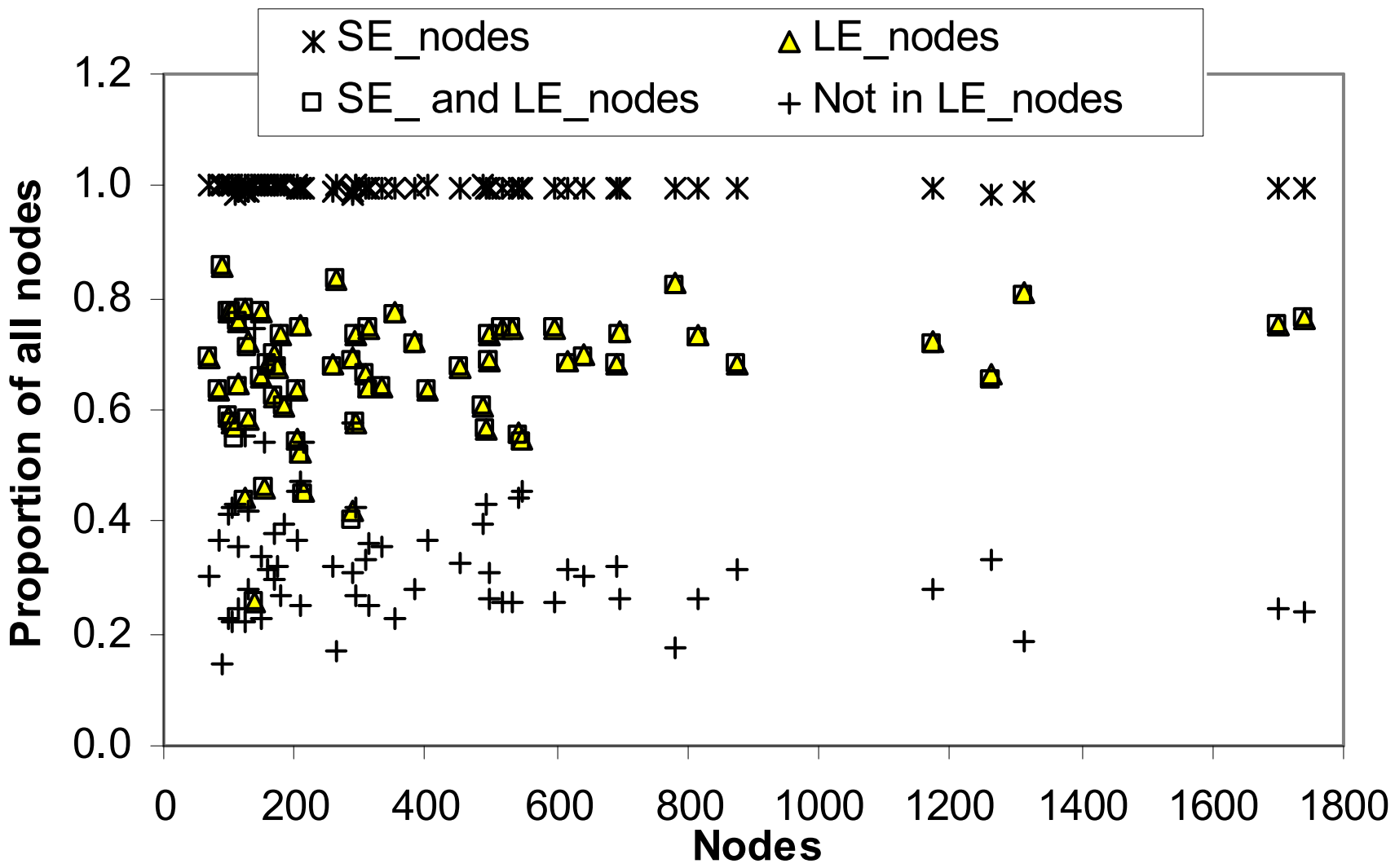

**Figure 5b Sizes of different subsets of nodes as a proportion of all nodes for GH64 PCNs. On average, 0.6628 ± 0.1103 of all nodes are in both SE_nodes and LE_nodes, while 0.3341 ± 0.1093 of all nodes are not in LE_nodes.**

Being able to distinguish LE_nodes for a protein without prior knowledge of its 3D structure could be an important step in predicting the 3D structure of proteins. Unfortunately, as we have observed above, more than half of the amino acids in a protein sequence are expected to have at least one long-range interaction (in the protein's native state). Perhaps a better strategy is to try to identify distinguishing characteristics of the 30% or so nodes which are not involved in long-range interactions.

What about the node degree characteristics of SE_nodes and of LE_nodes? We use two ways to do this comparison. First, we compute node degree statistics for nodes in SE_nodes (LE_nodes) using the node degree values from the PCN constructed with SE (LE) only. The result of this first comparison is displayed in Figure 6a. It yielded no significant difference at the 95% confidence level with the two-sided paired t test (p-value = 0.871) or with the two-sided paired Wilcoxon test (p-value = 0.3305). Nonetheless, we observed some differences between the degree distributions of PCN_SE (PCN constructed with SE only) and PCN_LE (PCN constructed with LE only). The narrow degree ranges of our PCNs limits our ability to observe long-tailed behavior over several scales (Figure 6b).

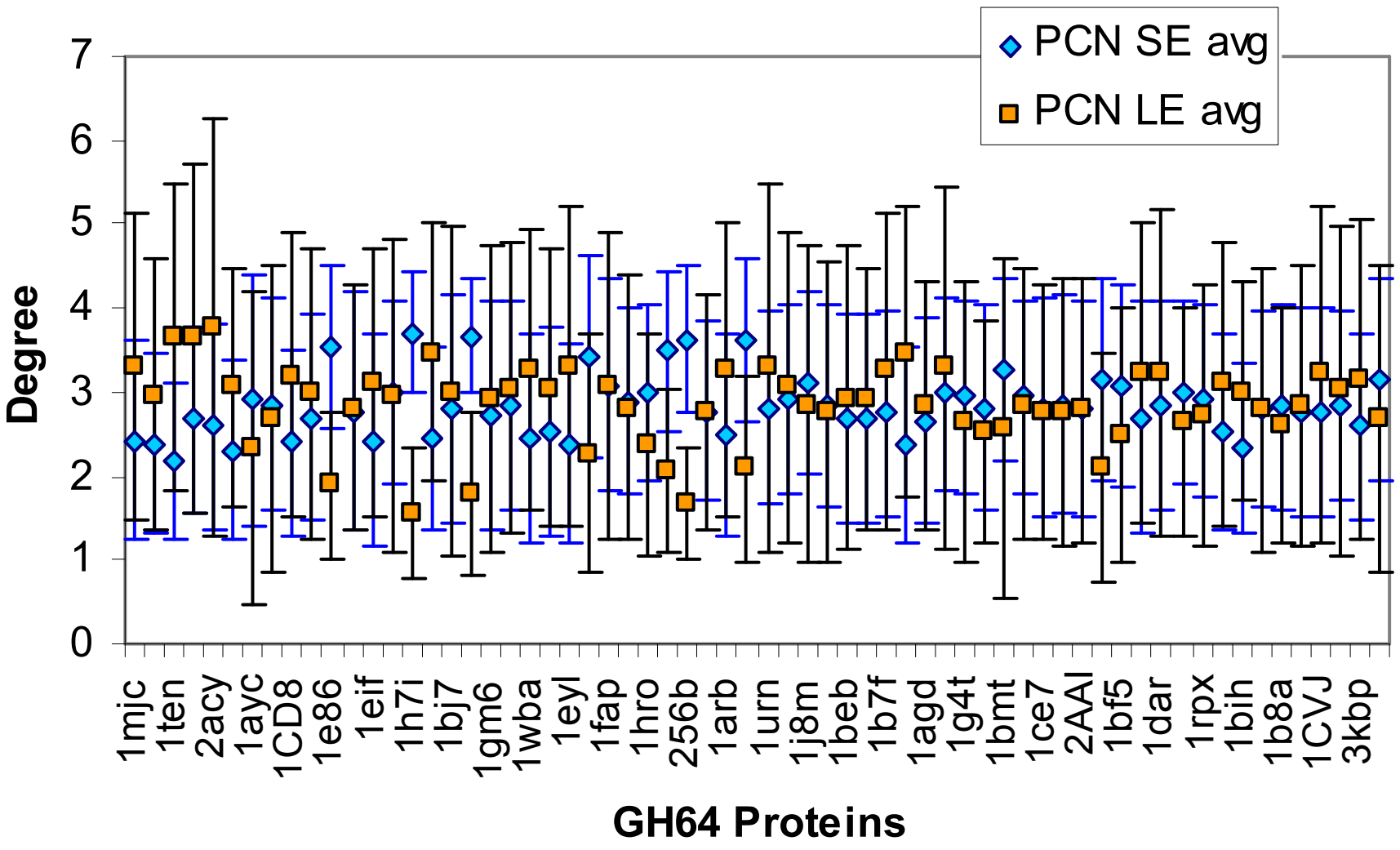

**Figure 6a Degree statistics for SE_nodes and for LE_nodes based respectively on PCNs constructed from short-range links (SE) only and from long-range links (LE) only. Error bars mark ± standard deviation.**

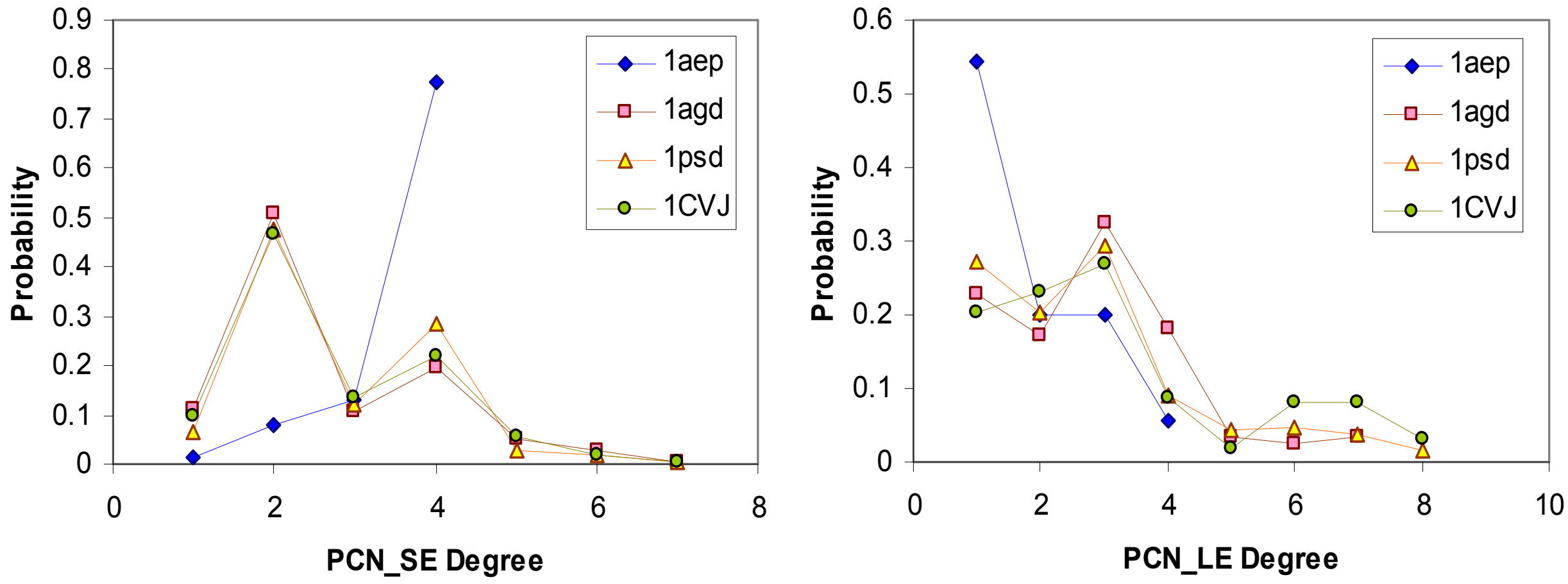

**Figure 6b Degree distributions of PCNs constructed with short-range (long-range) only links.**

Second, we compute node degree statistics for nodes in SE_nodes (LE_nodes) using the node degree values from the entire PCN. Figure 6c displays the result of this second comparison. According to the two-sided paired t test, and the two-sided paired Wilcoxon test, there is a significant difference at the 95% confidence level between the two vectors. However, given the high overlap between SE_nodes and LE_nodes, this difference though statistically significant, may be slight in magnitude.

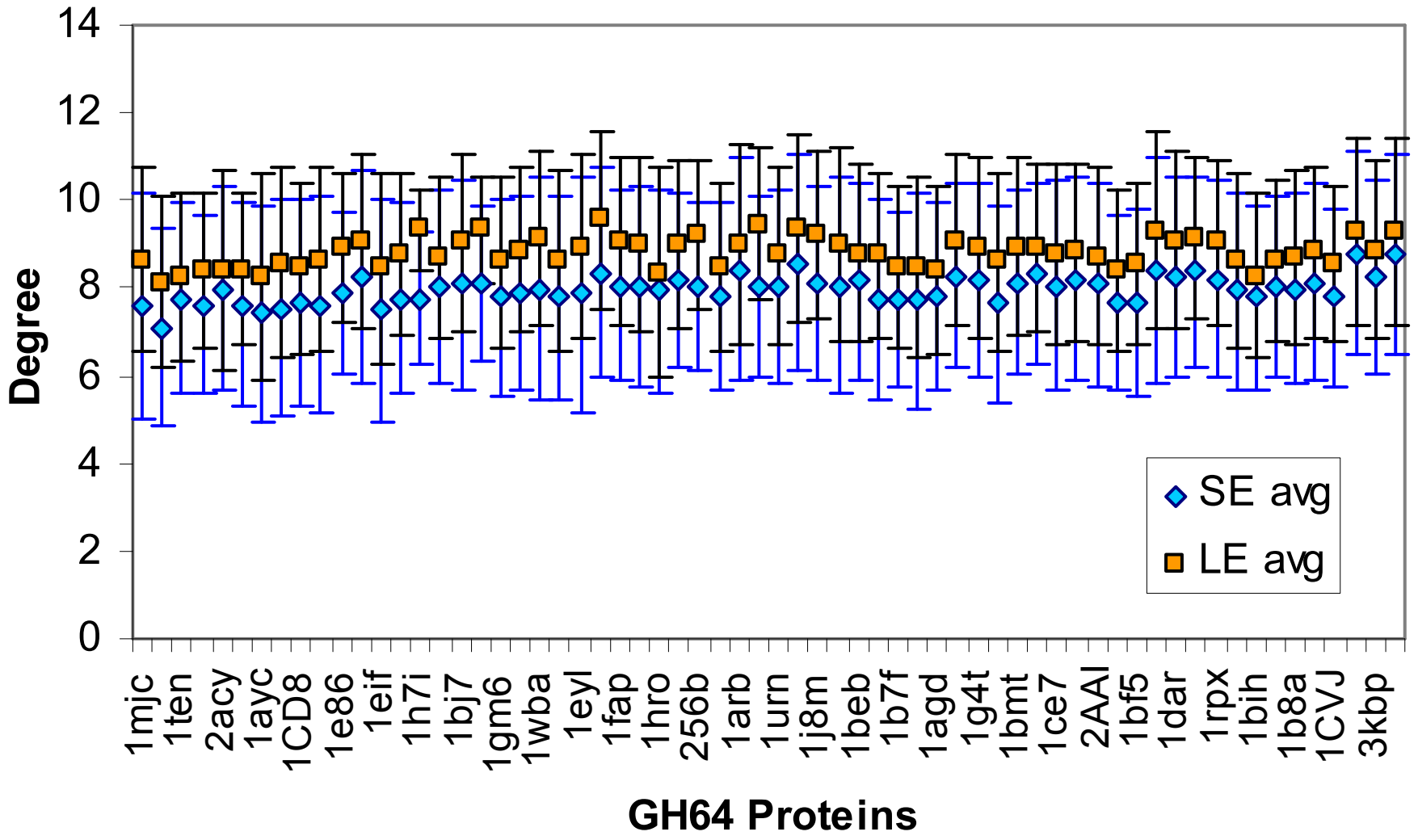

**Figure 6c Degree statistics for SE_nodes and for LE_nodes based on node degree values for the entire PCN. Error bars mark ± standard deviation.**

## 6. Clustering

Clustering or transitivity reflects the cliquishness of nodes in a network: if node X connects to node Y and to node Z, how likely is it that nodes Y and Z are connected to each other? There are different ways of measuring such transitive relationships, strongly indicated in graphs by the presence of triangles. One way is by taking the average clustering of all nodes in a network to yield the clustering coefficient: $C = \frac{1}{N}\sum_{i=1}^{N}\frac{2e_i}{k_i(k_i-1)}$ where $k_i$ is the degree of node $i$, and $e_i$ is the number of links that exist amongst the $k_i$ nodes (Watts and Strogatz 1998). Vendruscolo *et al* (2002) reports $C$ values of around 0.58 ± 0.04 for their PCN (which has a different construction than ours), and Bagler and Sinha (2005) found $C$ values of around 0.553 ± 0.027, irrespective of protein structural class as defined in SCOP (Murzin *et al* 1995).

For the GH64 PCNs, the $C$ values averaged at 0.5484 with a std. dev. of 0.0233, and they are independent of protein size. Figure 7a compares them with the $C$ values for Lattice8, Lattice4 and the theoretical $C$ values for regular ($C_{REGULAR}$) and random ($C_{RANDOM}$) networks of the same size (same number of nodes). $C_{RANDOM} \sim K / N$, and $C_{REGULAR} = 3 (K-2) / [4 (K-1)]$, where K is average degree and N is number of nodes (Watts 1999). The theoretical $C$ values for 1D regular networks ($C_{REGULAR}$) are indistinguishable from the $C$ values for Lattice8 (not surprising since mean degree of 8 is used). The $C$ values for GH64 PCNs are slightly smaller than $C_{REGULAR}$, and closer to the *C* values for Lattice4. This last observation is telling as SE includes protein backbone links, and the protein backbone is the main source of high clustering (Bartoli *et al* 2007). Further, Green and Higman (2003) observed that PCNs built with only long-range links (LE) have relatively lower *C* values than those for PCNs with all links. The same phenomenon is observed with GH64 proteins. Figure 7b shows the effect of randomizing PCN links. Unlike randomizing only long-range links (randLE), randomizing only short-range links (randSE) reduces clustering to levels similar to those of random graphs or randomizing all links (randAll).

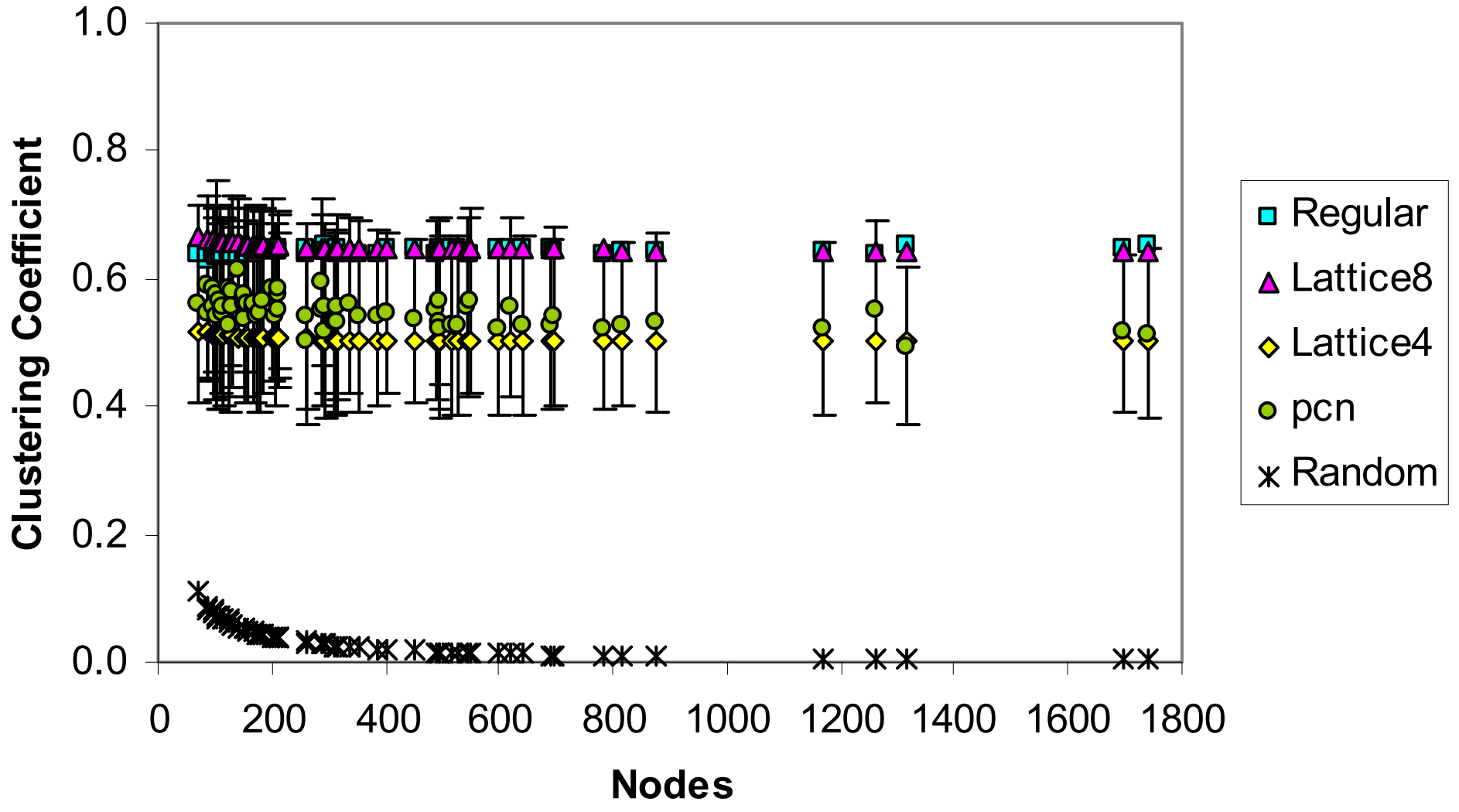

**Figure 7a Clustering coefficient values (± one standard deviation) for GH64 PCNs. Clustering coefficients average at 0.5484 with standard deviation of 0.0233.**

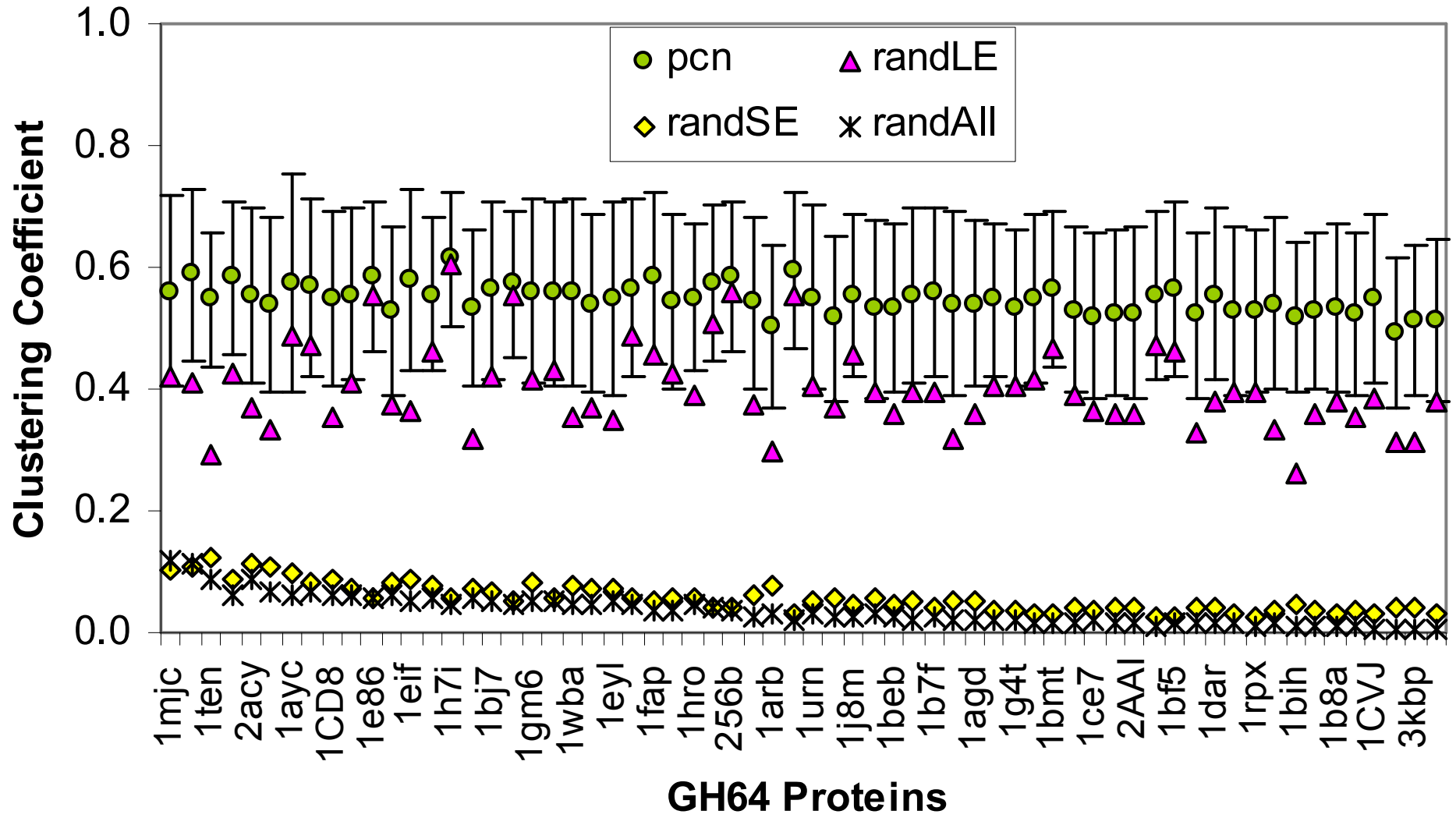

**Figure 7b Randomizing SE reduces clustering more than randomizing LE.**

Links are randomized in the usual manner (Maslov and Sneppen 2002), by rewiring nodes while preserving node degrees and without introducing multiple links between nodes. For example, to randomize the set of long-range links (randLE), two LE links *e1*(a, b) and *e2*(c, d) are picked uniformly at random with replacement from LE. If both *e3*(a, c) and *e4*(b, d) do not already exist in the network, then *e3* and *e4* replace *e1* and *e2*.

High clustering in PCNs is almost always discussed in terms of the small-world property (Watts and Strogatz 1998). Since the small-world property also involves the characteristic path length statistic (section 8), the importance of short inter-node distances to protein folding inevitably dominates the discussion. As a result, the role of high clustering (at the local interaction level) to protein folding is somewhat under explained. A small exception to this is Bagler and Sinha (2005; 2007). They suggest that high clustering in PCNs is indicative of a modular hierarchical organization as a result of the protein folding process, but called for further investigation into this hypothesis (Bagler and Sinha 2005). In (Bagler and Sinha 2007), they report a negative correlation between protein folding speed and high clustering levels in PCN_LE (PCNs that only take into account long-range links). However, they did not find any significant relationship between protein folding speed and clustering levels of entire PCNs.

Hence, the exact role clustering plays in protein folding is still an open question. Since most of the clustering comes from short-range links (SE), which comprise backbone and secondary structure links, it may be that high clustering of short-range links is necessary merely as a scaffolding device to create the “all-important” long-range contacts. Alternatively, it may be that the high clustering levels found in PCNs is a side effect of another network characteristic with a clearer relationship with some aspect of protein folding, for example assortativity and protein folding speed (section 7). Or, it could be that the local organization of short-range links really does play a role in some aspect of protein folding. In social networks, transitive relationships are assumed to be part of the nature of human social behavior, and the problem is to explain the establishment and subsequent efficient reuse of “random” short-cuts. An interesting example of such a study is Kleinberg (2000), although how far the analogy between human social networks and protein contact networks (or any other kind of residue “interaction” network) applies to protein folding remains to be seen.

## 7. Assortativity

Network assortativity refers to the extent that nodes associate or connect with their own kind. A common form of assortativity measured for PCNs is node degree. In this section (and throughout this paper), positive assortativity refers to the proclivity of nodes with small (large) degree to link with other nodes of small (large) degree. Using the method in (Newman 2002), Bagler and Sinha (2007) report degree-degree correlation coefficients up to 0.58, which is considered unusual for networks with biological origins (we discuss this point further ahead). Nonetheless, the positive assortativity values could be correlated in a positive manner to protein folding speeds (Bagler and Sinha 2007).

Similarly, we find that the GH64 PCNs have positive degree-degree correlations (Figure 8a). Their assortativity values average at 0.3387 with a standard deviation of 0.0536, and are independent of protein size. Figure 8b uses an alternative method (which is less sensitive to the effects of super-connected nodes) to assess degree assortativity (Pastor-Satorras *et al* 2001). In this other method, a positive correlation between degree *k* and the average degree of nodes directly neighbouring nodes of degree *k* is interpreted as evidence of positive degree-degree assortativity. In general, the plots in Figure 8b agree with the summary in Figure 8a. Although, the plots show some weakening (even reversal) in the relationship as node degree increases, which could be due to the limits on node degree (Bagler and Sinha 2007). This coupled with assortativity values close to those for Lattice4 (Figure 8a) is telling. Like clustering (section 6), assortativity is more sensitive to random rewiring of short-range links (randSE) than to random rewiring of long-range links (randLE) (Figure 8c). And when analyzed separately, long-range links (LE) show less positive assortativity than short-range links (SE). This means that like clustering, the high positive assortativity values of PCNs also mainly stem from SE.

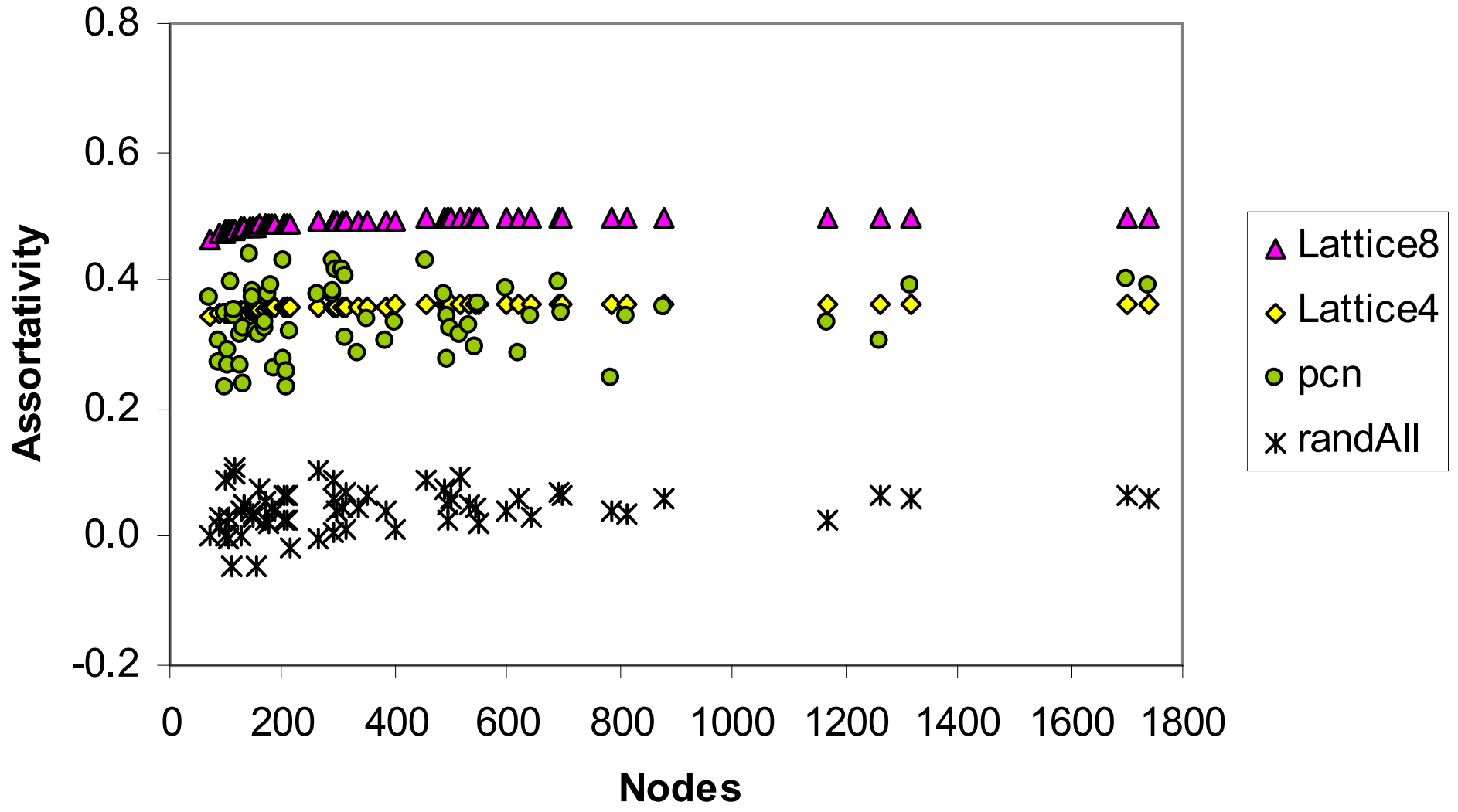

**Figure 8a The GH64 PCNs have positive degree assortativity, with values close to Lattice4. Mean Assortativity is 0.3387 with a standard deviation of 0.0536.**

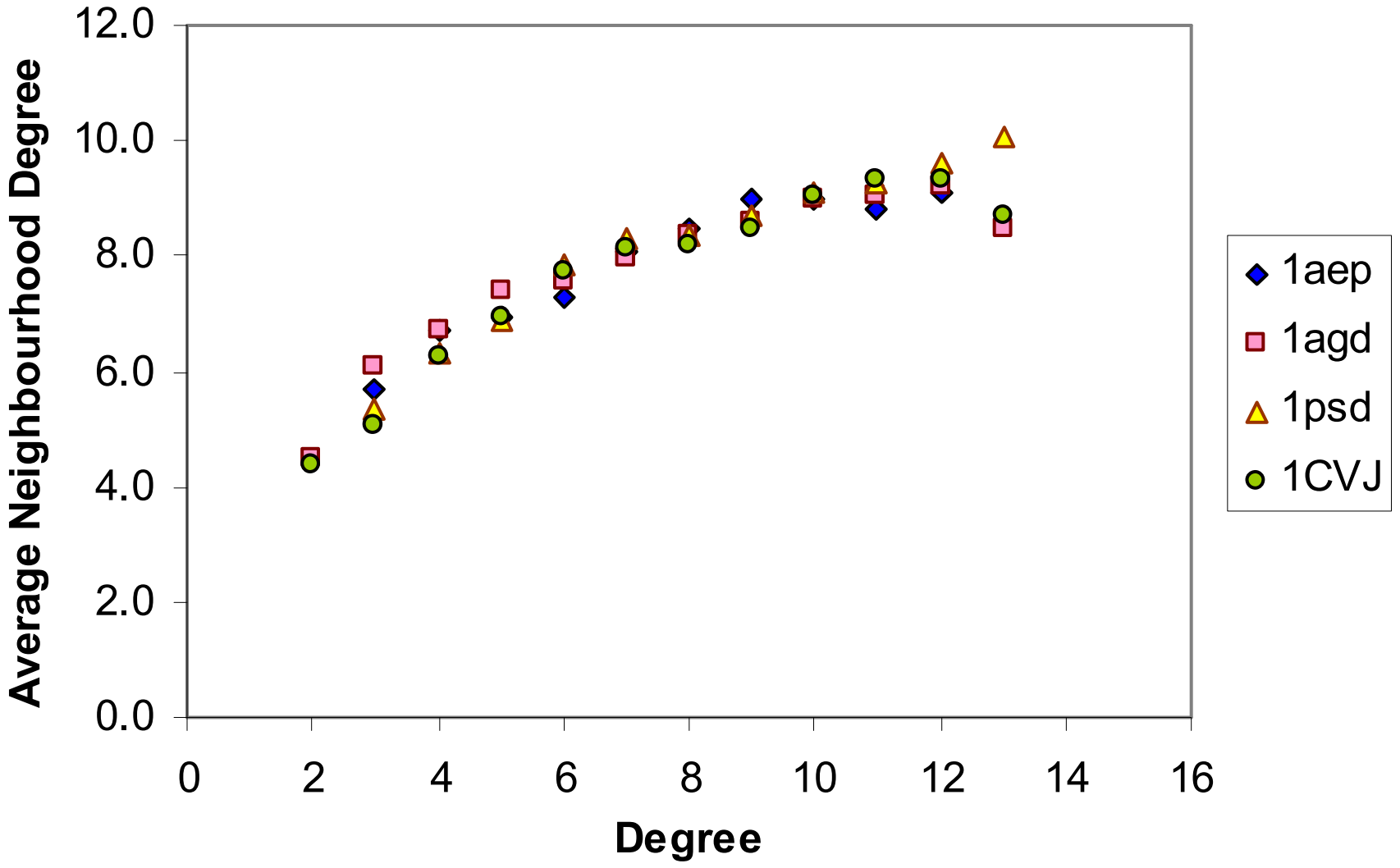

**Figure 8b Average neighbourhood degree for degree *k* is the average degree of all nodes adjacent to all nodes of degree *k*. A positive correlation implies positive degree-degree assortativity.**

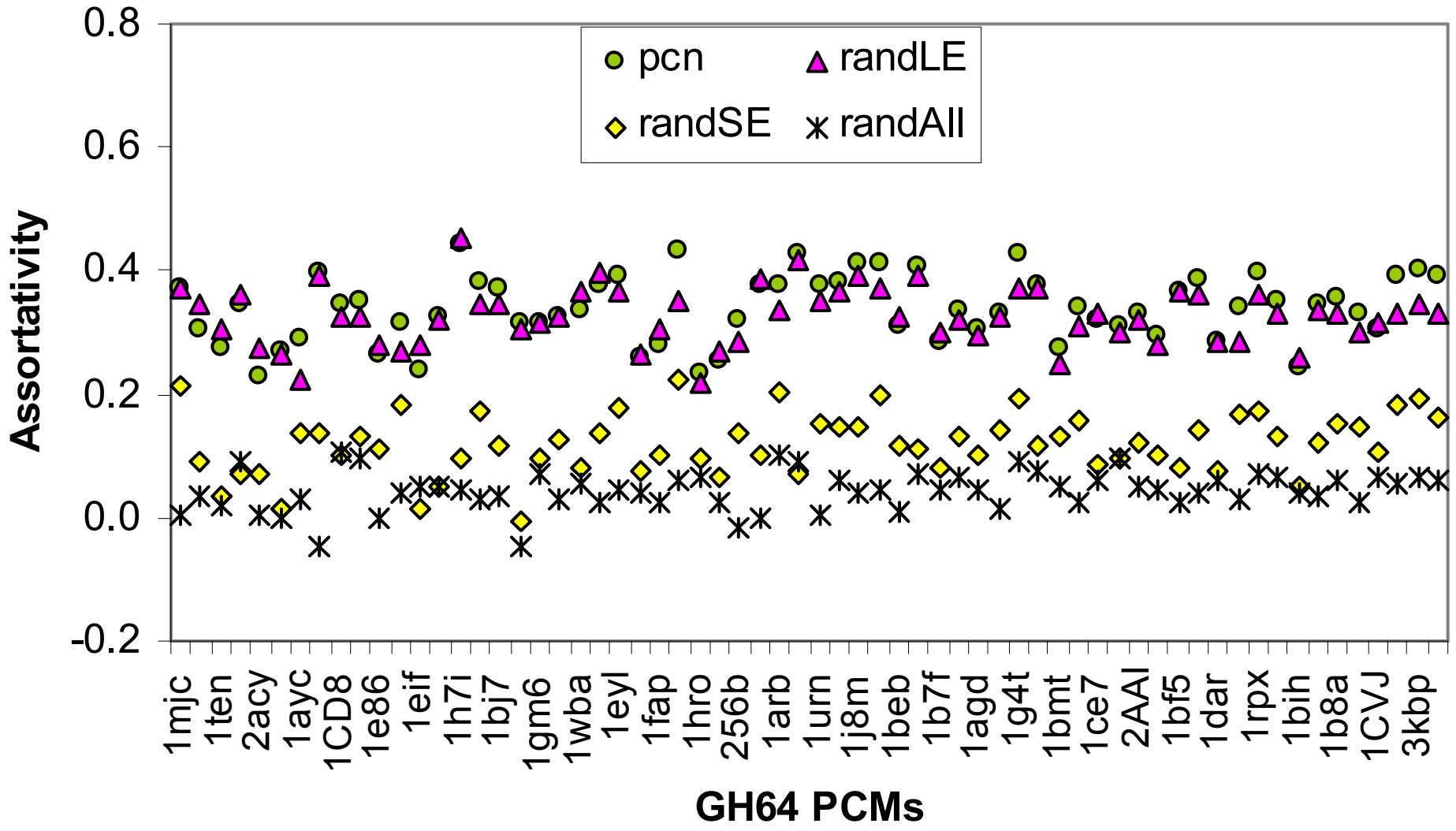

**Figure 8c Effect of randomizing links on Assortativity. randSE produces a larger effect than randLE.**

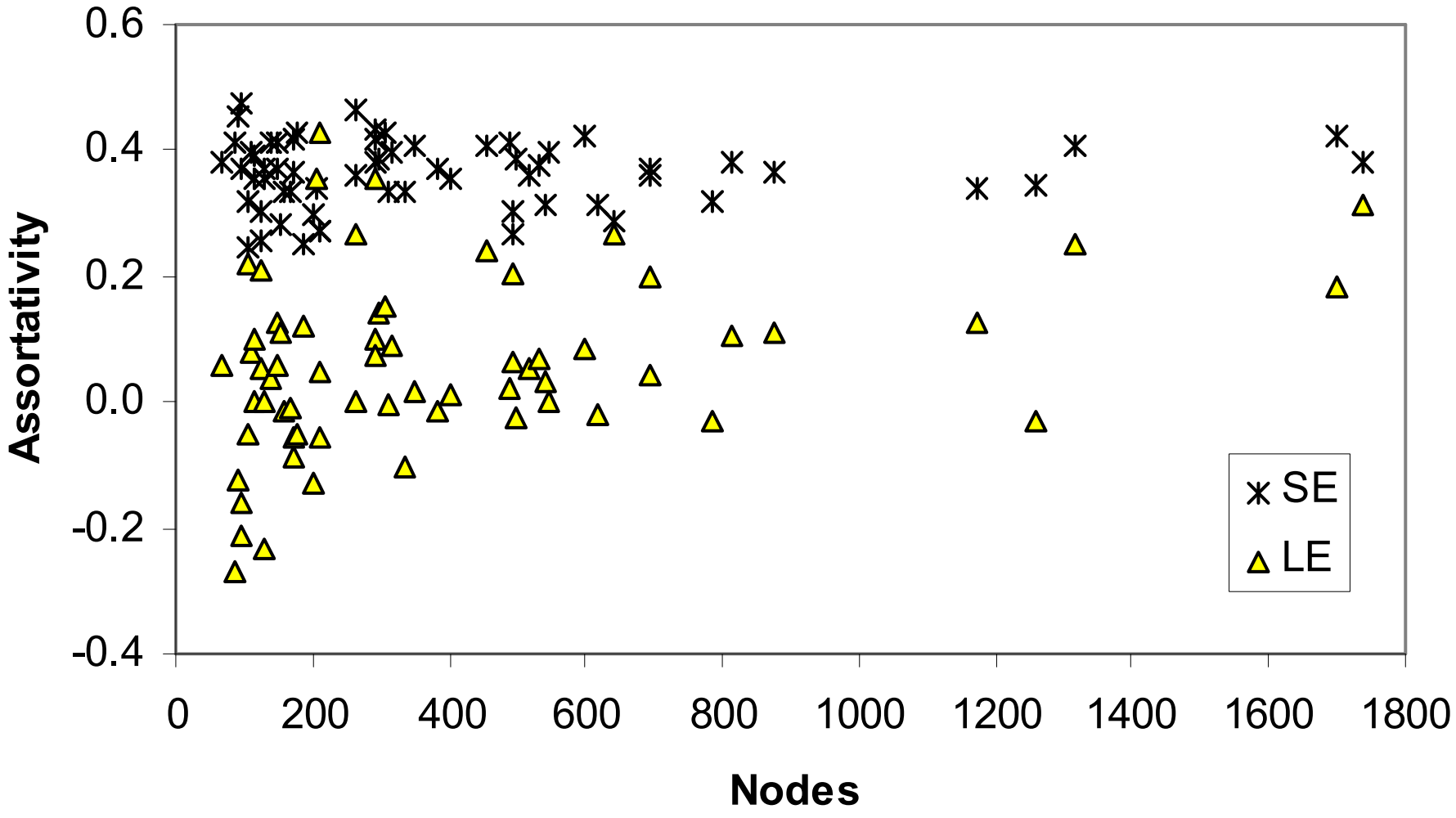

**Figure 8d Assortativity for long-range links (LE) and short-range links (SE).**

Previously, a number of biological networks including gene regulatory networks were described as disassortative, and it was suggested that this negative assortativity (specifically nodes of high degree or hubs are not directly connected with each other in the networks studied) is advantageous in the sense that the effects of harmful perturbations could be better localized (Maslov and Sneppen 2002). Brede and Sinha (2005) conclude that the stability of networks with assortative mixing by node degree (i.e. positive assortativity) declines more rapidly with increases in network size. A system with fragile stability is easily dislodged from its equilibrium state by small disturbances. However, the positive assortativity of PCNs may actually work to the benefit of the protein folding process by helping information to flow quickly and thus facilitate coordinated action crucial for correct and rapid protein folding. Further, native state proteins "adjust" themselves during ligand-binding, and thus some vulnerability to perturbations or marginal stability may be advantageous (Taverna and Goldstein 2002).

Indeed if percolation is important, Vazquez and Moreno (2003) conclude that node degree assortativity can make networks robust to random node or edge removals. Additionally, there could be some delegation of responsibilities between interacting biological networks in the sense that the amino acid chain is the product of gene translation and concerns about containing unwanted perturbations in a folding protein may be taken care of by the disassortative structure of genetic networks. Additionally, since LE is less assortative than SE (Figure 8d), there may still be some degree of isolation between parts (node subsets) of a PCN to buffer against unexpected perturbations. It appears that there are yet interesting discussions to be had on this topic.

## 8. Average Path Length

Average path length of a network is the average shortest path between all node-pairs. It represents the number of links that needs to be traversed on average when trying to move between nodes in a network. PCNs are known to have average path lengths characteristic of random graphs of the same size. For their respective protein sets and PCN constructions, Vendruscolo *et al* (2002) report an average path length of 4.1 ± 0.9, while Bagler and Sinha (2005) report an average path length of 6.88 ± 2.61. Path length is widely mentioned as a topological feature of PCNs that is highly relevant to protein folding (Vendruscolo *et al* 2002; Dokholyan *et al* 2002; Atilgan 2004; Del Sol *et al* 2006). The argument is: Short path lengths are conducive to rapid communication between amino acid molecules of a protein, which facilitates *interaction cooperativity* crucial for protein folding. We revisit this point elsewhere (manuscript submitted). Briefly, we find that shorter inter-node distances is no guarantee of finding a global optimum more easily if the shorter average path lengths is the result of more links and the fitness function is such that additional links increase frustration.

Figure 9a gives various path length statistics for GH64 PCNs. Both diameter (maximum shortest path length found) and average path length increase logarithmically as protein size (nodes) increases. When compared with average path lengths of other canonical networks, the average path lengths of PCNs are much shorter than regular graphs and approximate the average path lengths expected of random graphs of the same size (Figure 9b). Although the standard deviations indicate rather large dispersions of shortest path lengths in the PCNs, the average path length and median path length values are very close to each other (Figure 9a). We gather from this that there is little skew in the distribution of shortest path lengths for PCNs, and it is indeed the case. Figure 9c shows the Gaussian like distributions of shortest path lengths for four GH64 PCNs.

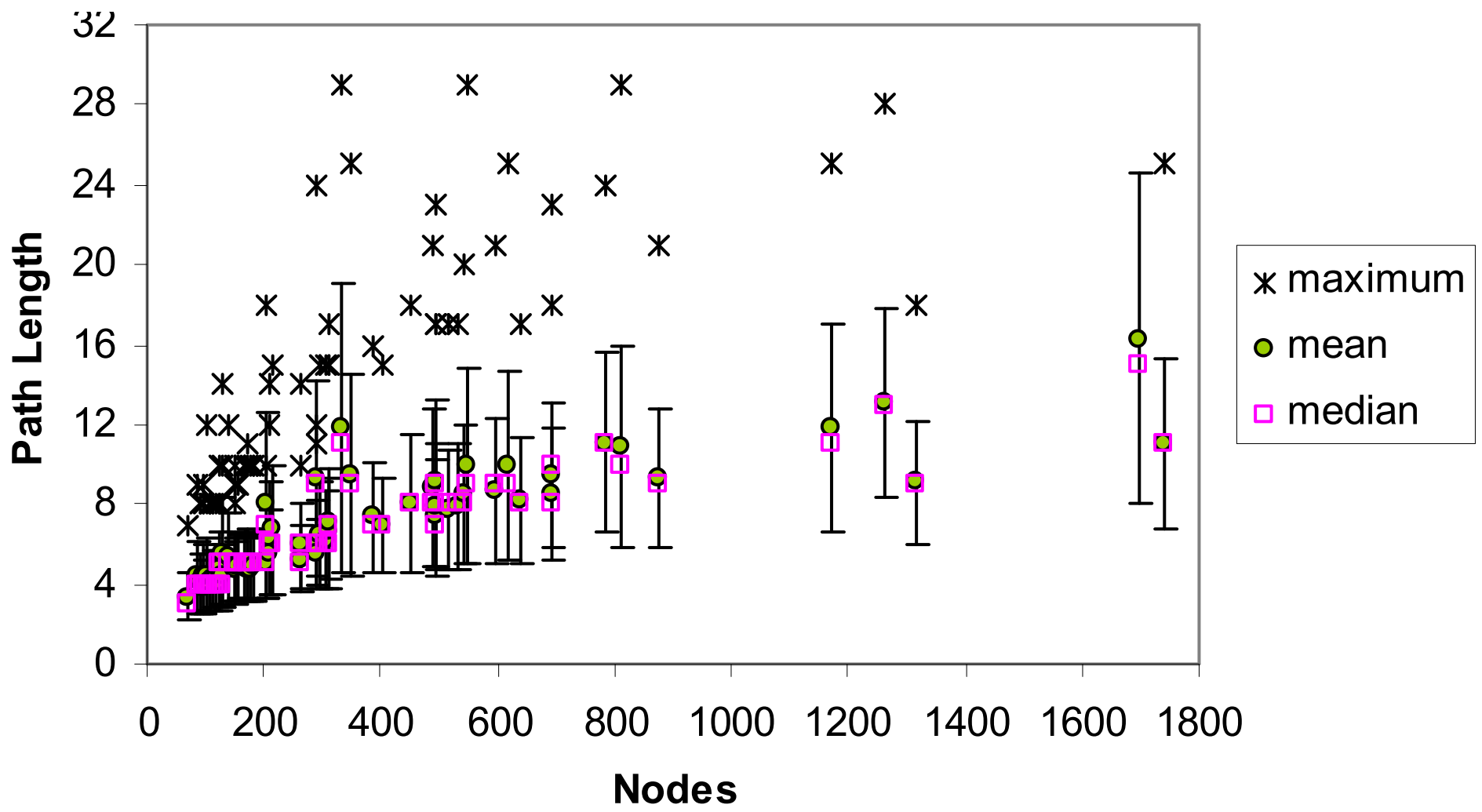

**Figure 9a Diameter, average (± one standard deviation) and median path length for GH64 PCNs.**

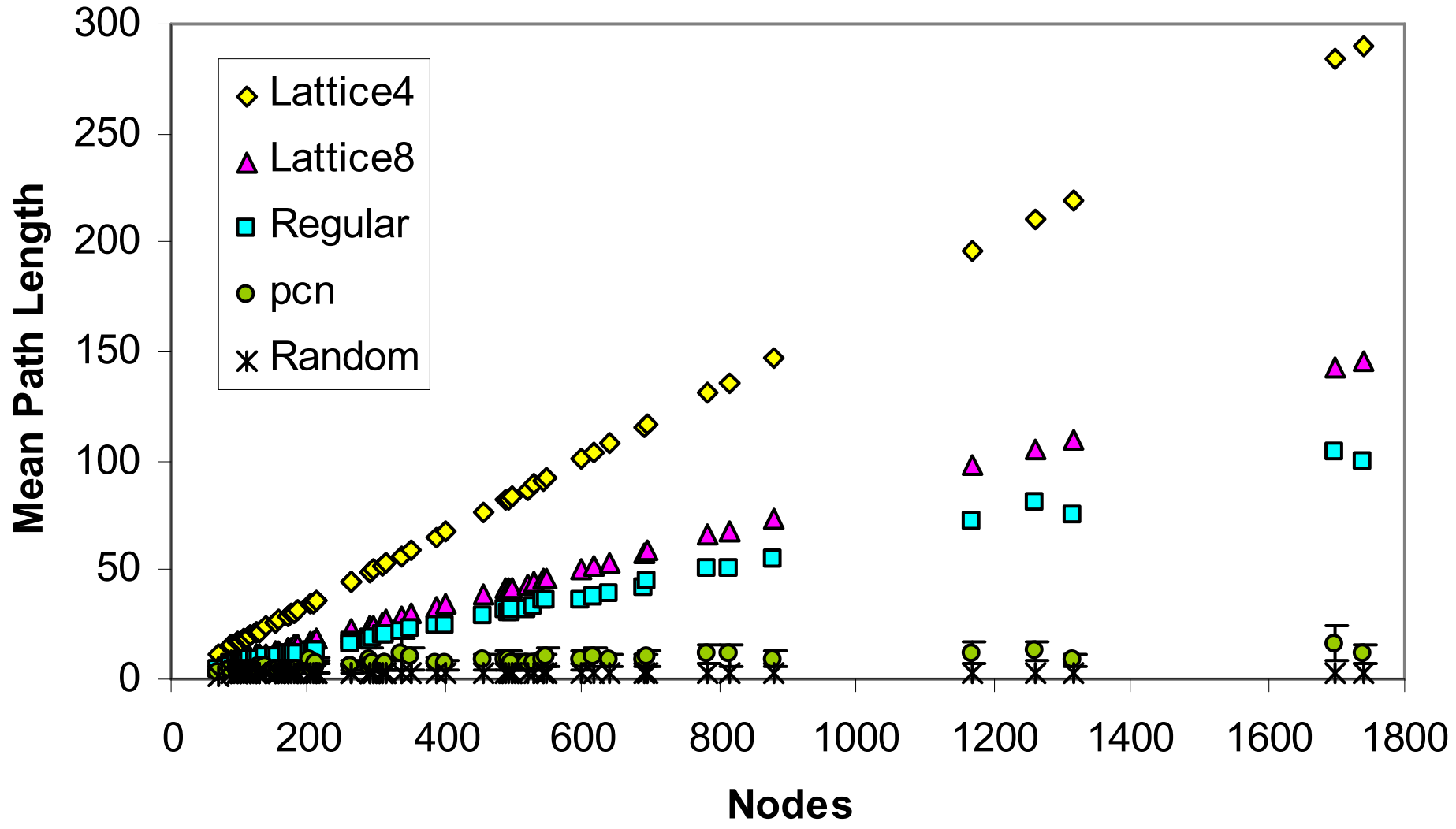

**Figure 9b Average path lengths of PCNs are much closer to average path lengths of random networks ($L_{RANDOM}$) than to average path length of regular networks ($L_{REGULAR}$). $L_{RANDOM} \sim \ln N / \ln K$, and $L_{REGULAR} = N (N + K-2) / [2K (N-1)]$, where K is average degree and N is number of nodes (Watts 1999).**

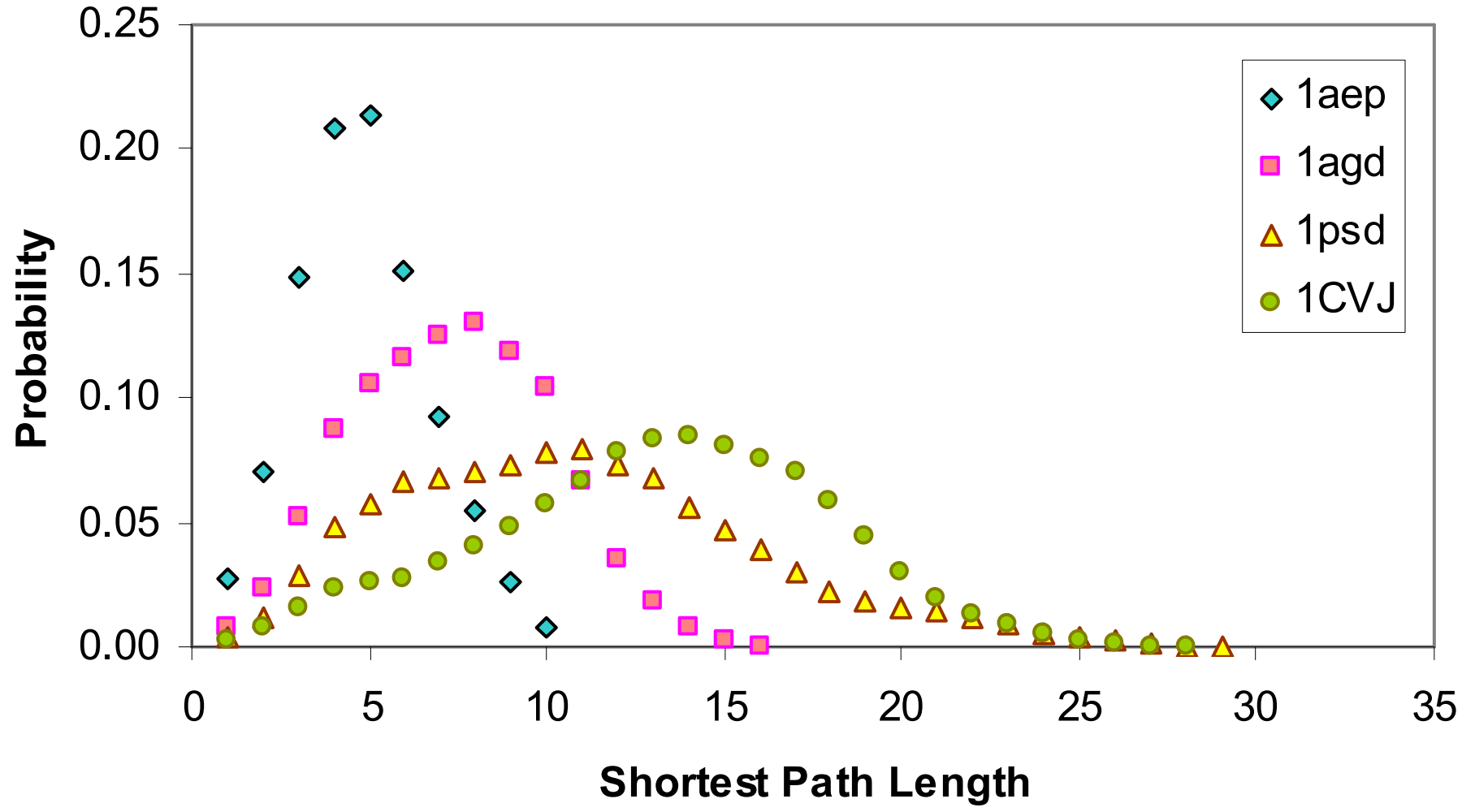

**Figure 9c Distribution of shortest path lengths for four GH64 PCNs.**

Figures 10a&b respectively show the effect of randomizing links on the average path lengths and on the diameters of PCNs. Like Clustering (section 6) and Assortativity (section 7), randomizing only long-range links (randLE) produces a smaller effect than randomizing only short-range links (randSE). Both the two-sided paired t test and the two-sided paired Wilcoxon test report a significant difference at the 95% confidence level between the randSE and the randLE vectors of mean path length values (Figure 10a). Indirectly, this shows that a larger proportion of the magnitude of a PCN's average path length is due to SE.

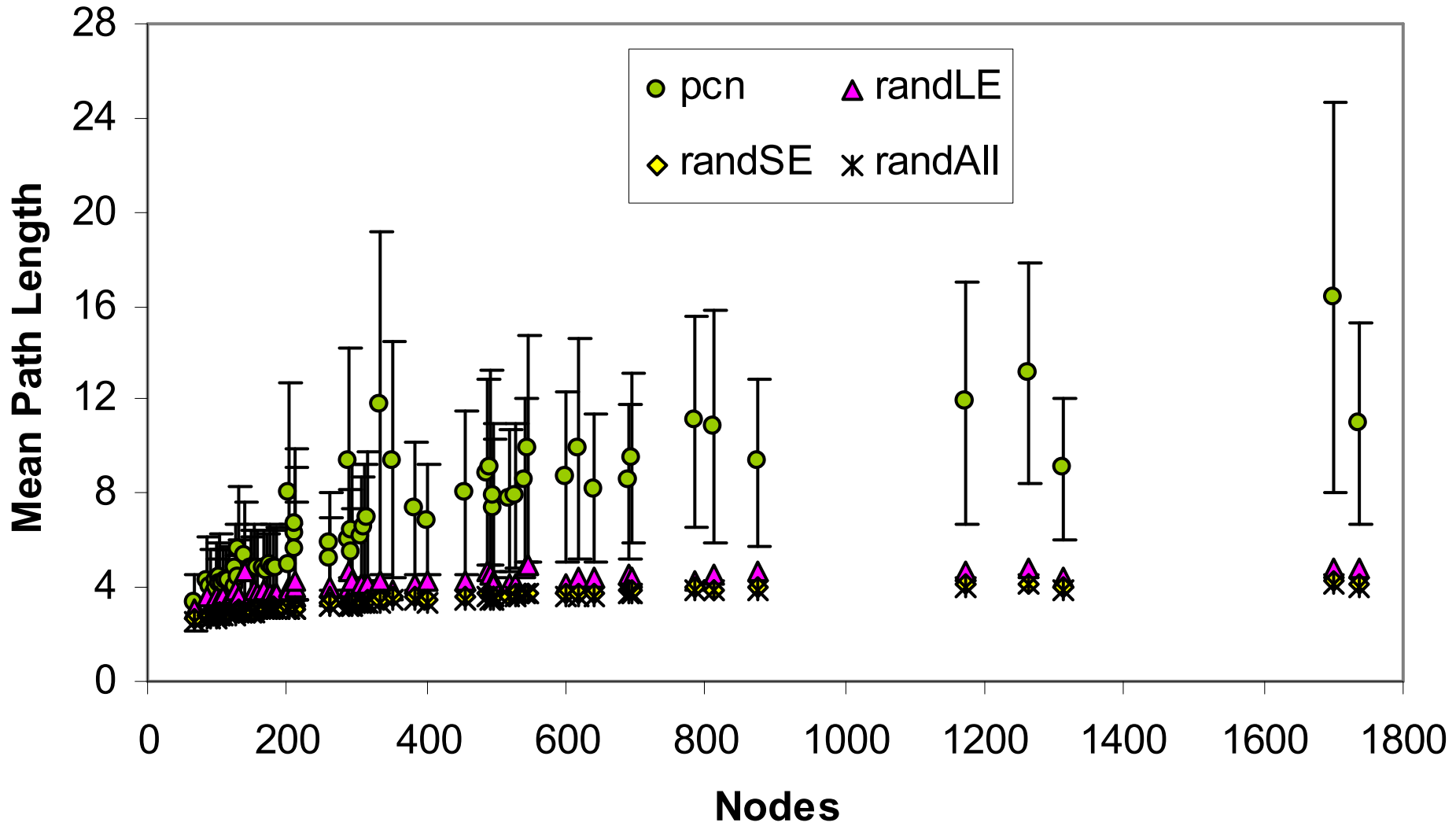

**Figure 10a Effect of randomizing links on the average path lengths of GH64 PCNs. randSE has a stronger effect than randLE.**

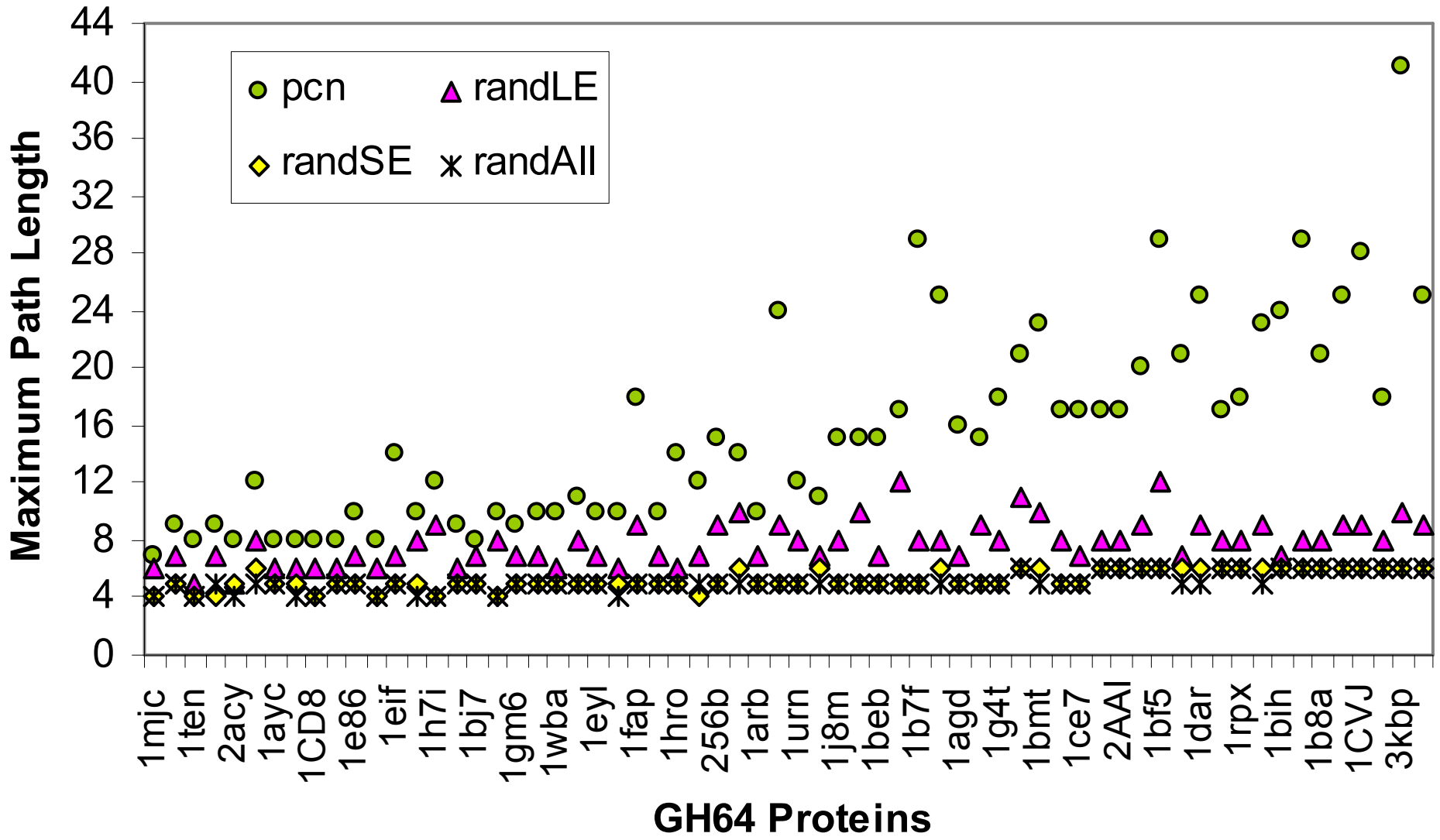

**Figure 10b Effect of randomizing links on the diameters of GH64 PCNs. randSE has a stronger effect than randLE.**

Randomization of links also reveals that it is possible to rearrange the links of a PCN so that the average path length is significantly reduced. Thus the question: if short inter-nodal distances are important for protein folding, why aren't the average path lengths of PCNs shorter than they are?

Preserving high clustering and positive assortativity (if they are important for protein folding) is not an adequate answer since it is still possible to maintain clustering and positive assortativity at levels higher than would be in random graphs, while significantly reducing the average path lengths of PCNs by randomizing the long-range links (randLE). We investigate this question elsewhere and find evidence showing that randLE can be less conducive to search than PCN (manuscript submitted).

## 9. Power-laws in PCNs

In the more recent history of the science of complex networks, power-law distribution or the lack of scale is often associated with the node degree variable. We have already touched upon this in section 5 in relation to PCNs. In this section we offer two other (non-node degree) places in PCNs where power-law distribution can be found. The first resides across proteins (PCNs) in the relationship between nodes and link density. This derives directly from the mean node degree constant, K (Figure 4a). The second resides within individual PCNs in terms of the distribution of link sequence distances.

### 9.1 Link Density Distribution across PCNs

Figure 11 (left) is the link density plot in Figure 3c on a log-log scale, and Figure 11 (right) is a similar plot for Figure A3 in Appendix A. Both show clear power-law relationships, but with an atypical exponent of -1. Number of links, M = K N / 2 where K is mean average node degree and N is number of nodes. Link density = (2 M) / [N (N-1)] ≈ K / N. Empirical data tells us that K = 8 (section 5).

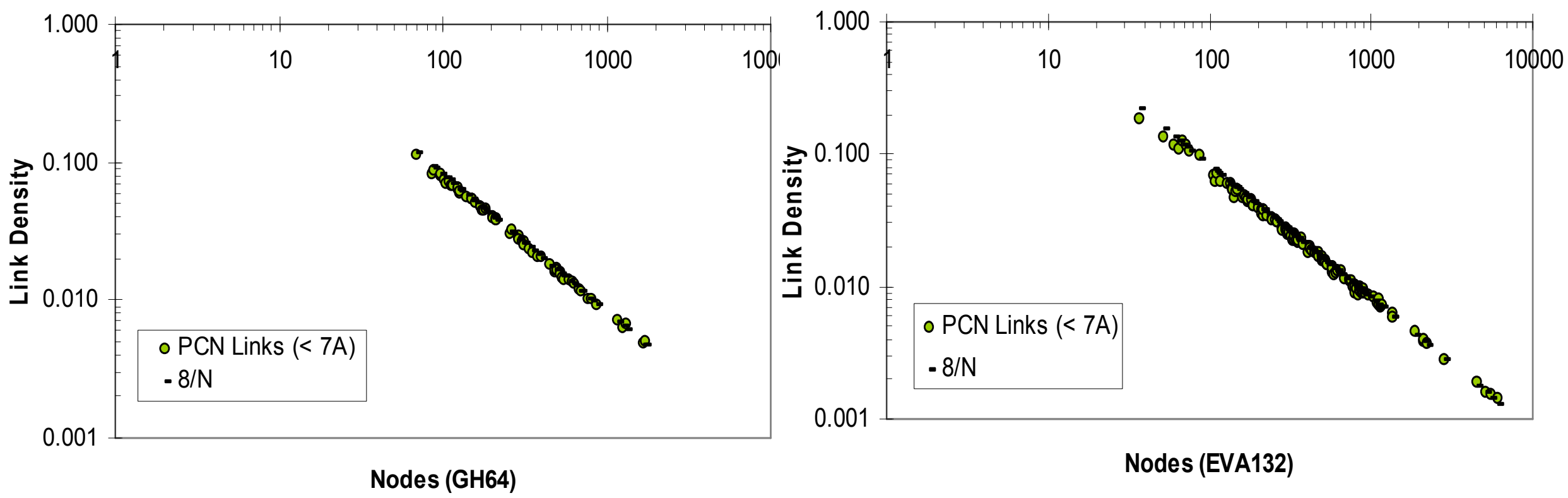

**Figure 11 Links density plots on logarithmic scales.**

### 9.2 Link Sequence Distance Distribution for a PCN

Figure 12a summarizes the link sequence distance for PCNs in GH64. Sequence distance is the absolute distance of a link on the protein sequence (section 4). Minimum sequence distance for all PCNs is 1. We find for N < 4000, the mean link sequence distance generally increases with protein size as a consequence of being pulled up by the maximum link sequence distance which generally increases with protein size. However, the median link sequence distance remains fairly constant with protein size and well below the average link sequence distance per PCN. Similar observations are made for PCNs in the EVA132 dataset (Figure A8). These summary statistics hint at a non-Gaussian distribution for link sequence distances. The link sequence distance distributions for four proteins on a log-log scale can be seen in Figures 12b and 13a. There is a straight line region with a negative slope and a sharp cut-off point in each plot, revealing traces of a power-law. Taking the maximum link sequence distance as a fraction of the number of nodes per PCN reveals that on average, links have a maximum span of about 0.7 N (Figures 12c and 13b).

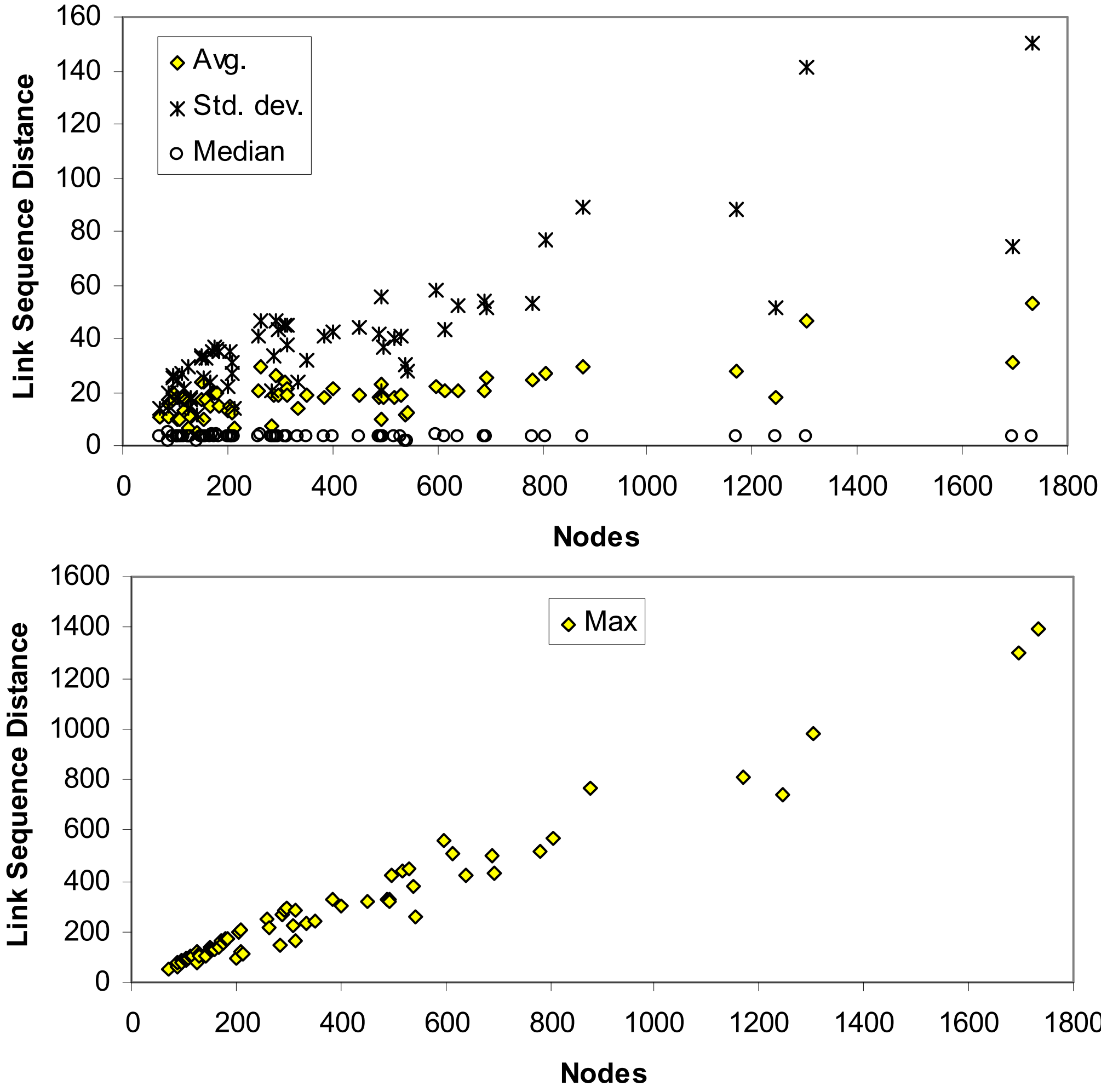

**Figure 12a Link sequence distance summary statistics for GH64 PCNs indicate right skewed distributions for link sequence distance.**

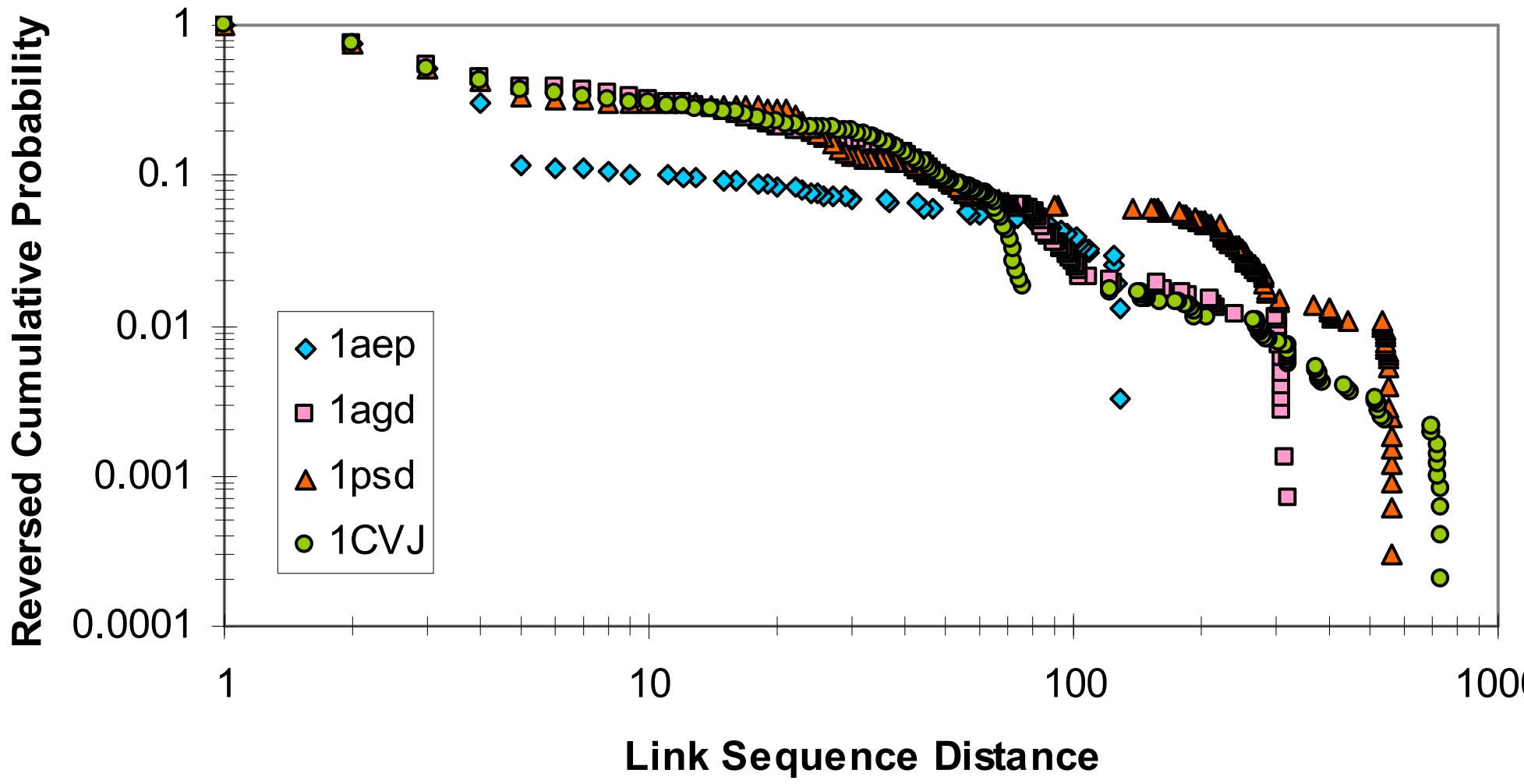

**Figure 12b Link sequence distance distributions for 4 GH64 PCNs on logarithmic scales.**

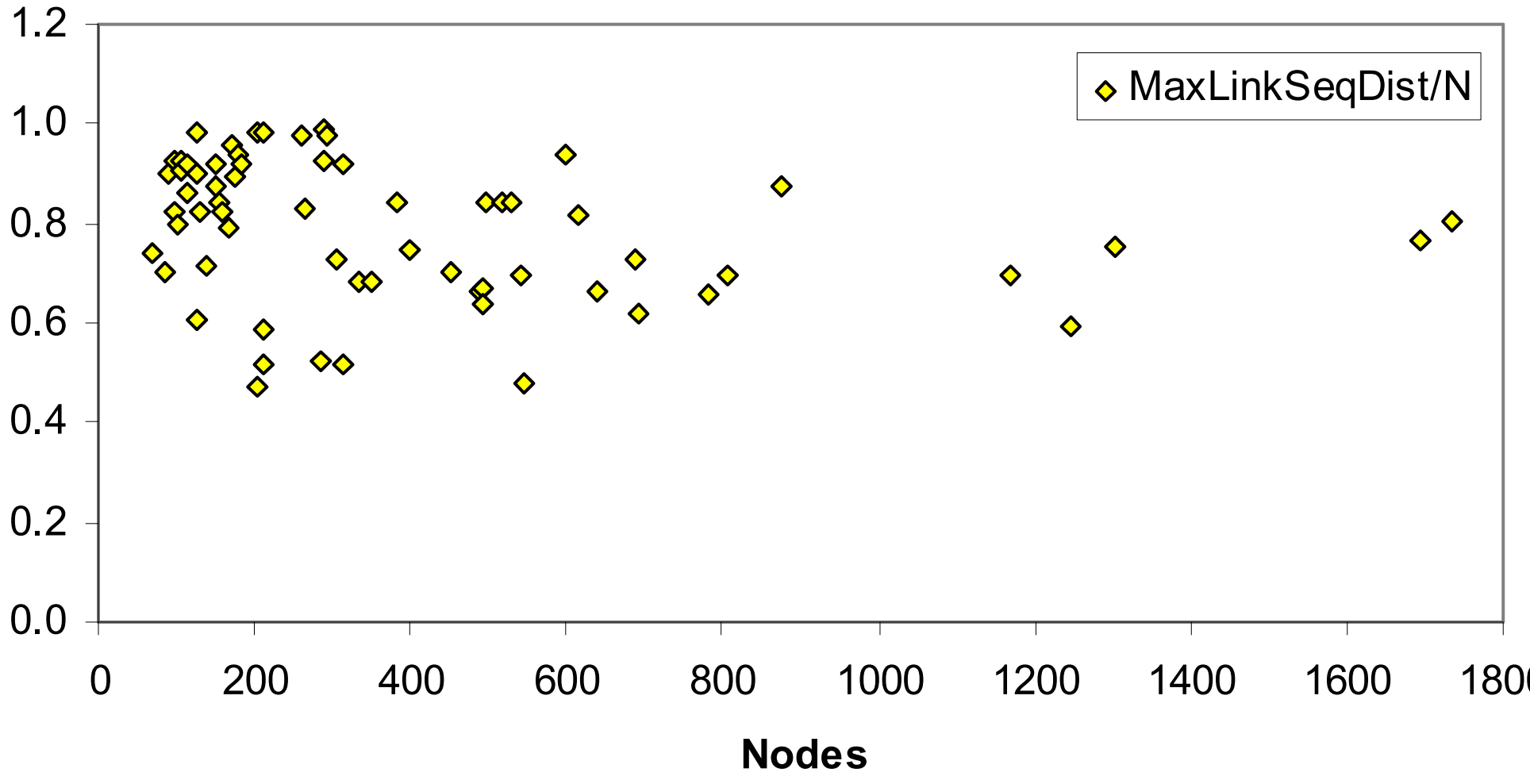

**Figure 12c Maximum link sequence distance as a fraction of the number of nodes in a PCN. MaxLinkSeqDist/N averages at 0.7874 ± 0.1402 for GH64.**

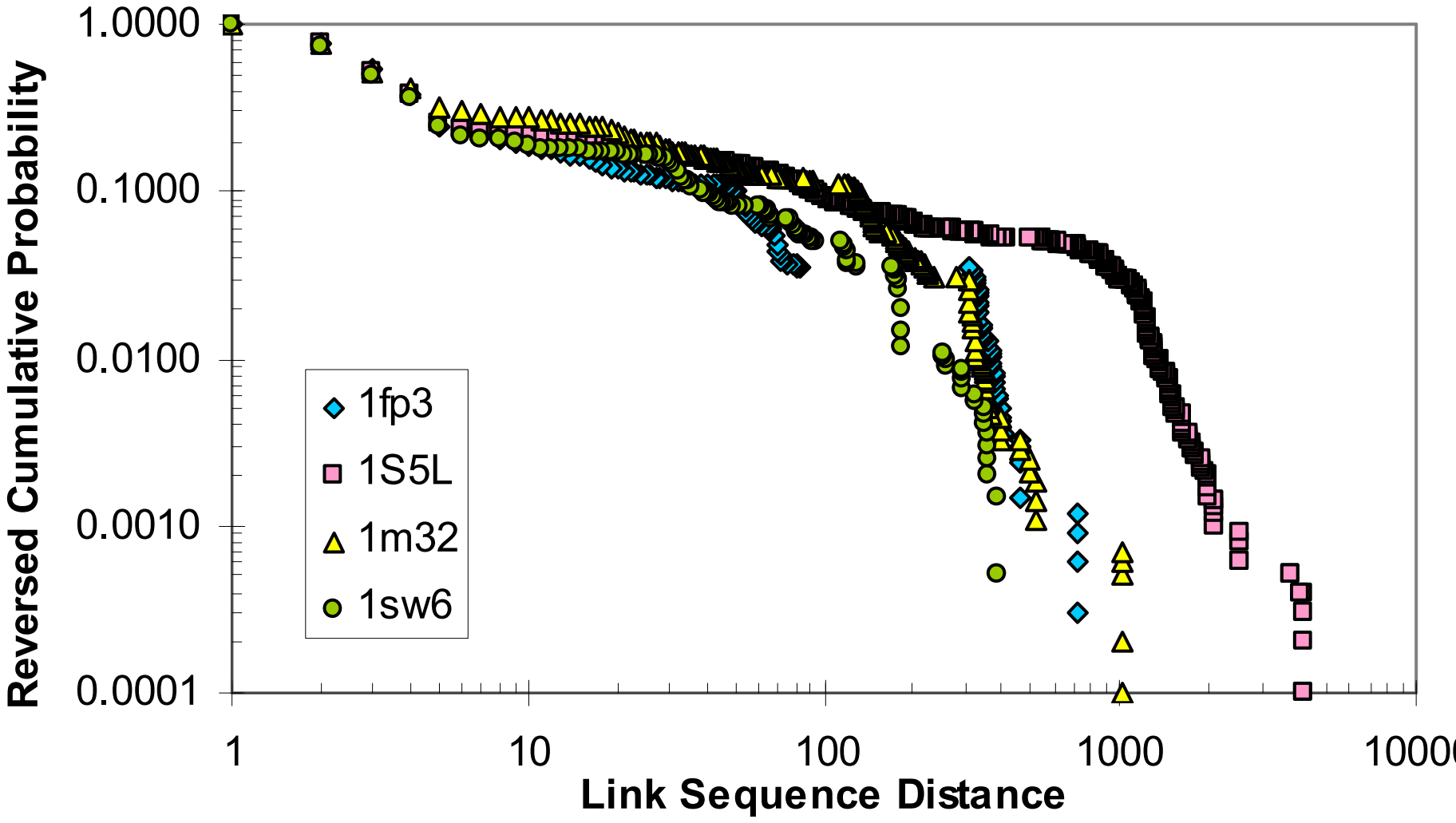

**Figure 13a Link sequence distance distributions for 4 EVA132 PCNs on logarithmic scales.**

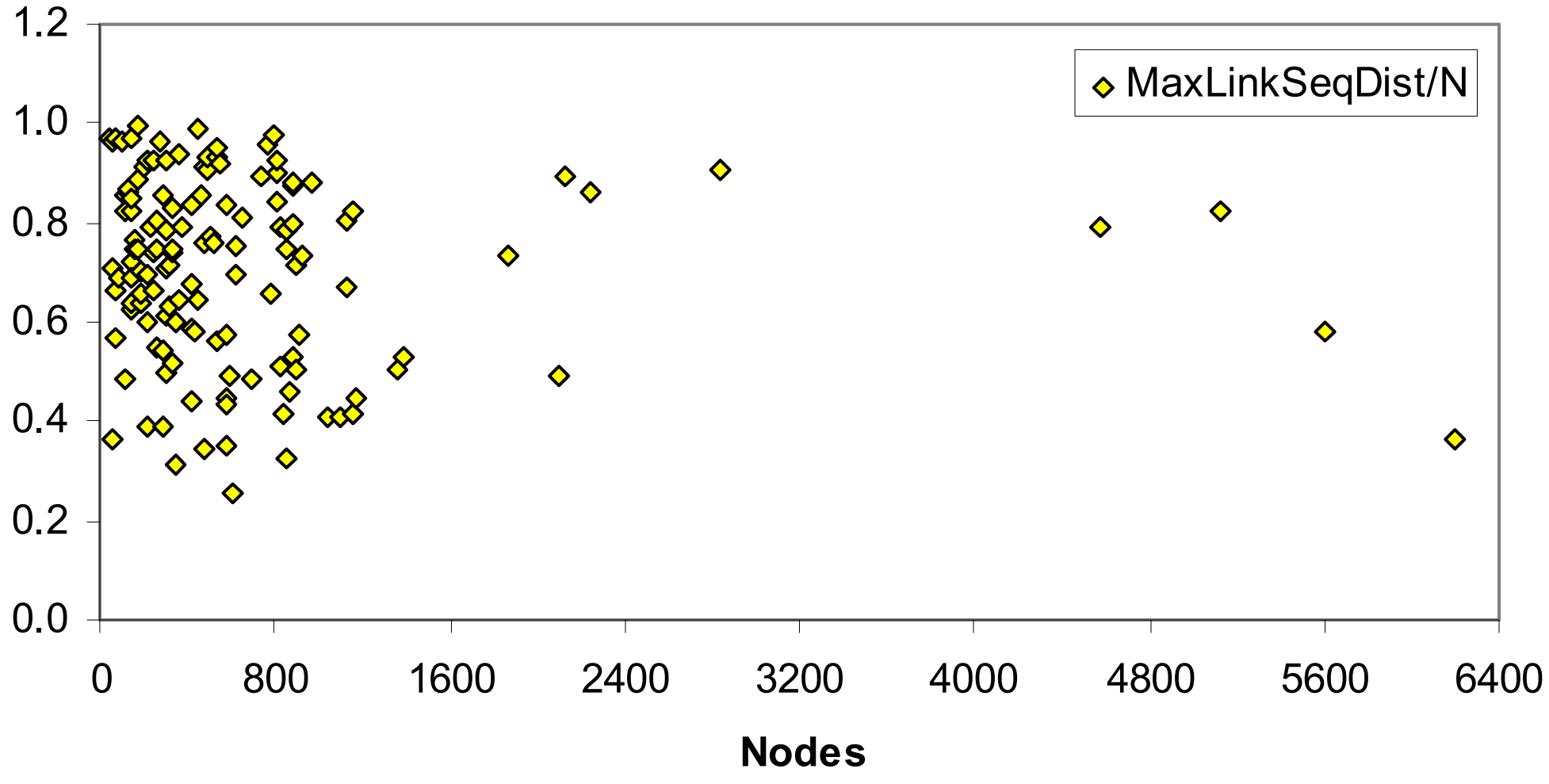

**Figure 13b Maximum link sequence distance as a fraction of the number of nodes in a PCN. MaxLinkSeqDist/N averages at 0.7068 ± 0.1860 for EVA132.**

## 10. A Dynamic View

In this section, we study the effects of adding long-range links to a PCN already containing short-range links. We assume links are independent of each other and that links of a shorter sequence distance form before links of a longer sequence distance. Long-range links are added one at a time, and in order of non-decreasing link sequence distance, and we do not pay attention to the order of links having the same sequence distance but use random selection instead. This is admittedly a crude way of simulating the folding process of a protein, but it captures its essence. We monitor the change over time of two network characteristics: node betweeness and average path length. Based on the literature surveyed, we expect a number of nodes to exhibit distinctively higher node betweeness, and we expect the average path length to decrease over time. That average path length will decrease is a given (although occasional small increases is possible), but what is more interesting and pertinent is the way the general decrease happens.

Betweeness of a node is a measure of the node's centrality, or importance in terms of path traversal between nodes in a network. More specifically, the betweeness of a node is the fraction of shortest paths found between all node pairs that pass through the node. In an effort to identify amino acids that play a key role in a protein's folding process, Vendruscolo *et al* (2002) observed that key residues often exhibit significantly higher node betweeness in a protein's transition state. However, this signal attenuates as a protein reaches its native state (due to compactness). Hence, looking at node betweeness in a protein's native state PCN is more likely to produce false positives of a protein's nucleation sites. Nonetheless, a topological (network structure) understanding of the protein folding process may help us understand how a protein's unique native conformation is chosen (Vendruscolo *et al* 2002).

Dokholyan *et al* (2002) used topological characteristics of protein conformation graphs (PCN) to distinguish conformations belonging to a protein's transition state ensemble (TSE) that will cross the energy barrier and reach the protein's native state versus those that will not. They found that post-transition conformations (TSE conformations which have their nucleus formed and are more likely to reach the native state) are more small-world like than pre-transitions conformations (TSE conformations which have not formed their nucleus and are less likely to reach the native state) in the sense that the average path length of post-transition conformation graphs (PCN) is significantly shorter than the average path length of pre-transition PCNs. In other words, there is a tightening of the small-world structure of proteins as they approach their native conformations (become more compact).

Figure 14a shows that as the 1agd protein "folds" (using the scheme described above), average node betweeness decreases slightly, but experiences a phase transition just after t=350, resulting in a sudden and substantial increase in average node betweeness accompanied also by an increase in node betweeness standard deviation, and a distinct shortening of average path length (Figure 14c). Both the median and the maximum node betweeness also register sudden and substantial increases at the same point in time (Figure 14b). These effects indicate a wider dispersion of node betweeness values amongst the nodes with some nodes exhibiting much higher node betweeness as observed in (Vendruscolo *et al* 2002). In Figure 15a, we compare the distribution of node betweeness for 1agd at three points in time: before the transition point (t=250), around the transition point (t=355) and at the end of the folding process (t=482). All three plots are highly right-skewed (the use of a log-log plot helps with visualization). Limiting the comparison to the three points in time, we find a definite widening of node betweeness distribution from t=250 to t=355, followed by a narrowing of node betweeness distribution from t=355 to t=482 (the tail of the plot swishes to the right then to the left). Betweeness of nodes become more evenly spread out amongst nodes in 1agd's native state.

However, the folding dynamics of 1aep shows a more subtle transition (Figure 14 right). At around the transition point (t=29) where average path length shortens at a faster rate (Figure 14f), there is no sudden or substantial increase in average node betweeness (Figure 14d) and maximum node betweeness is actually dipping rapidly (Figure 14e). But there is a spike in median node betweeness (Figure 14e). These effects are not indicative of a widening then narrowing range of node betweeness values (Figure 15b), but a convergence of some kind (average and median values get closer to each other in both 1aep and 1agd). It appears that as with 1agd, betweeness of nodes become more evenly spread out amongst nodes in 1aep's native state (and therefore become less indicative of the key residues). Analogous to the idea of crossing an energy barrier in protein folding kinetics, this change in node betweeness could be seen as crossing some kind of communication barrier.

Within the nucleation model of protein folding, time prior to the transition point could be regarded as a nucleation growth phase. After the nucleus is formed, the protein polymer becomes committed to its native conformation and quickly reaches it as indicated by the rapid shortening of average path lengths. It would be interesting to investigate this effect further: (i) to test the reliability of median node betweeness as an indicator of a nucleation event in protein folding, and (ii) to monitor the nucleation rates for proteins of different fold types and sizes to see whether there are constants or specific relationships.

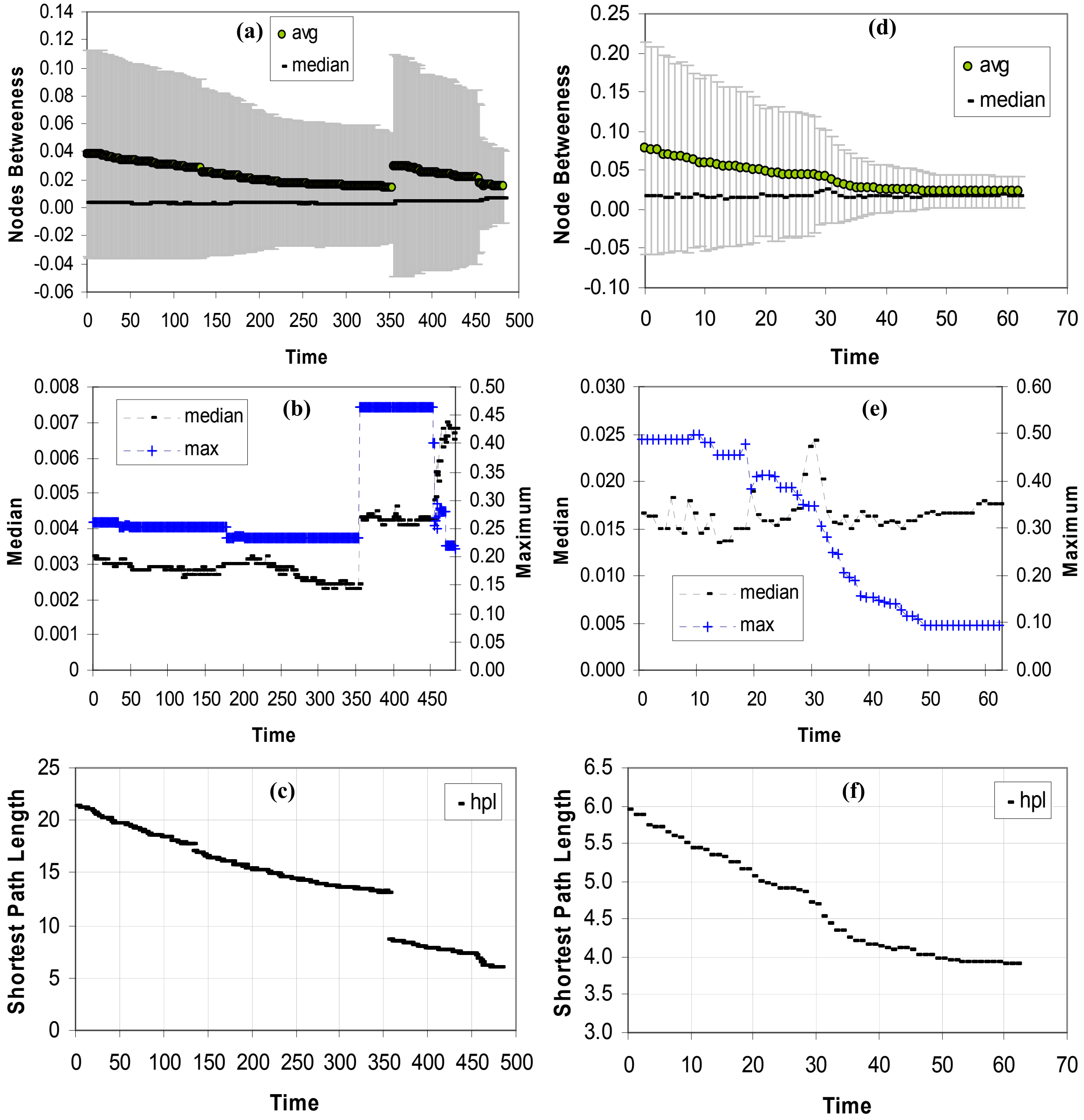

**Figure 14 Left: plots for 1agd. Right: plots for 1aep. Each time step marks the addition of a long-range link to the PCN. At time=0, the PCN only has short-range links. (a) and (d): Average node betweeness ± standard deviation. Median node betweeness lie below average values throughout. (b) and (e): Median and maximum node betweeness values. (c) and (f): Average path length measured using the harmonic mean method (Newman, 2003) to allow for possibility of disconnected PCNs. hpl is more reliable than apl when dealing with disconnected networks, e.g. 1agd (Figure 16).**

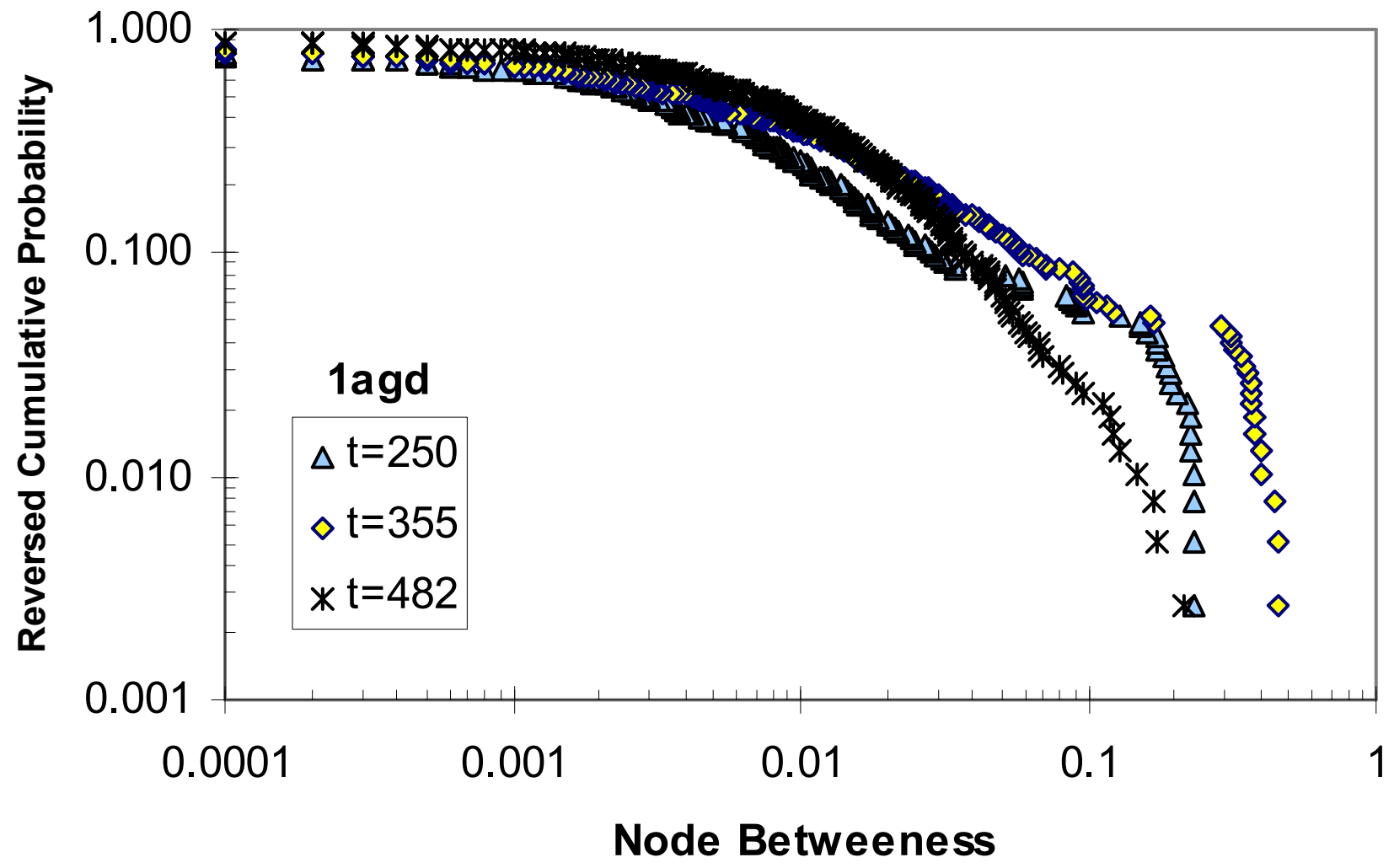

**Figure 15a Distribution of node betweeness for 1agd at three points in time: 250, 355 and 482, on logarithmic scales. The tail ends of the plots swishes to the right then to the left.**

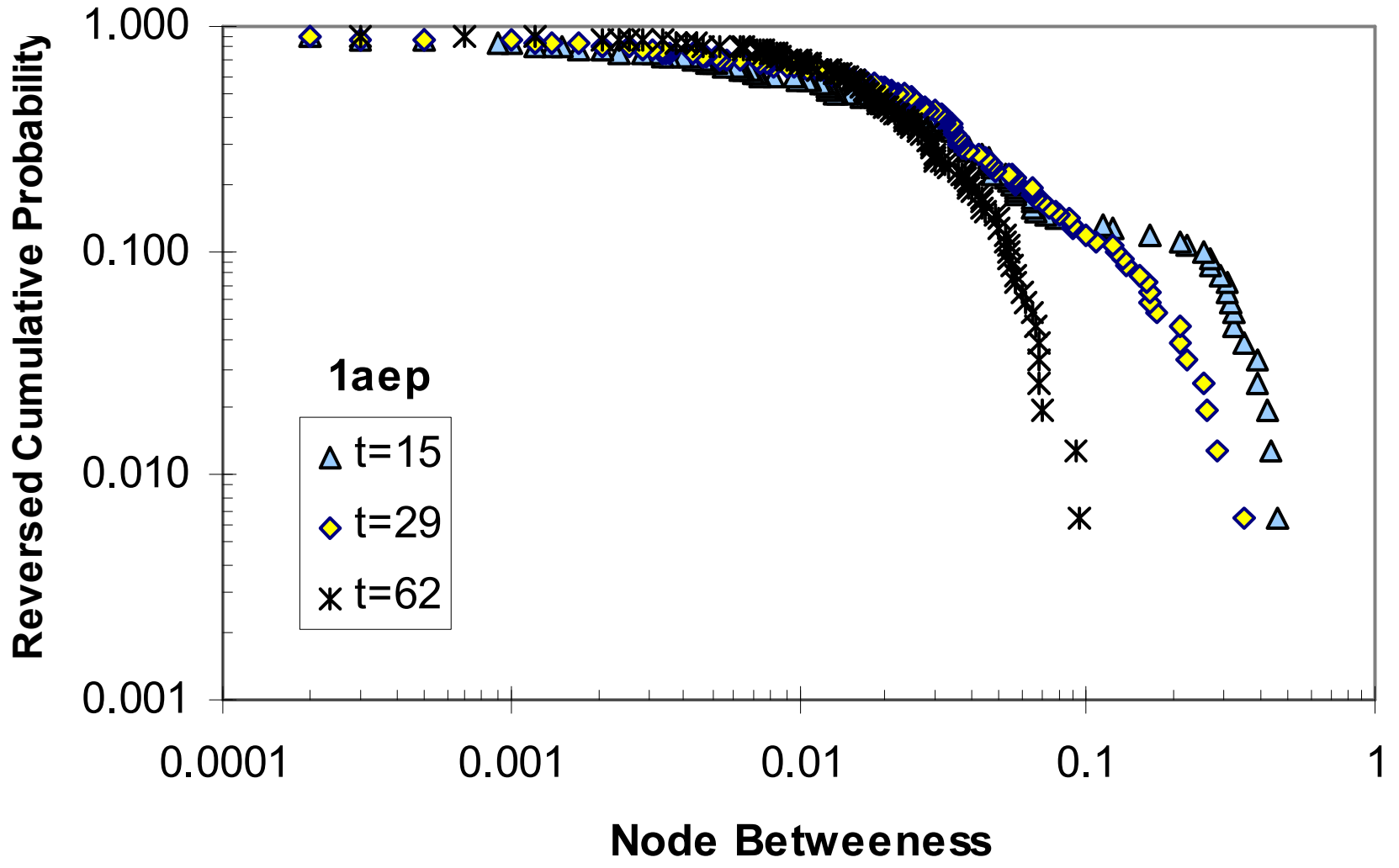

**Figure 15b Distribution of node betweeness for 1aep at three points in time: 15, 29 and 62, on logarithmic scales. The tail ends of the plots swishes only in one direction, towards the left.**

For comparison, Figures 16 and 17 show how average path length changes if long-range links are added in random order, i.e. not added in order of non-decreasing sequence distance. For both proteins, the drop in average path length occurs earlier when long-range links are added in random order, than when long-range links are added in order of non-decreasing link sequence distance order. This kind of decrease is similar to that observed in Figure 18 (section 11). Hence protein folding is a specific rewiring process (Dokholyan *et al* 2002).

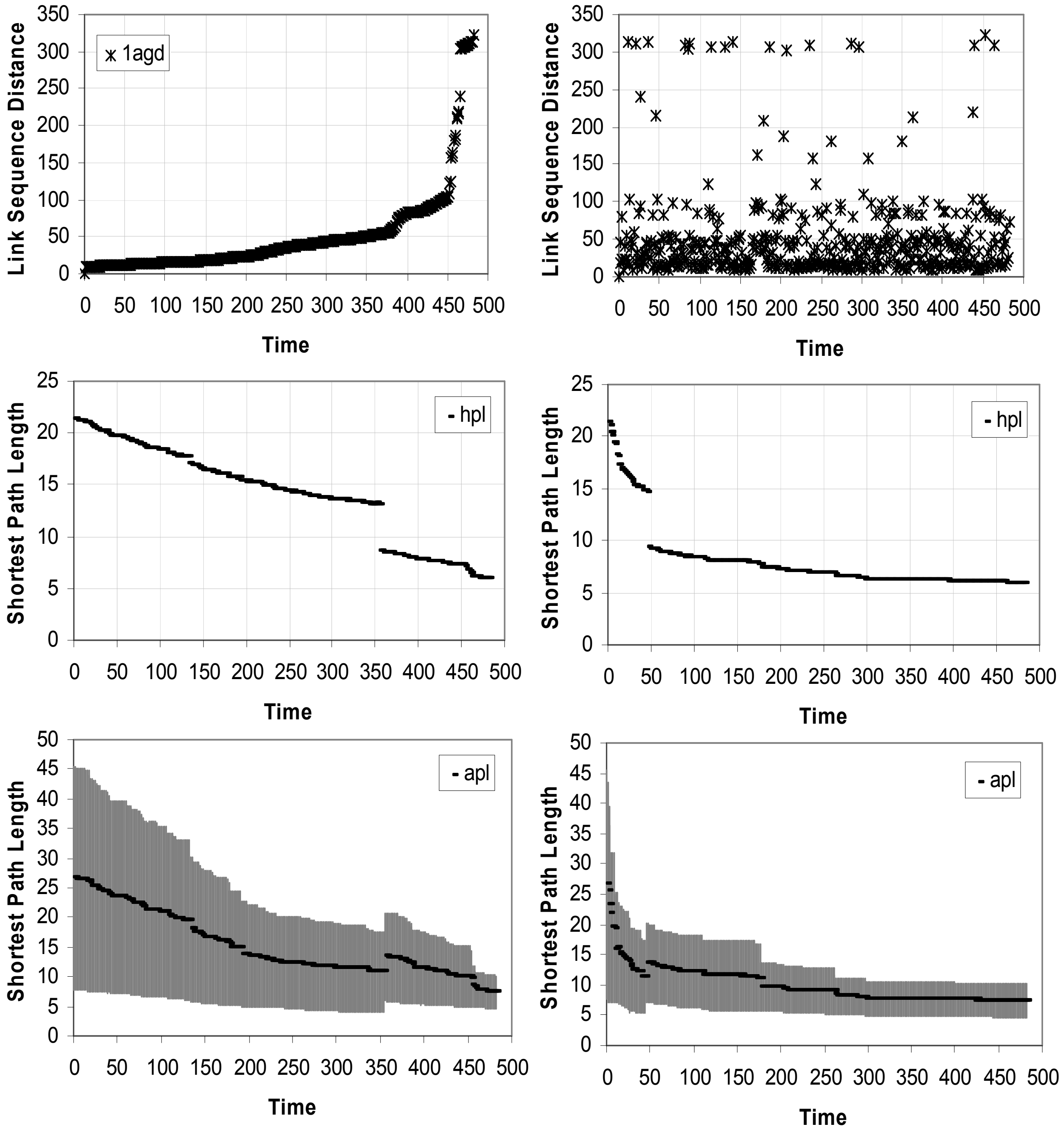

**Figure 16 Behavior of average path length as long-range links for 1agd are added in non-decreasing order (left) versus in random order (right).**

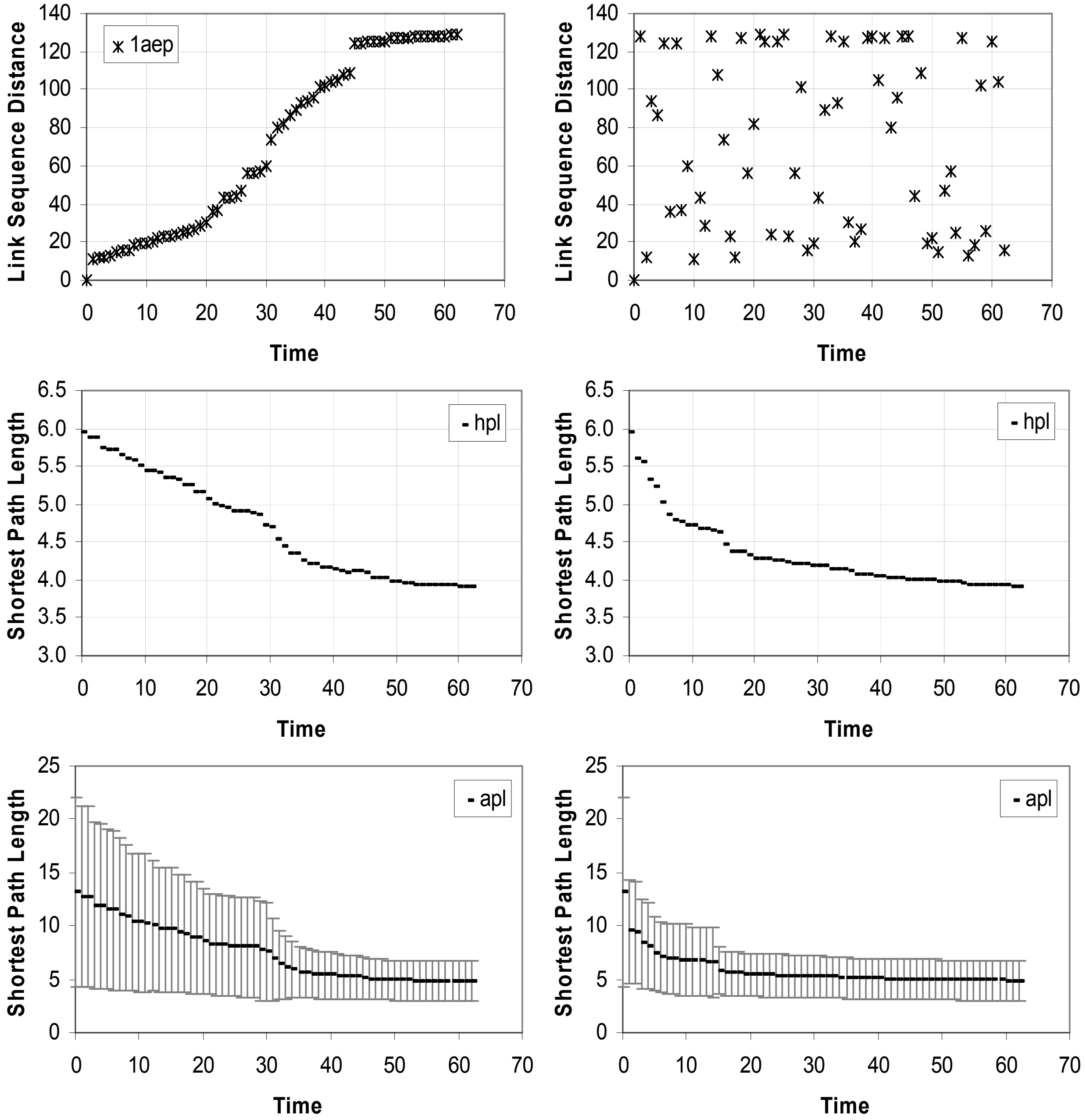

**Figure 17 Behavior of average path length as long-range links for 1aep are added in non-decreasing order (left) versus in random order (right).**

## 11. A Network Approach to Protein Folding

From a network perspective, the protein folding problem boils down to generating the right set of links between amino acid residues of a protein. While many network characteristics of PCNs have been discovered over the past decade, there is less (published) effort to integrate this knowledge into a single model and use it for the protein folding problem. Perhaps this is due to the current difficulty of creating networks to simultaneously fit a number of static and dynamic structural characteristics, still incomplete description of PCNs, the very specific and diverse arrangement of links in PCNs, and/or even the relative success of knowledge-based protein structure prediction methods. In this section, we outline an approach to protein folding that uses several of the network structural characteristics of PCNs mentioned in this paper. This model is still in its infancy, but we hope that it is a good first step in realizing a network approach to protein folding ab initio, or at least put to rest naïve conceptions of PCN models.

The model begins with short-range links arranged on a linear lattice with N nodes. We use Lattice6 with $N$ nodes, i.e. each node has 3 nearest neighbours to its left and to its right where possible. This gives approximately $\frac{V}{2}\left(N-\frac{V}{2}\right)$ links, or ≈ $3N$ when $V$=6. Since we found in section 4 that |SE| ≈ 0.7 $M$ and $M$ = 4 $N$, $V$=6 gives an empirically reasonable number of short-range links to start with. Long-range links are then added to the lattice. The endpoints of a long-range link is selected uniformly at random from the set of nodes such that the absolute distance between them are within [10, 0.7 N] (section 9.2).

We made two independent runs, and set N to 385. The statistics we monitored are link sequence distance distribution, node degree distribution, average path length, clustering, and assortativity. Figure 18 (specifically plots lin_11 and lin_12) present the results of our experiment with the model. The results are compared with statistics from the 1agd PCN, and runs made with the model on a ring topology (plots ring_11 and ring_12). For the experiments on a ring topology, the initial lattice is still linear, but long-range links are added as if the lattice is a ring. Results from both topologies showed no significant differences in that both failed to approximate the statistics for 1agd. Sorting the links generated by the model (plots slin_11 and slin_12) did not help to narrow this gap, although some difference in average path length behavior is observed (the initial descend is slower).

**11.1 The problem of identifying LE_nodes: some pitfalls**

What our model lacks, amongst other elements, is a reliable way to identify the endpoints of long-range links, or the set of LE_nodes. The last observation in section 5 (Figure 6c) implies that for the most part, long-range links (LE) more so than short-range links (SE) connect nodes with many contacts in a PCN. At this point, given the assortative nature of PCNs (section 7) and the frequent description of proteins with their domains or modules, and protein folding as a strict process from secondary to tertiary structures and beyond (see Gō 1983 for a messier view), it may be tempting to picture long-range links connecting hub nodes, i.e. nodes of high degree. However, this picture may not be entirely accurate for two possibly related reasons.

First, for PCNs of proteins in their native state, hubs do not necessarily occupy a central position (section 10). The correlation between node degree and node betweeness rapidly declines as protein size increases (Figure 19, Appendix B). This implies that many shortest paths in a PCN do not go through hub nodes, and since long-range links provide important short-cuts links in PCNs, it is contradictory to think of long- range links connecting hub nodes.

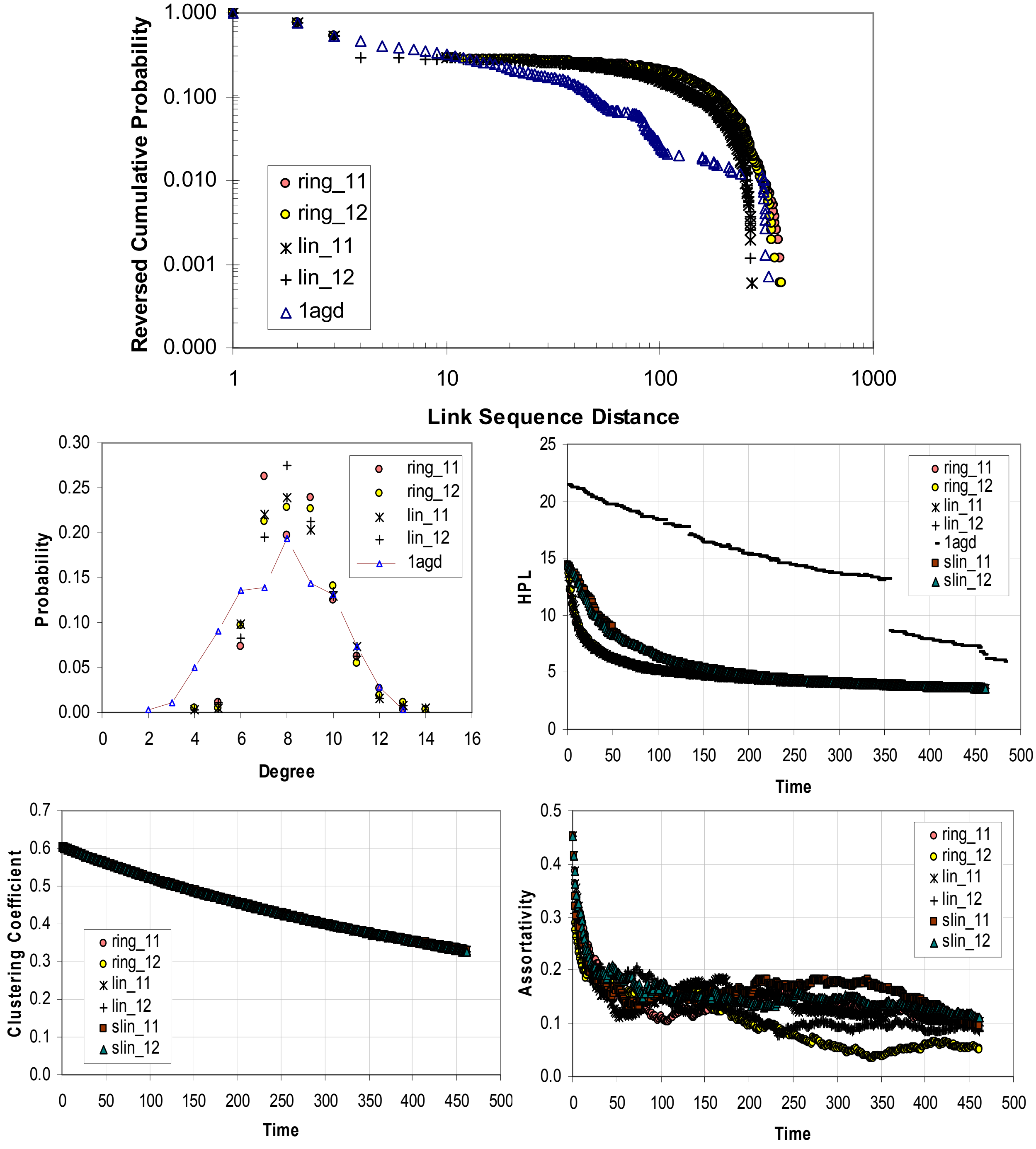

**Figure 18 Results from a network approach to protein folding compared with the network statistics of the PCN for 1agd which is of the same size. The PCN for 1agd has a clustering coefficient of 0.5397, an assortativity value of 0.3037, and about 72% of its nodes are endpoints of long-range links. In the model, for both ring and linear topology, about 91% of nodes are in LE_nodes.**

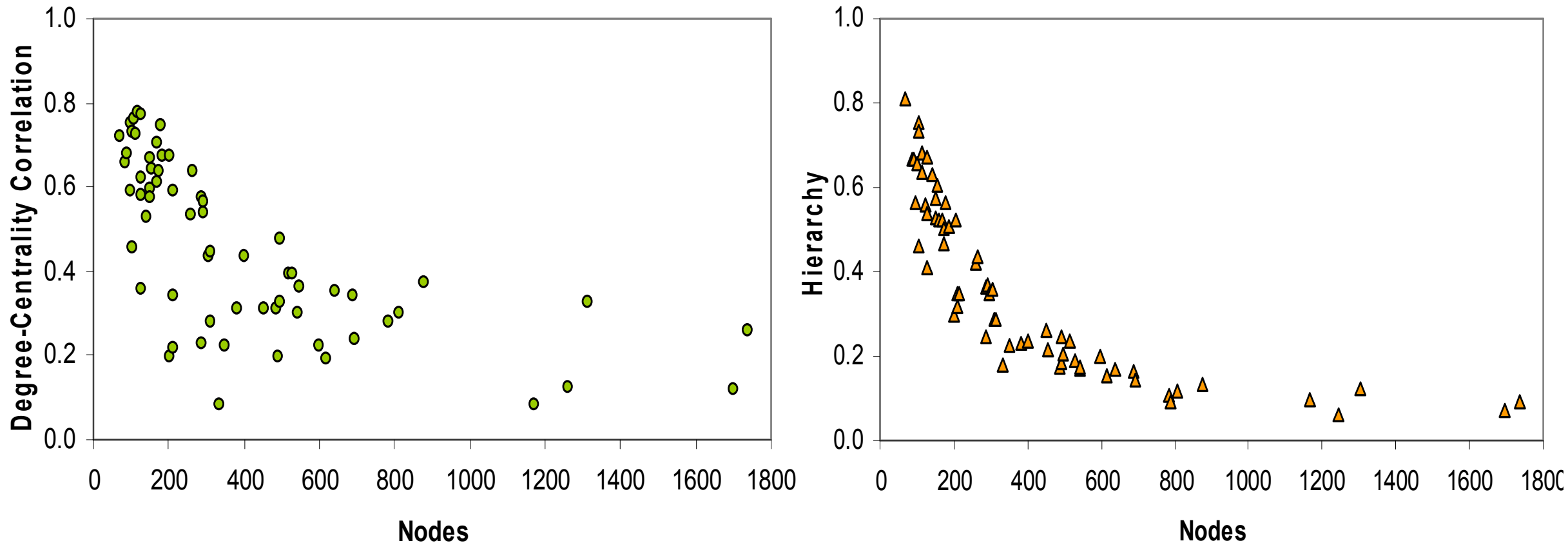

**Figure 19 (Left) Pearson's correlation between node degree and centrality for GH64 PCNs. (Right) Gao's hierarchy index for GH64 PCNs.**

Further, and as a possible consequence also, PCNs actually show declining hierarchical structure with increases in protein size (Figure 19). Following Trusina *et al* (2004), hierarchical organization is gauged with the formulae in (Gao 2001) which defines a hierarchical path as one with that satisfies one of the following conditions: monotonically increasing then monotonically decreasing node degree, monotonically increasing node degree, or monotonically decreasing node degree. A network where a high fraction of its shortest paths are also hierarchical paths will have a hierarchical index close to 1.0, signaling pronounced hierarchical organization. However, declining average clustering levels with increases in node degree has also been used to detect hierarchical organization (Ravasz *et al* 2002; Ravasz and Barabasi 2003). Using this method, Aftabuddin and Kundu (2006) report evidence of hierarchical structure for their weighted PCN, which is similar to the Greene and Higman (2003) PCN. We also observe hierarchical tendencies in the four randomly chosen GH64 PCNs (Figure B3). Evidently, there is some reconciliation to be performed between the two measures of hierarchy.

Second, there is evidence (depending on the scale used) of a positive correlation between node degree and hydrophobicity (Alves and Martinez 2007; Aftabuddin and Kundu 2007) which implies that hub nodes in PCNs tend to be hydrophobic. Hydrophobic residues shun water molecules and generally lie in the core of a protein. As a result, these residues come into close proximity with many other residues, and thus gain their hub status. The relationship(s) between hydrophobicity, high node degree, assortativity and average path length needs further study, keeping in mind that the relationship(s) may differ for transition versus native state proteins.

## 12. Summary and Outlook

Our main objective with this review paper is to integrate the many network characteristics of PCNs which have been reported in the literature, and to understand their respective roles and also the interplay between them. An attempt to apply the statistics to a network approach of protein folding was also made. In addition, we considered the dynamic aspects of PCNs, i.e. how their network statistics change as a protein folds. Our model is woefully simple, but we hope it provides good hints for building better models and integrating the dynamic and static aspects of PCNs. One can easily imagine several ways to modify the model by adding more constraints, e.g. restrict the set of LE_nodes to 0.6 N (section 5), use some kind of preferential selection mechanism to fit to the link sequence distance distribution (section 9) and node degree distribution (section 5). The model could also be extended to include link deletion and/or rewiring (Dokholyan *et al* 2002 observed changes in node degrees between post- and pre-transition states). It may also be interesting to extend the approach in section 11 to investigate how the dynamics of PCNs couples with the conformational landscape (Scala *et al* 2001; Doye 2002; Kovacs *et al* 2005; Auer *et al* 2007). Studies show that it is not only important for a protein to arrive at its final destination, a.k.a. its native state, but how it gets there, its journey or folding pathway(s), is also important (Chen *et al* 2008). In light of the exposition in this paper, a network approach may be the vehicle to address key aspects of the protein folding problem in an integrated way: the existence of a unique native structure, folding pathway(s) in conformation space to the native structure, protein folding rates, and information in an amino acid sequence to propel the attainment of all this.

In Table 2, we summarize the network structural statistics mentioned in this paper for PCNs in the GH64 and EVA132 data sets. We find a general agreement, both in qualitative and quantitative terms, between the observed network structural characteristics of these two data sets, and with those reported in literature. On this topic, there are of course other existing network characteristics and measurements, and other ways of constructing PCNs, that can and need to be explored. However it could also be useful to go further, and devise measures geared towards capturing the specific linkage patterns of PCNs.

**Table 2 Summary of quantitative values: mean ± standard deviation reported.**

| Section | Statistic | GH64 | EVA132 |
|---|---|---|---|
| 4 | \|LE\|/M<br>Fraction of links in LE (links with sequence distance > 9) | 0.2929<br>± 0.0834 | 0.2611<br>± 0.0986 |
| 5 | K<br>Average node degree | 7.9696<br>± 0.3126 | 7.9147<br>± 0.4300 |
| 5 | \|SE_nodes\|/N<br>Fraction of nodes which are endpoints to links in SE | 0.9970<br>± 0.0039 | 1.0000<br>± 0.0004 |
| 5 | \|LE_nodes\|/N<br>Fraction of nodes which are endpoints to links in LE | 0.6659<br>± 0.1093 | 0.6348<br>± 0.1367 |
| 6 | Clustering | 0.5484<br>± 0.0233 | 0.5538<br>± 0.0292 |
| 7 | Assortativity | 0.3387<br>± 0.0536 | 0.3457<br>± 0.0743 |
| 9.2 | (Maximum Link Sequence Distance) / N | 0.7874<br>± 0.1402 | 0.7068<br>± 0.1860 |

Throughout this paper, we have maintained a separation of short-range (SE) and long-range (LE) links. We did this to better understand their respective contributions to structural characteristics of PCNs. But ultimately, the prize is to predict the long-range links of PCNs to within a level of accuracy. Notwithstanding the physical, chemical and geometrical constraints, this is a challenging combinatorial task. Even if we could identify LE_nodes accurately, the probability that a link chosen at random from the set of all possible links with endpoints in LE_nodes is about 6/N (this includes links with sequence distances less than 10 or some LE_TH). Hence a pure network model of protein folding, like the one we attempted in section 11, is unlikely to be a winner. A more likely model could be one that incorporates multiple approaches.

Given the importance of short inter-node distances between residues (section 8), the sparseness of long-range links in PCNs could appear paradoxical. However, our model does show that in general, it only takes a few long-range links to reduce the average path length significantly, and to levels that is not improved upon by further addition of long-range links (Figure 18). According to Ngo and Marks (1992), long-range links (non-local interactions) actually increases the computational complexity of the protein folding problem (see also Ngo *et al* 1994). Thus, if a few well-placed long-range links is sufficient to reduce the average path length to a good enough level, this "least effort" (albeit, it could have taken evolution quite some time to figure the optimal wirings) strategy may well be adopted by proteins as a way out of the paradox.

## Appendix A. The EVA132 Protein Dataset

The EVA132 protein dataset was extracted from the list of 3477 unique chains archived by EVA (Rost, 1999). 200 proteins were selected at random from this list, with no overlap with GH64. PCNs for the proteins in EVA132 were constructed and selected in the manner as described in sections 2 and 3, producing 132 valid PCNs (listed in Table A1) for us to work with.

**Table A1. PID of proteins in the EVA132 dataset**

1dfe, 1bi6, 1a7i, 2ezh, 1ctf, 1cfw, 1tif, 1gh5, 1apc, 1cto, 1by2, 1d0d, 1AX8, 1gak, 1jgs, 1j74, 1iko, 2a1j, 1ECY, 1af3, 1grj, 1f8a, 1lf7, 1faj, 1awp, 1qhn, 1r5p, 1CZ4, 1ylx, 1j77, 1bwp, 1lkv, 1mfq, 1jfx, 1hek, 1gy5, 1o9g, 1e7k, 1F9Z, 1awc, 1sml, 1boo, 1tfr, 1ltq, 1e8o, 1pvl, 1bob, 1sbp, 2aql, 1xiz, 1pnf, 1fkm, 1btk, 1c28, 1nrj, 1e6c, 1hh2, 1dpt, 1g2q, 1a1r, 1a59, 1f45, 1i5e, 1esg, 1h3e, 1wdu, 1i85, 1gk4, 1bkj, 1dj8, 1oft, 1jzt, 1fgg, 1rh5, 1sw6, 1bfd, 1rhz, 1b44, 1g7r, 1aui, 1adu, 1h7z, 1zrl, 1dhp, 1e3h, 1ojh, 1bp7, 1kke, 2fmt, 1K3I, 1bwd, 1d8d, 1a5k, 1bih, 1bev, 1fp3, 1aoi, 1tmf, 1l1o, 1KQ4, 1upt, 1h4l, 1ngw, 1b4u, 1ee4, 1a2z, 1eqz, 1jct, 2bpt, 1e8t, 1zhq, 1ewt, 1eo8, 1eum, 1g2c, 1fp7, 1e6e, 1l8w, 1ndb, 1g8m, 1knz, 1gg1, 1b37, 1bcp, 1m32, 1fz6, 1jb0, 1hqm, 1h6d, 1S5L, 1mjg, 1M34

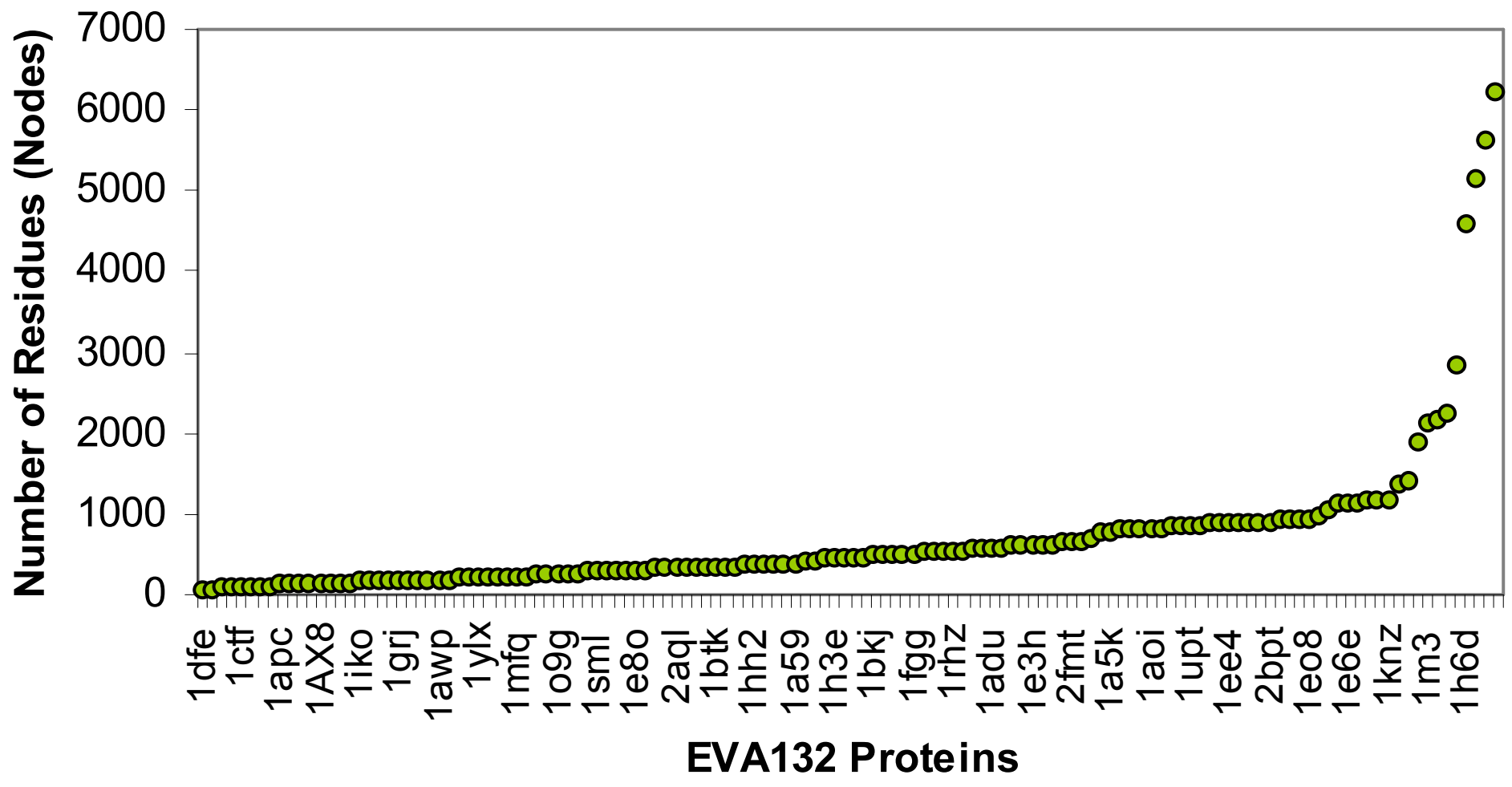

**Figure A1 Size of EVA132 PCNs in terms of the number of Cα atoms documented their PDB files.**

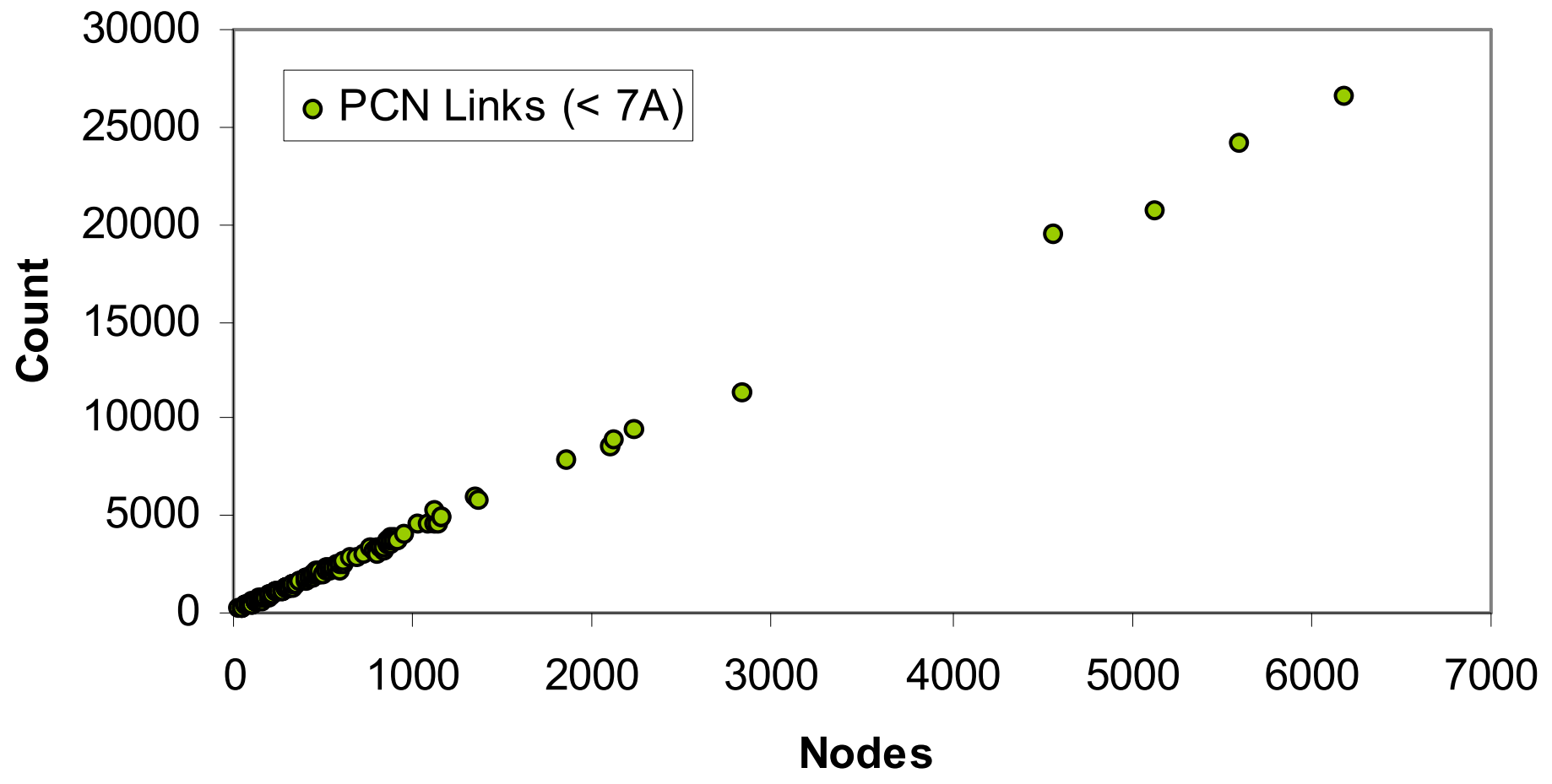

**Figure A2 Link count M, by protein size for EVA132 proteins.**

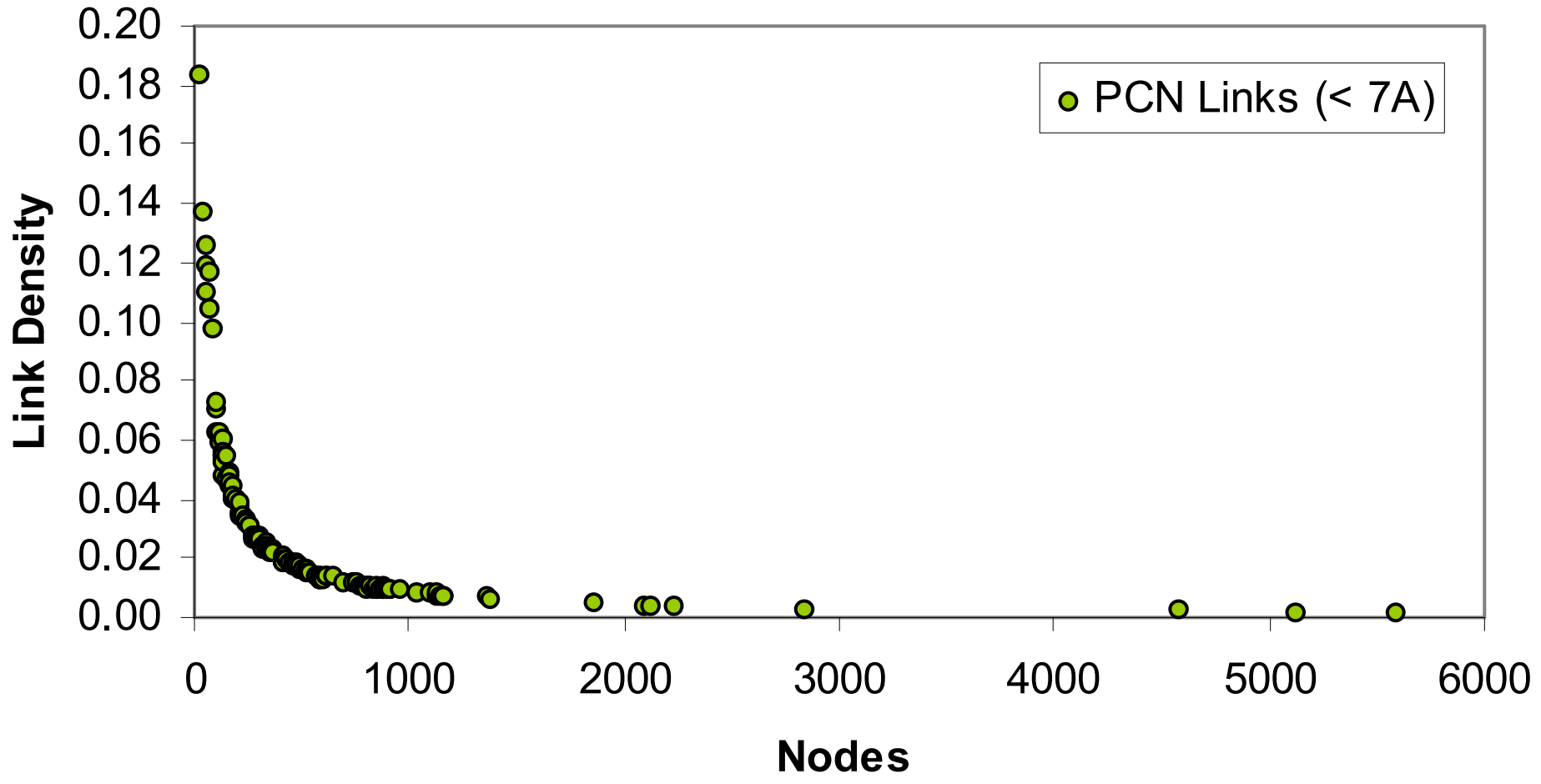

**Figure A3 Link density by protein size for EVA132 proteins.**

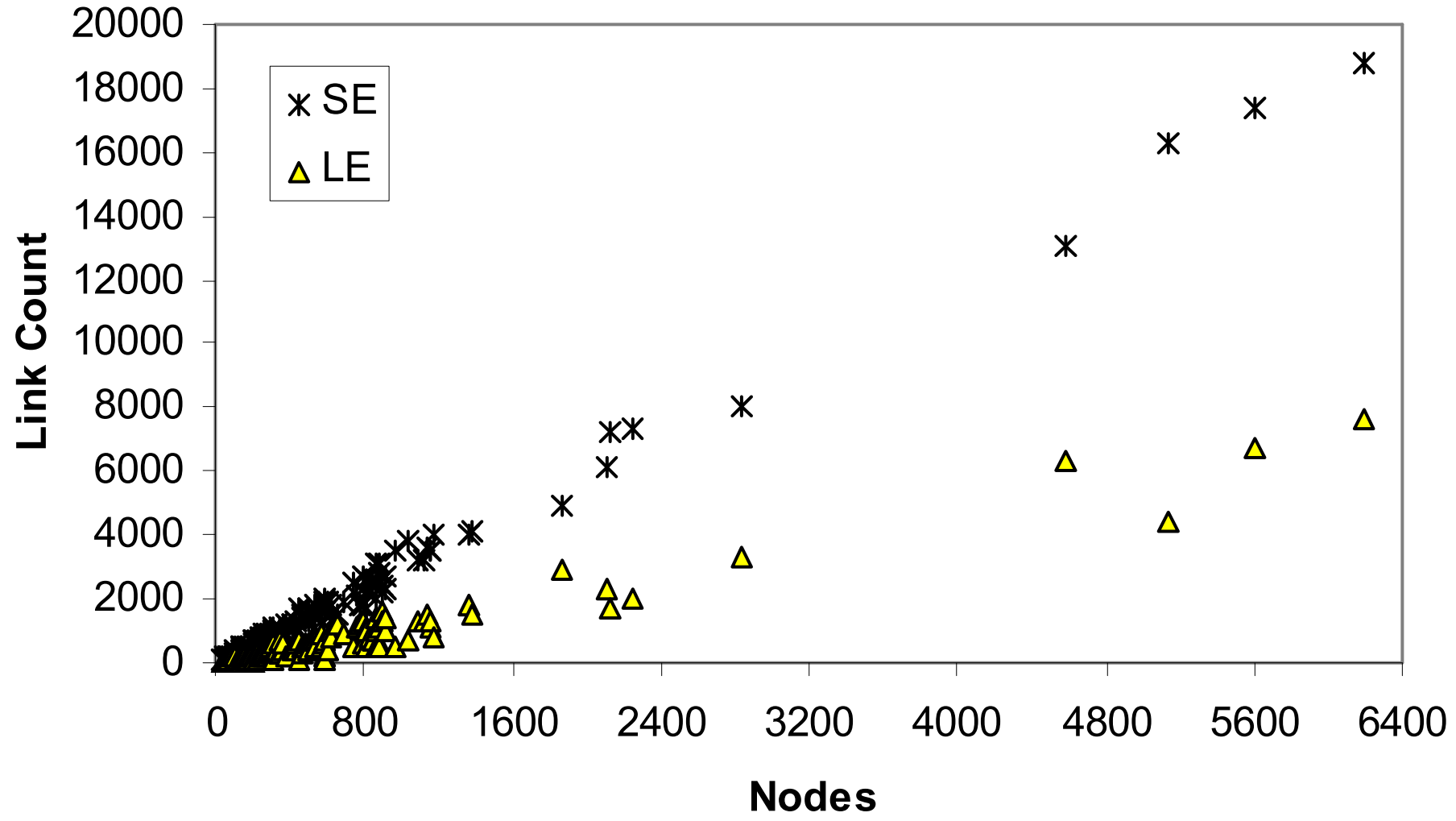

**Figure A4a Number of short-range (SE) and long-range (LE) links, by protein size for EVA132. The fraction of links in LE is 0.2611 ± 0.0986. The fraction of links in SE is 0.7389 ± 0.0986.**

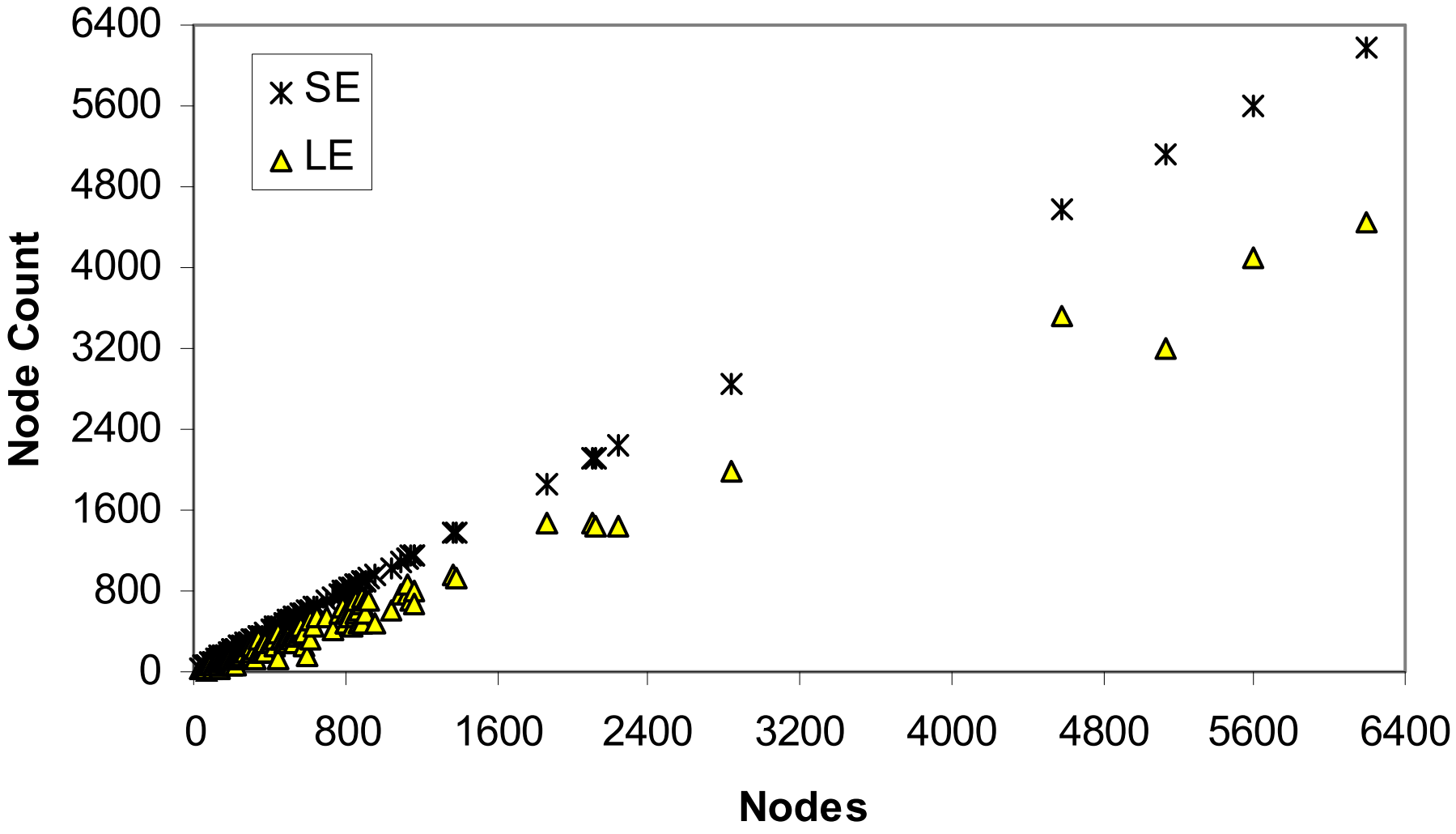

**Figure A4b Number of nodes which are endpoints for links in SE (SE_nodes) and nodes which are endpoint for links in LE (LE_nodes), by protein size for EVA132. The fraction of nodes in LE_nodes is 0.6348 ± 0.1367.The fraction of nodes in SE_nodes is 1.0000 ± 0.0004.**

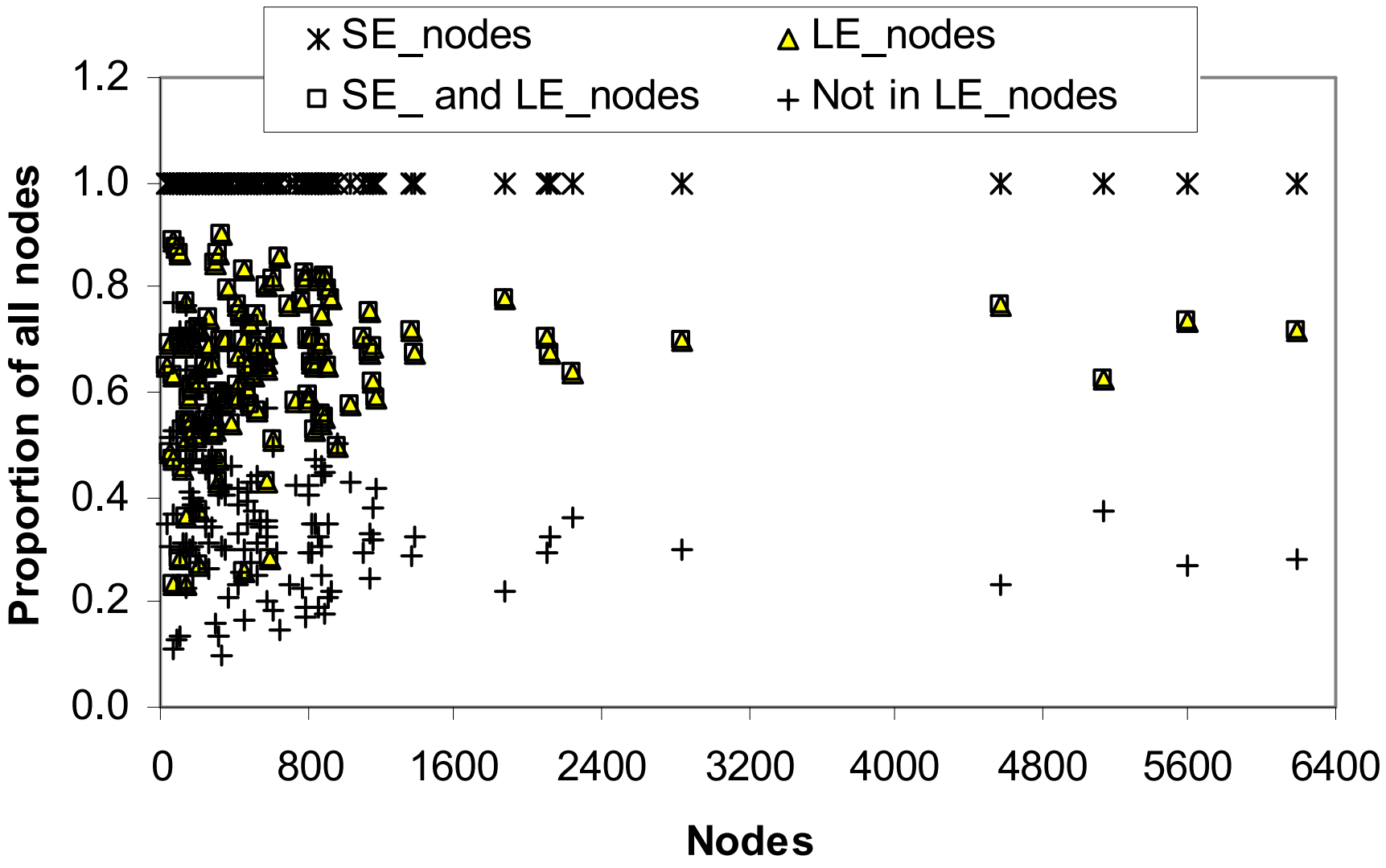

**Figure A4c Sizes of different subsets of nodes as a proportion of all nodes for EVA132 PCNs. On average, 0.6347 ± 0.1367 of all nodes are in both SE_nodes and LE_nodes, while 0.3652 ± 0.1367 of all nodes are not in LE_nodes.**

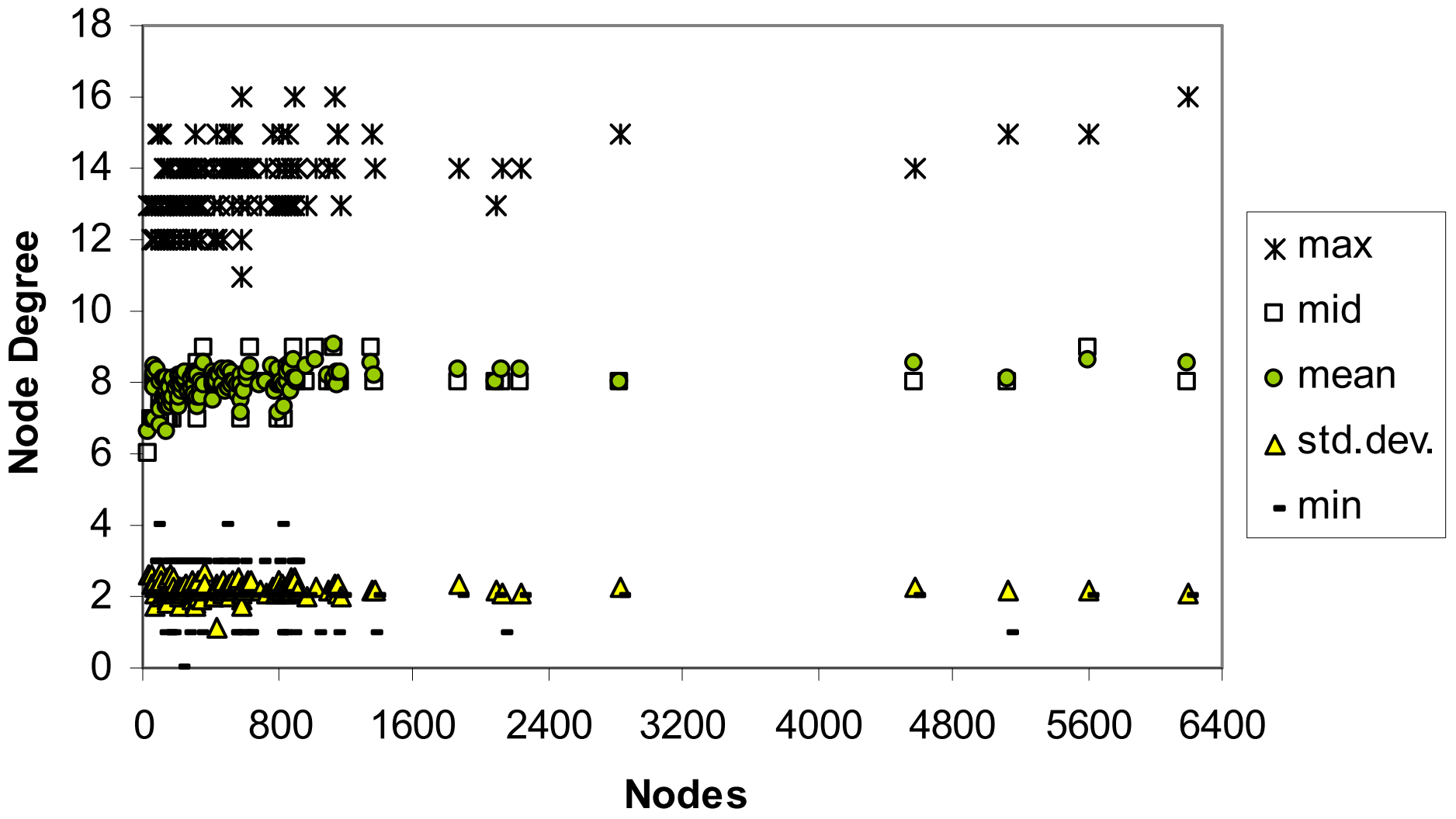

**Figure A5a Node degree summary statistics for EVA132 PCNs. Mean node degree average at 7.9147 with a std. dev. of 0.4300.**

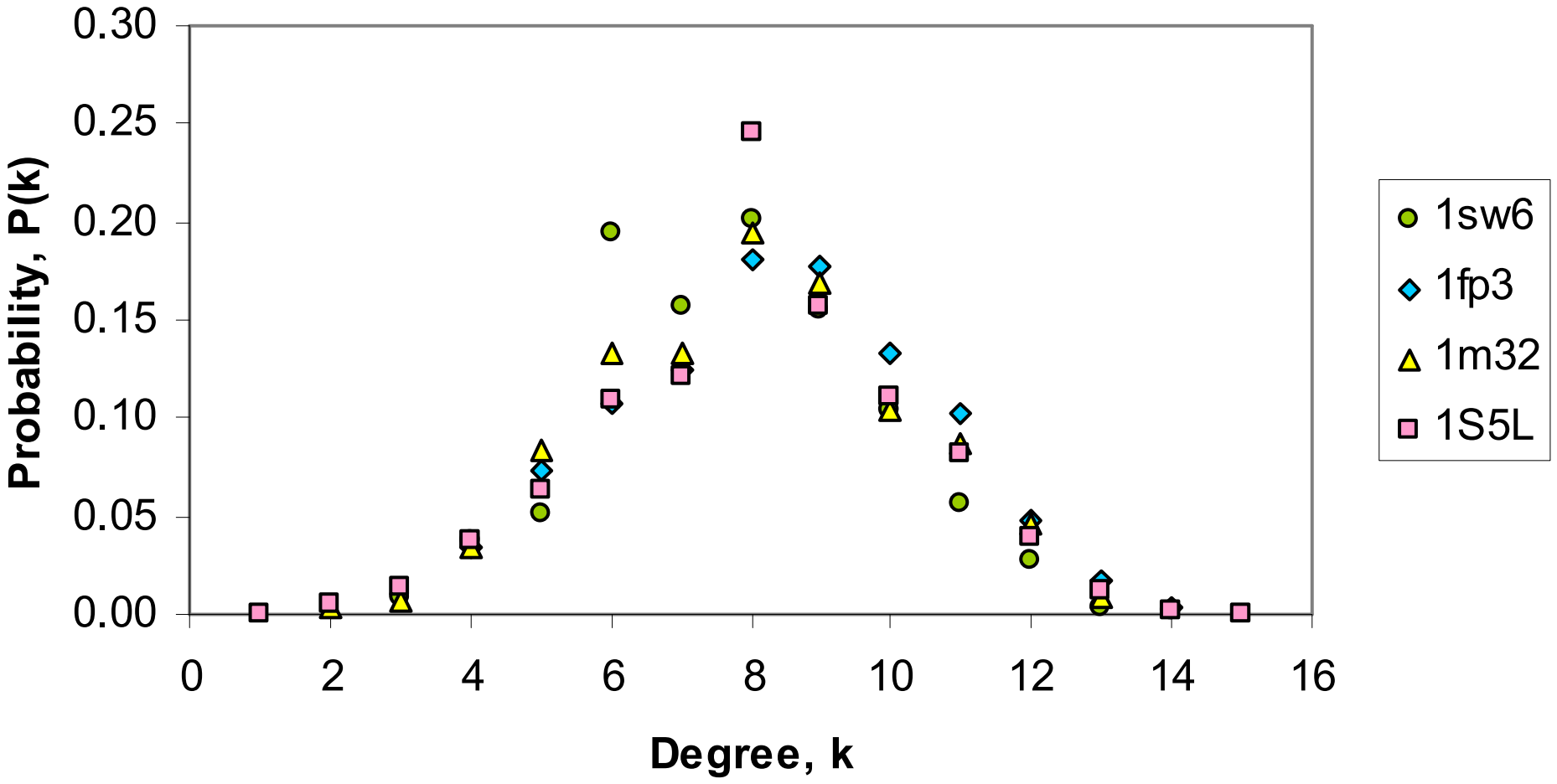

**Figure A5b Node degree distributions for four PCNs in the EVA132 dataset. The respective PCNs of 1sw6, 1fp3, 1m32 and 1S5L has 508, 804, 2103 and 5130 nodes respectively.**

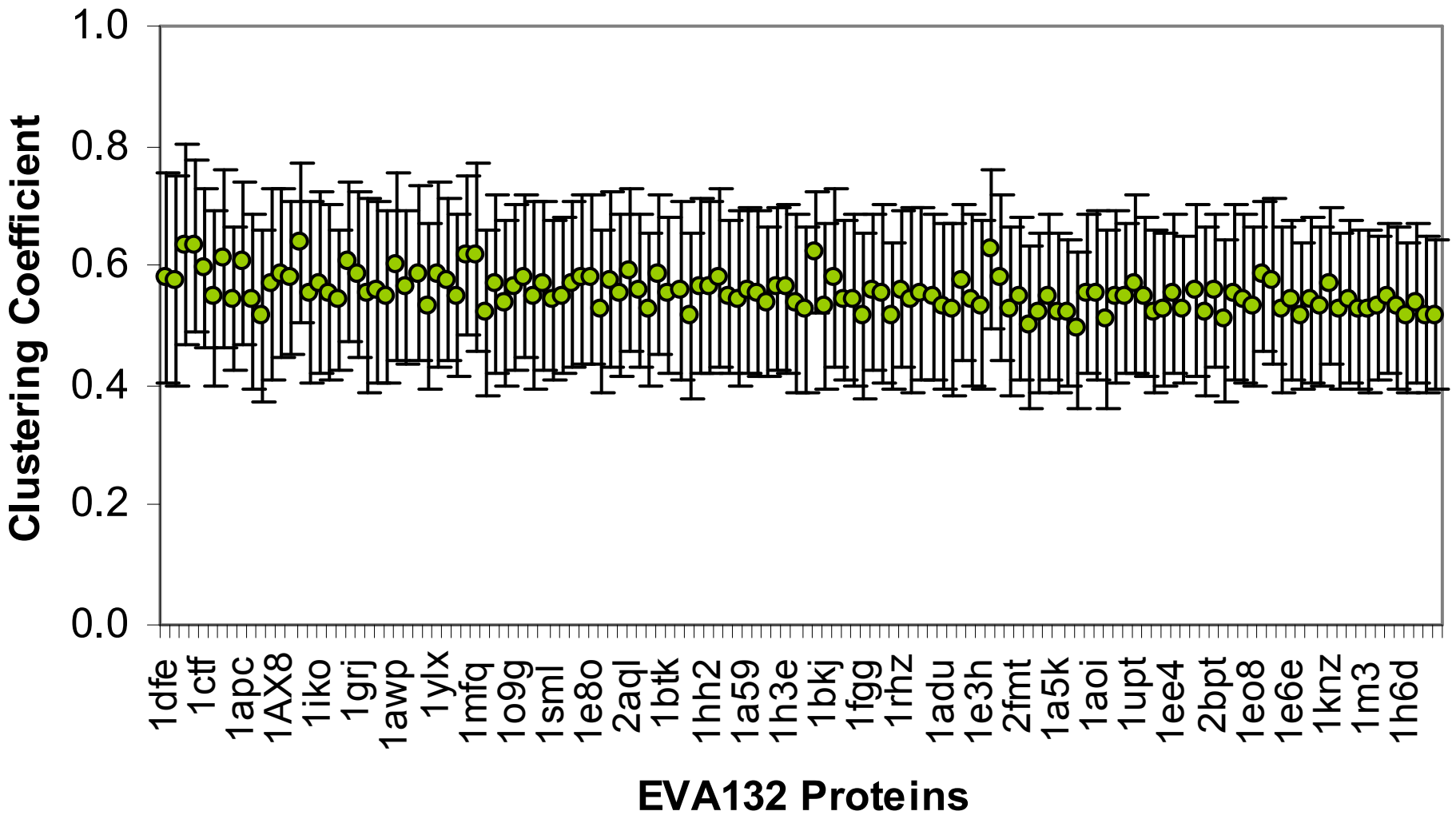

**Figure A6 Clustering coefficients ± standard deviation. Mean is 0.5538 with a std. dev. of 0.0292.**

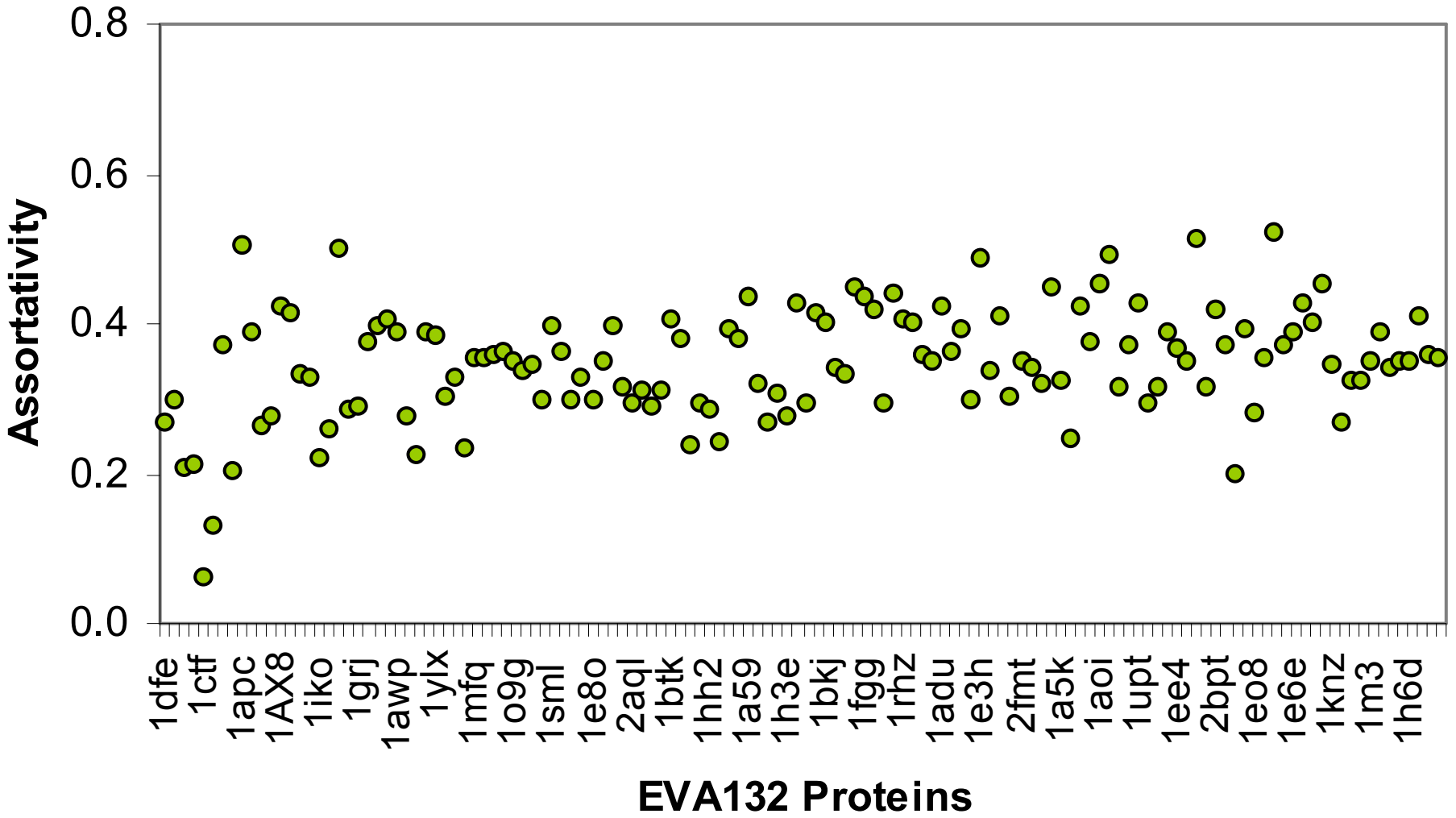

**Figure A7 Assortativity values. Mean is 0.3457 with a std. dev. of 0.0743.**

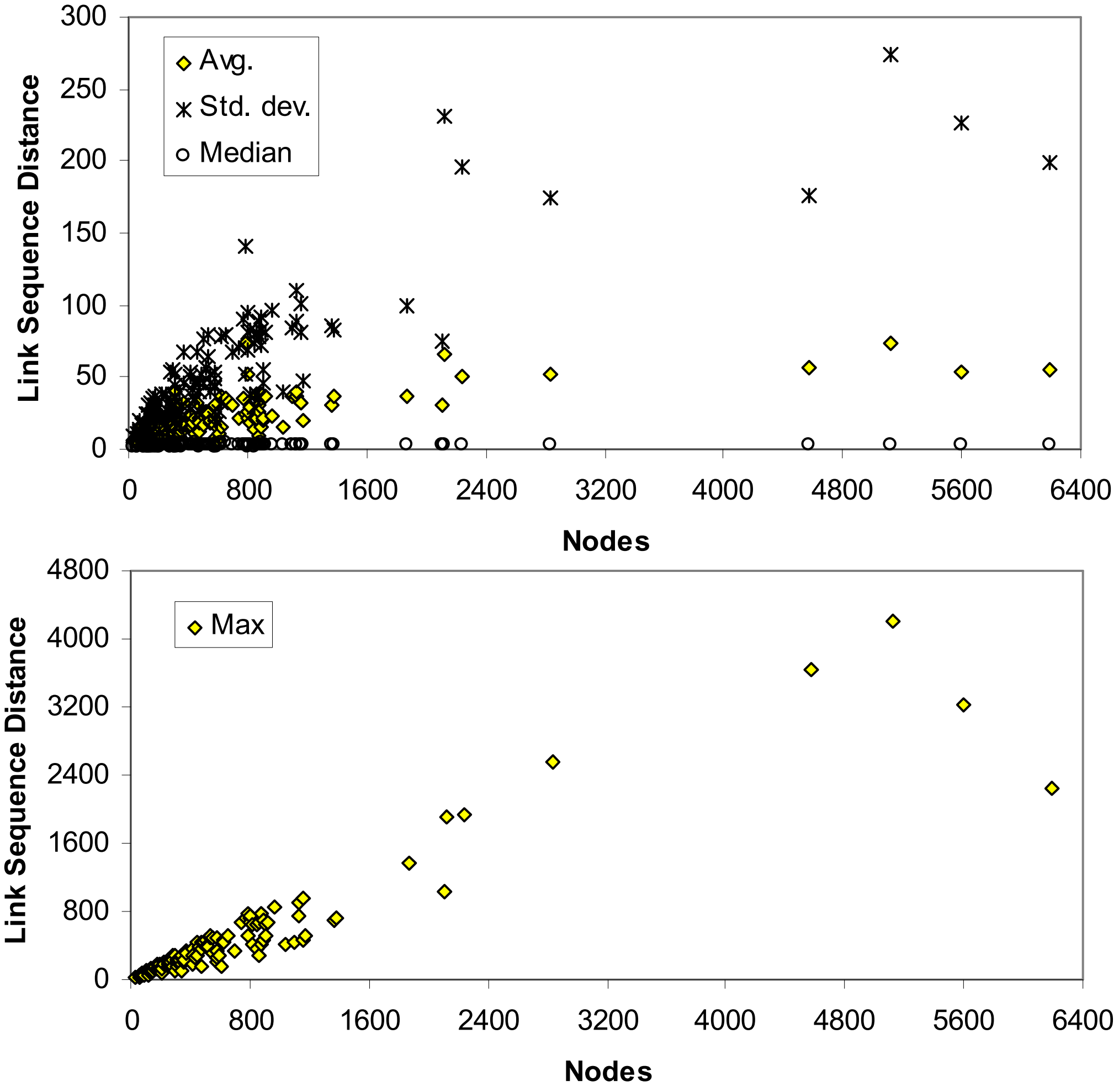

**Figure A8 Link sequence distance summary statistics for EVA132 PCNs indicate right skewed distributions for link sequence distance.**

## Appendix B

Additional material for Figure 19 (section 11).

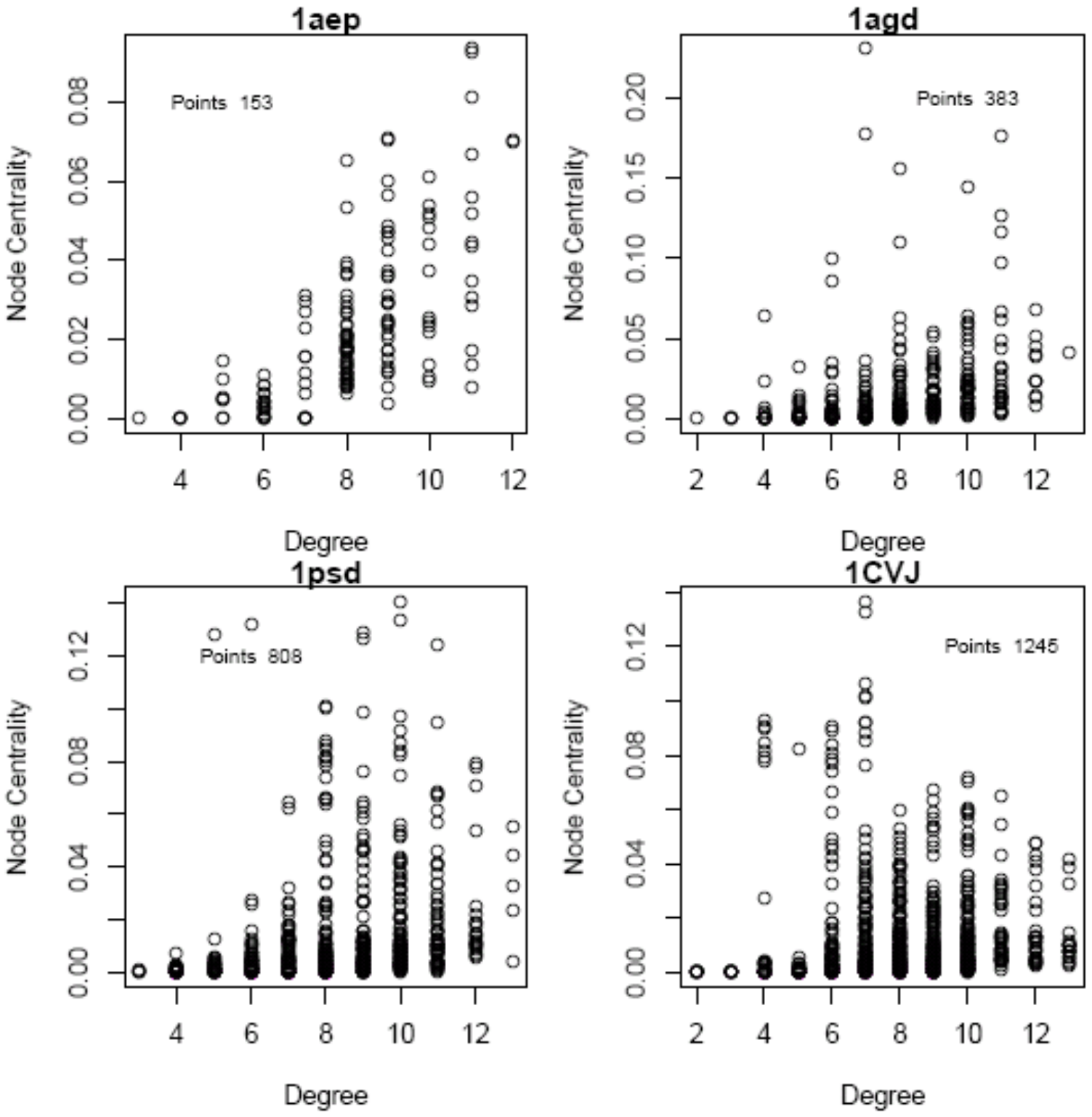

**Figure B1 Degree-node centrality scatter plots for four GH64 PCNs.**

Correlation coefficients are given in the table below. Both methods show weakening positive correlation with increase in N. We did not report Spearman's rho values as their significance could not be computed due to ties.

| Correlation method | 1aep | 1agd | 1psd | 1CVJ |
|---|---|---|---|---|
| Pearson | 0.6691 | 0.3112 | 0.3010 | 0.1222 |
| Kendall's tau | 0.5689 | 0.4661 | 0.4114 | 0.3616 |

All correlation coefficients are significant at the 95% confidence level.

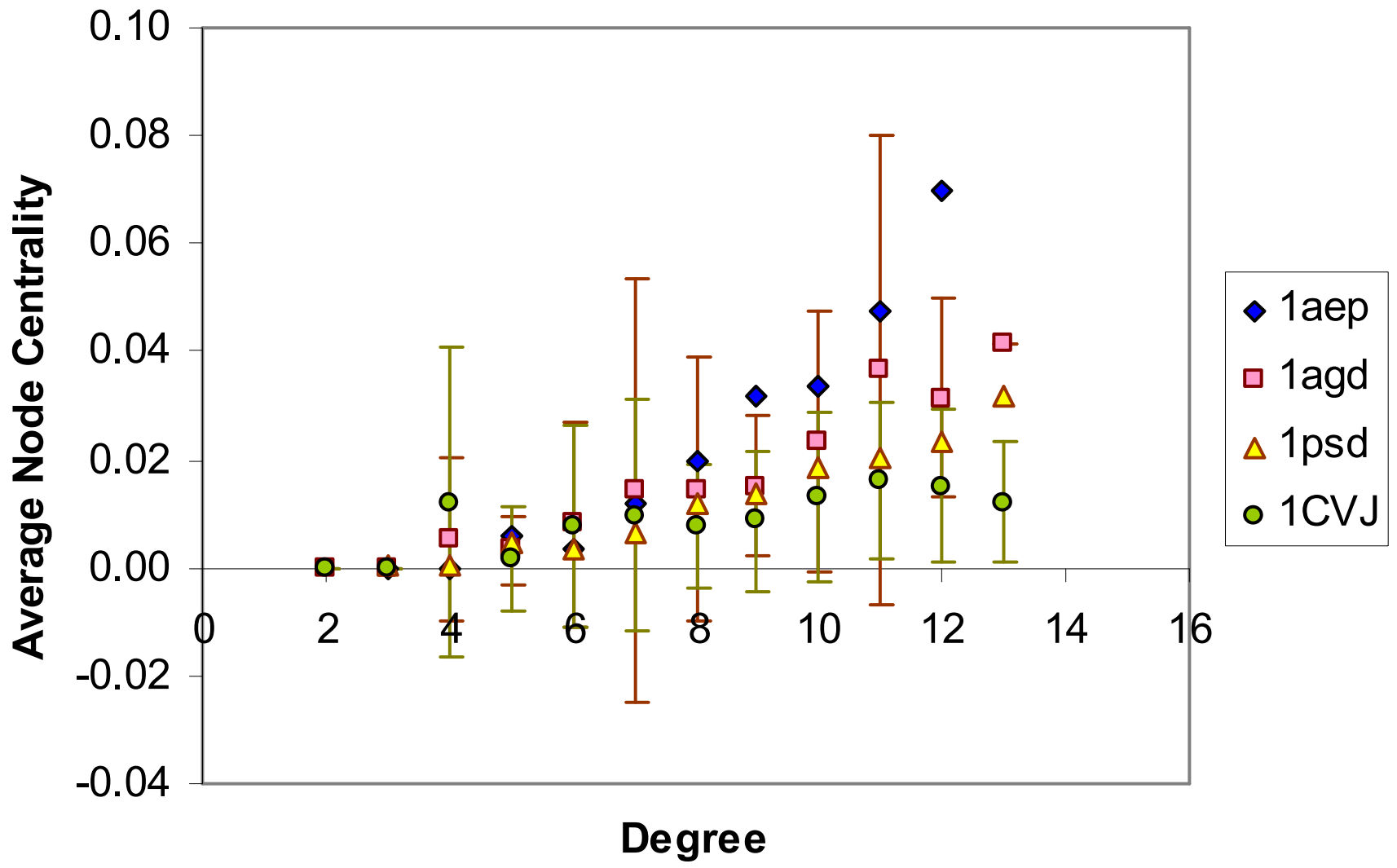

**Figure B2 Node centrality for nodes with degree k are averaged and reported with their standard deviations. The relationship is marginally positive and grows weaker with increase in protein size. The large standard deviations point to wide dispersion in node centrality values, as seen in the scatter plots in Figure B1.**

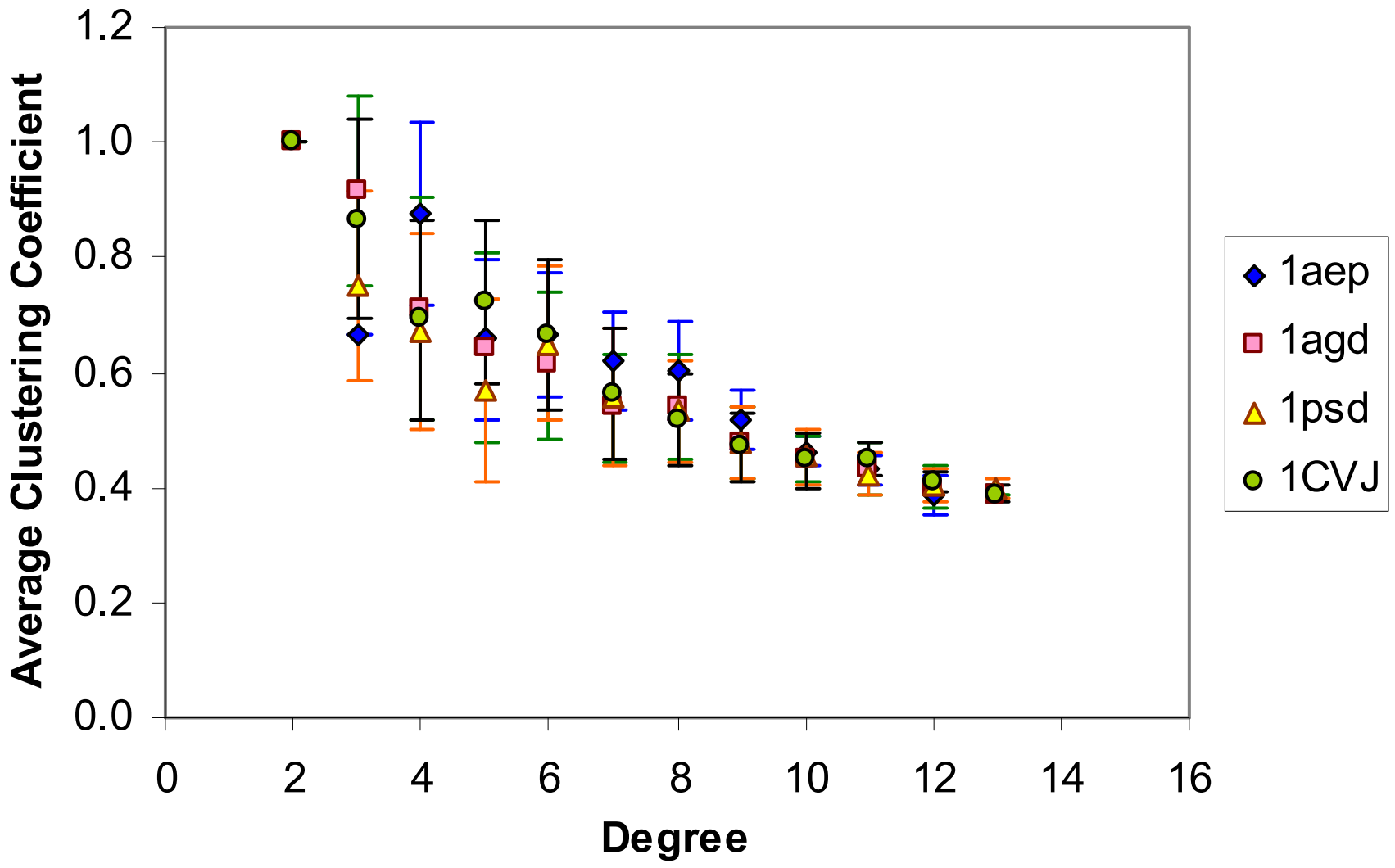

**Figure B3 Some evidence of hierarchical organization in PCNs.**